\documentclass[aps,prb,superscriptaddress,amsfonts,amsmath,amssymb,showpacs,floatfix,reprint,longbibliography,footinbib]{revtex4-2}

\usepackage{natbib}
\usepackage[english]{babel}
\usepackage{letltxmacro}
\usepackage{latexsym}
\LetLtxMacro{\ORIGselectlanguage}{\selectlanguage}
\makeatletter
\DeclareRobustCommand{\selectlanguage}[1]{%
 \@ifundefined{alias@\string#1}
 {\ORIGselectlanguage{#1}}
 {\begingroup\edef\x{\endgroup
 \noexpand\ORIGselectlanguage{\@nameuse{alias@#1}}}\x}%
}
\newcommand{\definelanguagealias}[2]{%
 \@namedef{alias@#1}{#2}%
}
\makeatother
\definelanguagealias{en}{english}
\definelanguagealias{English}{english}
\usepackage{graphicx}
\usepackage{amsmath}
\usepackage{amsfonts}
\usepackage{amssymb}
\usepackage{bm}
\usepackage{color}
\usepackage[percent]{overpic}
\usepackage{soul} 
\usepackage{amssymb}
\usepackage{wasysym}
\usepackage{dsfont}
\usepackage{float}
\usepackage[mathscr]{euscript}
\usepackage{lipsum}
\usepackage{hyperref}
\hypersetup{
 pdfstartview={FitH}, 
 colorlinks=true, 
 linkcolor=blue, 
 citecolor=blue, 
 filecolor=magenta, 
 urlcolor=blue 
}

\newcommand{\pdagger}{\phantom{\dagger}}
\newcommand{\be}{\begin{equation}}
\newcommand{\ee}{\end{equation}}
\newcommand{\bea}{\begin{eqnarray}}
\newcommand{\eea}{\end{eqnarray}}

\newcommand{\mc}{\mathcal}

\newcommand{\vect}[1]{\boldsymbol{#1}}

\usepackage{graphicx}
\usepackage[colorinlistoftodos]{todonotes}
\usepackage{verbatim}
\usepackage[normalem]{ulem}
\usepackage{enumitem}
\usepackage{multirow}

\begin{document}
\title{Gapped and gapless quantum spin liquids on the ruby lattice}

\author{Atanu Maity}
\affiliation{Institut f\"ur Theoretische Physik und Astrophysik and W\"urzburg-Dresden Cluster of Excellence ctd.qmat, Julius-Maximilians-Universit\"at W\"urzburg, Am Hubland, Campus S\"ud, W\"urzburg 97074, Germany}
\affiliation{Department of Physics and Quantum Centre for Diamond and Emergent Materials (QuCenDiEM), Indian Institute of Technology Madras, Chennai 600036, India}
\author{Rhine Samajdar}
\email{rhine\_samajdar@princeton.edu}
\affiliation{Department of Physics, Princeton University, Princeton, New Jersey 08544, USA}
\affiliation{Princeton Center for Theoretical Science, Princeton University, Princeton, New Jersey 08544, USA}
\author{Yasir Iqbal}
\email{yiqbal@physics.iitm.ac.in}
\affiliation{Department of Physics and Quantum Centre for Diamond and Emergent Materials (QuCenDiEM), Indian Institute of Technology Madras, Chennai 600036, India}

\begin{abstract}
The ruby lattice has been the subject of much interest recently due to its realization in Rydberg atom arrays, where its rich variety of frustrated interactions gives rise to topologically ordered quantum spin liquids. Similarly, numerical studies of ruby-lattice spin models, with both isotropic and anisotropic interactions, have provided evidence of gapped and gapless spin-liquid ground states with different low-energy gauge structures. Motivated by these findings, we perform a projective symmetry group (PSG) classification of U(1) and $\mathbb{Z}_{2}$ fermionic spinon mean-field theories---respecting space-group and time-reversal symmetries---for $S=1/2$ spins. We obtain a total of 50 U(1) and 64 $\mathbb{Z}_{2}$ PSGs, and upon restricting their realization via mean-field \textit{Ans\"atze} with up to second-nearest-neighbor singlet amplitudes (relevant to the models studied here), only 8 U(1) and 18 $\mathbb{Z}_{2}$ distinct states are obtained. We present the singlet fields for all \textit{Ans\"atze} up to third-nearest-neighboring bonds and discuss their spinon dispersions as well as their dynamical spin structure factors. Building on this information, we also obtain the phase diagram of the Heisenberg model in the presence of first- ($J_{1}$), second- ($J_{1}'$), and third- ($J_{2}$) neighbor antiferromagnetic couplings within a self-consistent mean-field approximation.
\end{abstract}

\date{\today}
\maketitle
\section{Introduction}
Lattice geometries which induce frustration for systems of antiferromagnetically interacting spins are much sought after due to the possibilities they offer in realizing enigmatic quantum spin liquids (QSLs)~\cite{Savary-2017}. The spin $S$\,$=$\,$1/2$ antiferromagnetic Heisenberg models on the kagome and triangular lattices serve as canonical examples in this regard, where there is mounting evidence in support of an exotic QSL that is putatively identified to be a U(1) Dirac spin liquid~\cite{Iqbal-2013,He-2017,Iqbal-2016,Hu-2019}. In a similar spirit, another highly frustrated Archimedean semiregular tiling, the ruby (also called bounce) lattice (Fig.~\ref{fig:fig1}), has recently garnered much interest owing to its implementation in synthetic platforms, with the scope of engineering novel phases of quantum matter. Indeed, recent experiments on a programmable quantum simulator based on Rydberg atoms arrayed on the ruby lattice have demonstrated the preparation of a topologically ordered $\mathbb{Z}_2$ QSL~\cite{Semeghini-2021,Giudici-2022,Samajdar-2023,tarabunga2022gauge}, which corresponds to a resonating valence bond (RVB) phase of an underlying quantum dimer model \cite{Samajdar.2021,Veressen-2021}. Moreover, there is also an intimate connection between phases of anisotropic spin Hamiltonians on the ruby lattice and those of the Kitaev honeycomb model~\cite{Veressen-2022}. Interestingly, the anisotropic Kitaev model on the ruby lattice was shown to host two gapless QSLs~\cite{Jahromi-2016,Jahromi-2018} and a $\mathbb{Z}_{2}\times\mathbb{Z}_{2}$ topologically ordered gapped QSL~\cite{Kargarian-2010,Bombin-2009}, while the $S=1/2$ Heisenberg antiferromagnet on the same lattice has long been considered a candidate for a nonmagnetic ground state due to its low ($\mc{Z}$\,$=$\,$4$) coordination number. For the latter, descriptions based on topological RVB wave function, coupled-cluster, and exact diagonalization approaches have not been able to unambiguously resolve the delicate competition between conventional magnetic and quantum paramagnetic ground states~\cite{Farnell2011,Farnell2014,Richter2004,Jahromi-2020,Nakano2026}. However, a recent state-of-the-art variational infinite tensor-network calculation has ruled out the presence of magnetic order and provided strong evidence in favor of a gapless symmetric QSL ground state of the spin-$1/2$ Heisenberg antiferromagnet on the ruby lattice~\cite{Schmoll-2024}. In particular, inclusion of frustrated further-neighbor interactions~\cite{Ghosh2025,Mello2026,Vu2025} or mixed ferromagnetic and antiferromagnetic couplings~\cite{Ebert2026} has been shown to yield unconventional nonmagnetic phases. Finally, the ruby
lattice also finds solid-state realizations in the layered material Bi$_{14}$Rh$_3$I$_9$, wherein the bismuth ions form a ruby structure~\cite{Rasche-2013,Pauly-2015}, as well as in the layered transition-metal fluorides CsBaFe$_3$F$_{12}$ and CsBaCr$_3$F$_{12}$~\cite{Jeschke2026}.

These findings naturally invite us to ask the questions: what are the symmetry-allowed QSL states on the ruby lattice, how can we identify them from spectroscopic signatures, and what are their connections to the QSLs on the maple-leaf lattice~\cite{Sonnenschein-2024}? A powerful framework to answer these questions, by systematically classifying spinon mean-field theories of QSLs with different low-energy gauge groups, is provided within a parton representation by the method of projective symmetry groups~\cite{Wen-2002,Wenbook}. This formalism has been applied to a wide range of two- and three-dimensional lattices. In two dimensions, PSG classifications have been carried out for the canonical frustrated geometries---the triangular~\cite{Wang-2006,Lu-2016a} and kagome~\cite{Lu-2011a,Yang-2012,Messio-2013,Bieri-2015,Bieri-2016} lattices---as well as for the honeycomb~\cite{Lu-2011b}, square~\cite{Yang-2016}, star~\cite{Yang-2010,Choy-2009}, stuffed honeycomb~\cite{Sahoo-2020}, square-octagon~\cite{Maity-2023}, trellis~\cite{Chatterjee-2026}, Shastry-Sutherland~\cite{Liu-2024,maity2026}, and maple-leaf~\cite{Sonnenschein-2024} lattices, with chiral spin liquids in particular receiving systematic attention~\cite{Messio-2013,Bieri-2015,Bieri-2016,Benedikt-2022}. In three dimensions, the framework has been extended to the simple cubic, body-centered cubic, face-centered cubic~\cite{Sonnenschein-2020}, pyrochlore~\cite{Benedikt-2022,Liu-2019,Liu-2021,liu2026,Chern-2022}, hyperkagome~\cite{Lawler-2008,Huang-2017,Jin-2020}, hyperhoneycomb~\cite{Huang-2018}, hyper-hyperkagome~\cite{Chern-2021}, diamond~\cite{Chauhan-2023}, and trillium~\cite{Li-2024} lattices, while symmetry-protected gapless $\mathbb{Z}_{2}$ spin liquids have been classified in a lattice-independent setting~\cite{Lu-2018}. Building on these classifications, the framework has met with considerable success in identifying the ground states of frustrated quantum spin models, particularly for the kagome~\cite{Iqbal-2011a,Iqbal-2011b,Iqbal-2012,Iqbal-2013,Iqbal-2014,Iqbal-2018_bk,Kiese-2023,Iqbal-2021} and triangular~\cite{Iqbal-2016} Heisenberg antiferromagnets, the $J^{}_{1}$--$J^{}_{2}$ Heisenberg model on the square lattice~\cite{Hu-2013}, and in the analysis of dynamical signatures of frustrated magnets~\cite{Ferrari-2019,Ferrari-2023}.

To this end, we employ the framework of projective symmetry groups~\cite{Wen-2002} for fermionic spinons to classify mean-field \textit{Ans\"atze} of fully symmetric $S$\,$=$\,$1/2$ QSLs with U(1) and $\mathbb{Z}_{2}$ low-energy gauge groups on the ruby lattice. We find a total of 50 U(1) and 64 $\mathbb{Z}_{2}$ distinct algebraic PSGs on the ruby lattice. Upon restricting the (singlet) mean-field \textit{Ans\"atze} to first- and second-neighbor amplitudes only, a total of 8 U(1) and 18 $\mathbb{Z}_{2}$ states can be realized, while if amplitudes up to third neighbors are included, a total of 18 U(1) and 22 $\mathbb{Z}_{2}$ distinct states are realizable. In light of tensor-network results~\cite{Schmoll-2024} which point to a gapless QSL in the Heisenberg model with purely antiferromagnetic couplings, our treatment here principally focuses on singlet QSLs. We identify the gapless and gapped spin liquids in the parameter regime of interest, presenting their spinon band structures and dynamical spin structure factors at the mean-field level. Furthermore, we obtain the global phase diagram of the $J_{1}$-$J_{1}'$-$J_{2}$ model within a self-consistent mean-field approximation and identify parameter regimes hosting extended regions of two distinct QSL phases.

\section{Model and methods}
\label{sec:method}
\begin{figure}[t]	\includegraphics[width=1.0\linewidth]{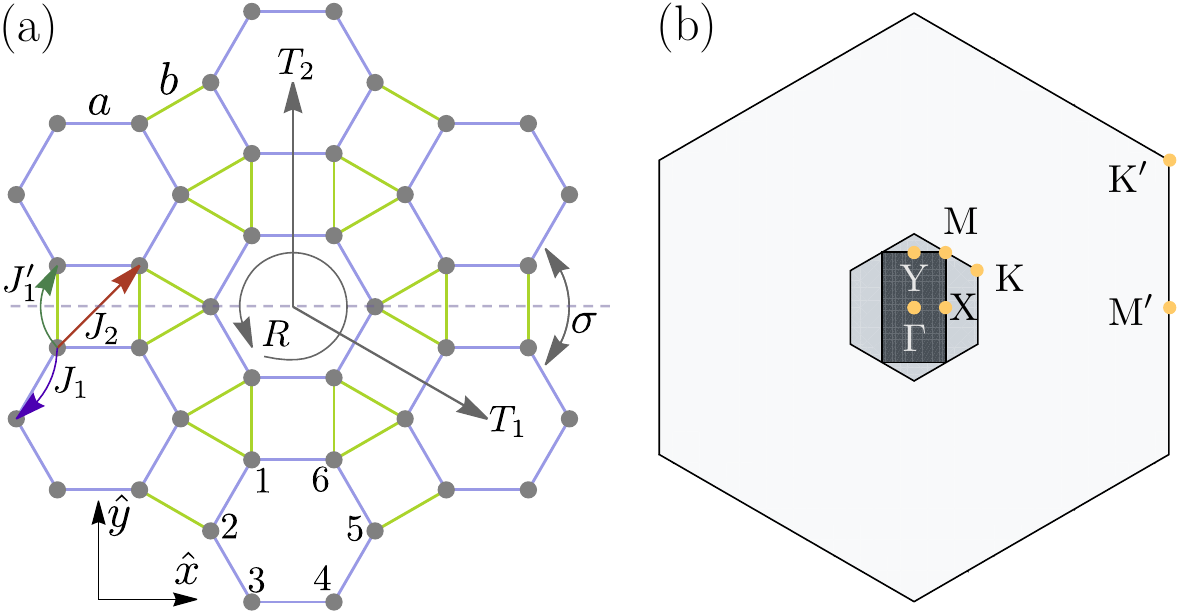}
	\caption{(a) Illustration of the ruby lattice, marking the underlying triangular Bravais lattice, the six sublattice indices, the exchange interactions, and the space-group symmetry elements. (b) The extended Brillouin zone (white) and its relation to the first Brillouin zone (light gray) for the choice $b/a=1/\sqrt{3}$. The high-symmetry points in the extended Brillouin zone are given by $\mathrm{M}'=(2\pi/a,0)$ and $\mathrm{K}'= (2\pi/a,2\pi/\sqrt{3}a)$, while those inside the first Brillouin zone include $\Gamma = (0,0)$, $\mathrm{M}=(\pi/4a,\sqrt{3}\pi/4a)$, and $\mathrm{K}=(\pi/2a,\pi/2\sqrt{3}a)$. In the reduced Brillouin zone (dark gray) defined for unit-cell doubling, the additional high-symmetry points are $\mathrm{X}=(\pi/4a,0)$, and Y$=(0,\sqrt{3}\pi/4a)$.}
	\label{fig:fig1}
\end{figure}

To begin, we consider the $S=1/2$ Heisenberg model, with nearest- ($J_1$) and next-nearest ($J_2$) neighbor antiferromagnetic couplings on the ruby lattice
\begin{equation}
\hat{\mathcal{H}}=J^{}_1\sum_{\langle i,j\rangle}\hat{\mathbf{S}}_{i} \cdot \hat{\mathbf{S}}_{j} +J^{}_2\sum_{\langle\langle i,j\rangle \rangle}\hat{\mathbf{S}}_{i} \cdot \hat{\mathbf{S}}_{j},
\label{eq:ham}
\end{equation}
where $\hat{\mathbf{S}}_i$ denotes the SU(2) spin operator acting on the $S=1/2$ representation at site $i$. The ruby lattice consists of alternating square and triangular plaquettes centered around a hexagonal motif, as illustrated in Fig.~\ref{fig:fig1}. One can also consider a generalization of this lattice, where the square plaquettes are deformed into rectangles while leaving the symmetry unchanged. In fact, this version of the ruby lattice is precisely what is obtained by constructing the medial lattice (or line graph) of the kagome lattice. In this case, nearest-neighbor interactions along the two axes of each rectangle can be manifestly inequivalent,
\begin{equation}
\hat{\mathcal{H}}=J^{}_1\sum_{\langle i,j\rangle_{1}}\hat{\mathbf{S}}_{i} \cdot \hat{\mathbf{S}}_{j}+J'_1\sum_{\langle i,j\rangle_{2}}\hat{\mathbf{S}}_{i} \cdot \hat{\mathbf{S}}_{j}+J^{}_2\sum_{\langle i,j\rangle_{3}}\hat{\mathbf{S}}_{i} \cdot \hat{\mathbf{S}}_{j},
\label{eq:mod-ham}
\end{equation}
as sketched in Fig.~\ref{fig:fig1}. Hereafter, we will refer to sites connected by bonds with couplings $J^{}_1$, $J'_1$, and $J^{}_2$ as first-, second-, and third-nearest neighbors, respectively, even for the case when $J^{}_1 = J'_1$.

Our interest here lies in identifying candidate QSL ground states of this model motivated by a recent tensor-network study~\cite{Schmoll-2024} claiming a gapless spin liquid phase of the isotropic ($J^{}_1=J'_1, J^{}_2 = 0$) model. However, due to the absence of any local physical order parameters in a QSL phase, it is challenging to construct a theoretical description of these states in terms of the native spin degrees of freedom. Instead, the properties of the state are better understood in terms of the emergent fractionalized quasiparticles of the QSL at low energies; this motivates the fermionic parton construction~\cite{Abrikosov-1965,Affleck-1988,Affleck-1989} that we now outline. 

\subsection{Fermionic parton construction}
\label{sec:afmft}

In the SU(2) parton formalism, the physical spin operators on each site are represented in terms of two flavors of complex spin-$1/2$ charge-neutral fermionic quasiparticles (called {\it spinons}), which are commonly termed Abrikosov fermions. Labeling the two species of partons with the pseudospin index $\sigma=\{\uparrow,\downarrow$\}, we can write
\begin{equation}
\label{eq:abrikosov}
\hat{S}^{\alpha}_{i}=\frac{1}{2}\hat{f}^\dagger_{i\sigma}\tau^{\alpha}_{\sigma\sigma^\prime}\hat{f}^{\pdagger}_{i\sigma^\prime},
\end{equation}
where $\tau^\alpha$ ($\alpha=1,2,3$) denote the Pauli matrices, and repeated indices are summed over. While the operators on both sides of Eq.~\eqref{eq:abrikosov} follow the same SU(2) commutation relations, this construction enlarges the on-site Hilbert space from $\mathbb{C}^{2}$ to a four-dimensional space. Hence, to remain within the physical subspace, one must impose the additional local constraint~\cite{Baskaran-1987,Baskaran-1988}
\begin{align}
\label{eq:constraint} \hat{f}_{i\sigma}^{\dagger}\hat{f}^{\pdagger}_{i\sigma} = 1, \; \hat{f}^{\pdagger}_{i\sigma}\hat{f}^{\pdagger}_{i\sigma'}\epsilon^{\pdagger}_{\sigma\sigma'} = 0.
\end{align}
This ensures that there is exactly one fermion per site, with the second constraint actually being a consequence of the first one.

In the fermionic representation, the spin operator in Eq.~\eqref{eq:abrikosov} can be reorganized in terms of a doublet of spinors $\hat{\psi}_i=(\hat{\phi}_i,\hat{\bar{\phi}}_i)$, with $\hat{\phi}_i=(\hat{f}_{i,\uparrow},\hat{f}_{i,\downarrow})^\textsc{T}$ and $\hat{\bar{\phi}}_i=(\hat{f}^\dagger_{i,\downarrow},-\hat{f}^\dagger_{i,\uparrow})^\textsc{T}$, as~\cite{Affleck-1988b}
\begin{equation}
\label{eq:abrikoshov_doublet}
\hat{S}^{\alpha}_{i}=\frac{1}{2}\text{Tr}\left[\hat{\psi}^\dagger_i\tau^\alpha\hat{\psi}^{\pdagger}_i\right].
\end{equation}
The most striking implication of such a description is an \textit{emergent} local gauge symmetry which is not present in the original spin space. It is easy to see from Eq.~\eqref{eq:abrikoshov_doublet} that the spin operators remain invariant under a site-dependent transformation $\hat{\psi}_i\rightarrow\hat{\psi}_i W^{}_i$ with $W^{}_i\in$ SU(2). Furthermore, in this language, the constraint~\eqref{eq:constraint} can be compactly recast as
\begin{equation}\label{eq:constraint_doublet} \hat{\psi}^{\pdagger}_{i}\tau^{\alpha}\hat{\psi}^{\dagger}_{i}=0.
\end{equation}

Plugging the fermionized spin operators~\eqref{eq:abrikoshov_doublet} back into the Heisenberg model~\eqref{eq:ham}, we arrive at a Hamiltonian which is quartic in fermion operators. The equivalence between the quartic Hamiltonian and the original Heisenberg Hamiltonian is valid only if the constraint~\eqref{eq:constraint} is imposed exactly. For this model to be solvable, however, one proceeds with a mean-field treatment, in which we first perform a Hubbard-Stratonovich decomposition in terms of the fields $\hat{U}_{ij}=\hat{\psi}^\dagger_i\hat{\psi}^{\pdagger}_j$ and $\hat{U}^{(\alpha)}_{ij}=\hat{\psi}^\dagger_i\tau^\alpha\hat{\psi}^{\pdagger}_j$, thereby bringing the Hamiltonian to a quadratic form. Among these fields, $\hat{U}_{ij}$ is invariant under global SU(2) spin rotations---which act as $\hat{\psi}_i$\,$\rightarrow$\,$G\hat{\psi}_{i}$, $G$\,$\in$ SU(2)---thus representing a singlet field, while $\hat{U}^{(\alpha)}_{ij}$ correspond to triplet fields. Since, in this work, we are interested in QSL ground states of a purely antiferromagnetic and SU(2)-spin-rotation invariant Hamiltonian~\eqref{eq:ham}, we restrict our treatment to singlet fields only. When the Hamiltonian lacks spin-rotation symmetry or has competing ferromagnetic couplings, it is important to take the triplet terms into account; these are worked out and assessed in the context of an anisotropic model in a companion work~\cite{mis}.
A generic singlet field $\hat{U}_{ij}$ is composed of two building blocks, to wit, a hopping field $\hat{\chi}^{\pdagger}_{ij}=\hat{f}^\dagger_{i,\uparrow}\hat{f}^{\pdagger}_{j,\uparrow}+\hat{f}^\dagger_{i,\downarrow}\hat{f}^{\pdagger}_{j,\downarrow}$, and a pairing field $\hat{\Delta}_{ij}=\hat{f}_{i,\downarrow}\hat{f}_{j,\uparrow}-\hat{f}_{i,\uparrow}\hat{f}_{j,\downarrow}$:
\begin{eqnarray}
\hat{U}^{\pdagger}_{ij} =
\begin{bmatrix}
\hat{\chi}^{\pdagger}_{ij} & \hat{\Delta}^\dagger_{ij} \\
\hat{\Delta}^{\pdagger}_{ij} & -\hat{\chi}^\dagger_{ij}
\end{bmatrix}.
\label{eq:ansatz}
\end{eqnarray}

Now, we formally construct a mean-field theory by defining $u_{ij}=\langle\hat{U}_{ij}\rangle$, i.e., $\chi_{ij}=\langle\hat{\chi}_{ij}\rangle$ and $\Delta_{ij}=\langle\hat{\Delta}_{ij}\rangle$, on the links and considering 
 the constraints of Eq.~\eqref{eq:constraint_doublet} to be fulfilled on average, $\langle\hat{\psi}_{i}\tau^{\alpha}\hat{\psi}^{\dagger}_{i}\rangle=0 \,\forall~i$. The promotion of the link operators to their $c$-number expectation values, together with the imposition of the local single-occupancy constraint only on average, constitutes the mean-field \emph{approximation} that we adopt in the rest of this work. The incorporation of the constraint at a mean-field level requires the introduction of three site-dependent and time-independent Lagrange multipliers $a_\mu$. Putting these ingredients together, the zeroth-order mean-field Hamiltonian obtained for a generic antiferromagnetic Heisenberg Hamiltonian~\eqref{eq:ham} is
\begin{align}\label{eq:mf_ham}
 \hat{H}^{\pdagger}_{\textsc{mf}} &= \sum_{ i,j} \frac{3}{8} J^{\pdagger}_{ij} \left[\frac{1}{2}\text{Tr}\left(u_{ij}^{\dagger} u^{\pdagger}_{ij}\right) - \text{Tr}\left(\hat{\psi}^{\pdagger}_{i} u^{\pdagger}_{ij} \hat{\psi}^{\dagger}_{j} + \text{h.c.}\right) \right] \; \notag \\
 & + \sum_i a^{\pdagger}_{\mu} \text{Tr}\left(\hat{\psi}^{\pdagger}_{i} \tau^{\mu} \hat{\psi}^{\dagger}_{i} \right),
\end{align}
where we have retained only singlet fields. The expectation values $(u_{ij},a_\mu)$ define a so-called \textit{Ansatz} for QSL states. Note that as per Eq.~\eqref{eq:ansatz} above, the link field $u_{ij}$ can be reparametrized as 
\begin{equation}\label{eq:link_singlet}
 u^{\pdagger}_{ij}=\dot{\iota}\chi^0_{ij} \tau^0 + \chi^3_{ij} \tau^3+\Delta^1_{ij} \tau^1+\Delta^2_{ij} \tau^2,
\end{equation}
where $\tau^0$ is the $2\times2$ identity matrix, and $\chi^0_{ij}$, $\chi^3_{ij}$, $\Delta^1_{ij}$, $\Delta^2_{ij} \in \mathbb{R}$. Here, $\chi_{ij}$ and $\Delta_{ij}$ are related to these real parameters as $\chi_{ij}=\langle\hat{\chi}_{ij}\rangle=\dot\iota\chi^0_{ij} + \chi^3_{ij}$ and $\Delta_{ij}=\langle\hat{\Delta}_{ij}\rangle=\Delta^1_{ij} +\dot\iota\Delta^2_{ij}$, respectively. We remark that fluctuations beyond the mean-field saddle point can be partially incorporated by Gutzwiller projection, which restores the local one-fermion-per-site constraint exactly. For gapped $\mathbb{Z}_{2}$ spin liquids, in particular, the discrete invariant gauge group renders the gauge sector itself gapped, so that gauge fluctuations mediate only short-range interactions between spinons and the mean-field description is parametrically stable beyond the saddle-point level. The main goal of the projective symmetry group approach described in the following section is to systematically classify quadratic spinon Hamiltonians of the form of Eq.~\eqref{eq:mf_ham} with desired symmetries.

\subsection{Projective symmetry group}
As mentioned above, the representation of spin operators in terms of fermions introduces a gauge redundancy in spinon space leading to additional freedom in how lattice, time-reversal, and spin symmetries act in the spinon Hilbert space. More concretely, a given symmetry transformation $\mathcal{S}$ can now be accompanied by an SU(2) gauge transformation $W_{\mathcal{S}}$ provided that $W_{\mathcal{S}}$ respects the same algebraic relations as obeyed by the symmetry transformations. 

To make this manifest, we observe that the mean-field Hamiltonian~\eqref{eq:mf_ham} remains invariant under a local SU(2) gauge transformation acting as
\begin{equation}
 \hat{\psi}^{\pdagger}_i\rightarrow \hat{\psi}^{\pdagger}_i W^{\pdagger}_i, \; u^{\pdagger}_{ij}\rightarrow W^\dagger_i u^{\pdagger}_{ij}W^{\pdagger}_j,\;a^{\pdagger}_\mu\tau^\mu \rightarrow a^{\pdagger}_\mu W^\dagger_i\tau^\mu W^{\pdagger}_i.
 \label{eq:gauge_symmetry}
\end{equation}
As a result, two \textit{Ans\"atze} $u_{ij}$ and $u'_{ij}=W^\dagger_iu^{\pdagger}_{ij}W^{\pdagger}_j$, which are related by such a gauge transformation, are simply different labels for the same physical QSL state~\cite{Wen-2002}. This property carries deep implications for how symmetries act in the spinon Hilbert space. For instance, consider an element $\mathcal{O}$ of the space group of a given lattice. When acting on a given \textit{Ansatz}, $\mathcal{O}(u_{ij})=u_{\mathcal{O}(i)\mathcal{O}(j)}$, generically, $u_{ij}\neq u_{\mathcal{O}(i)\mathcal{O}(j)}$. It would thus appear, at first sight, that the \textit{Ansatz} is not invariant under the lattice symmetry operation $\mathcal{O}$. This is indeed true if one only considers the symmetries to act linearly. However, the gauge degrees of freedom provide us with a way to reinstate the symmetry: if one associates a local SU(2) gauge transformation $W^{}_i$ such that 
\begin{align}\label{eq:sym_con^{}_gauge}
W_{\mathcal{O}(i)}^\dagger u^{\pdagger}_{\mathcal{O}(i)\mathcal{O}(j)} W^{\pdagger}_{\mathcal{O}(j)}= u^{\pdagger}_{ij},
\end{align}
then, owing to the gauge redundancy, the symmetry is actually preserved. Thus, we say that the symmetry acts \textit{projectively} in the spinon Hilbert space. In general, there can be several different projective actions of the symmetries considered. Accordingly, \textit{Ans\"atze} corresponding to different gauge-inequivalent projective symmetry actions represent distinct QSL states at the mean-field level. This approach provides us with a powerful mathematical tool to classify all possible QSL mean-field \textit{Ans\"atze} for a given set of symmetries. The combined operation of the symmetry element $\mathcal{O}$ and its associated local SU(2) gauge transformation $W^{}_{\mathcal{O}}(i)$ constitutes a symmetry group, called the projective symmetry group (PSG)~\cite{Wen-2002}.

Of course, any symmetry group also has an identity element $\mathds{1}$, so one has to define the corresponding projective extension of $\mathds{1}$ too. This projective action is specified by a gauge group $\mathcal{G}$ such that for $G_i, G_j \in \mc{G}$, 
\begin{equation} G_{i}^{\dagger}u^{\pdagger}_{ij}G^{\pdagger}_{j}=u^{\pdagger}_{ij}.
 \label{eq:IGG}
\end{equation}
These operations constitute a subgroup of the PSG, which is known as the invariant gauge group (IGG). Therefore, in projective space, the identity can be defined up to an element of $\mathcal{G}$. Note that according to Eq.~\eqref{eq:IGG}, $\mathcal{G}$ is the symmetry group of the ground-state mean-field {\it Ansatz}, i.e., it is the low-energy symmetry group. Importantly, the IGG is distinct from the SU(2) symmetry group of the fermionic representation of Eq.~\eqref{eq:gauge_symmetry}. In order to realize an SU(2) IGG, one needs a bipartite lattice~\cite{Wen-2002}, which the ruby lattice is clearly not. We are thus left with {\it Ans\"atze} with either a U(1) or a $\mathbb{Z}_{2}$ IGG as the only possibilities in our case. 
 
In the canonical gauge~\cite{Wen-2002}, the IGG can be represented as the group of global transformations of the generic form $\mathcal{G}$\,$=$\,$\{e^{\dot{\iota}\phi\hat{n}\cdot\hat{\tau}}\}$; for U(1) and $\mathbb{Z}_2$ IGGs, these reduce to $\mathcal{G}=\{e^{\dot{\iota}\phi\tau^3}\}$ and $\mathcal{G}=\{\pm1\}$, respectively. An {\it Ansatz} featuring only {\it real} hoppings, i.e., $\chi^{3}_{ij}$, will have a U(1) IGG, while the inclusion of {\it real} singlet pairings, i.e., $\Delta^{1}_{ij}$, lowers the IGG to $\mathbb{Z}_{2}$.

\begin{table*}
	\begin{ruledtabular}
		\begin{tabular}{cccccccc}
$w^{}_R$ & $w^{}_\sigma$ & $w^{}_\mathcal{T}$ & $\xi$ & $\rho^{}_{R,s}$ & $\rho^{}_{\sigma,s}$ & $\rho^{}_{\mathcal{T},s}$ & \# of PSGs\\
			\hline
$0$ & $0$ & $1$ & $0,\pi$ & $\{0,n^{}_{\mathcal{T}R}\pi,0,n^{}_{\mathcal{T}R}\pi,0,n^{}_{\mathcal{T}R}\pi\}$ & $\{0,n^{}_{R\sigma}\pi,0,n^{}_{R\sigma}\pi,0,n^{}_{R\sigma}\pi\}$ & $\{0,0,0,0,0,0\}$ & 8\\
$0$ & $1$ & $1$ & $0,\pi$ & $\{0,0,0,0,0,0\}$ & $\{0,n^{}_{R\sigma}\pi,0,(n^{}_{R\sigma}+n^{}_{\mathcal{T}R})\pi,n^{}_{\mathcal{T}R}\pi,(n^{}_{R\sigma}+n^{}_{\mathcal{T}R})\pi\}$ & $\{0,0,0,0,0,0\}$ & 8\\
$1$ & $0$ & $1$ & $0,\pi$ & $\{0,0,0,0,0,n^{}_R\pi\}$ & $\{0,n^{}_{R\sigma}\pi,0,n^{}_{R\sigma}\pi,n^{}_R\pi,n^{}_{R\sigma}\pi\}$ & $\{0,0,0,0,0,0\}$ & 8\\
$1$ & $1$ & $1$ & $0,\pi$ & $\{0,0,0,0,0,n^{}_R\pi\}$ & $\{0,n^{}_{R\sigma}\pi,0,n^{}_{R\sigma}\pi,n^{}_R\pi,n^{}_{R\sigma}\pi\}$ & $\{0,0,0,0,0,0\}$ & 8\\
\hline
$0$ & $0$ & $0$ & $0,\pi$ & $\{0,0,0,0,0,0\}$ & $\{0,n^{}_{R\sigma}\pi,0,n^{}_{R\sigma}\pi,0,n^{}_{R\sigma}\pi\}$ & $\{0,\pi,0,\pi,0,\pi\}$ & 4\\
$0$ & $1$ & $0$ & $\mathbb{Q}\, \pi $ & $\{0,0,0,0,0,0\}$ & $\{0,n^{}_{R\sigma}\pi,0,n^{}_{R\sigma}\pi,0,n^{}_{R\sigma}\pi\}$ & $\{0,\pi,0,\pi,0,\pi\}$ & 2\\
$1$ & $0$ & $0$ & $0,\pi$ & $\{0,0,0,0,0,n^{}_R\pi\}$ & $\{0,n^{}_{R\sigma}\pi,0,n^{}_{R\sigma}\pi,n^{}_R\pi,n^{}_{R\sigma}\pi\}$ & $\{0,\pi,0,\pi,0,\pi\}$ & 8\\
$1$ & $1$ & $0$ & $0,\pi$ & $\{0,0,0,0,0,n^{}_R\pi\}$ & $\{0,\xi^{}_{R\sigma},0,\xi^{}_{R\sigma},n^{}_R\pi,\xi^{}_{R\sigma}\}$ & $\{0,\pi,0,\pi,0,\pi\}$ & 4\\
		\end{tabular}
	\end{ruledtabular}
 	\caption{\label{table:u1_psg}All possible gauge-inequivalent choices of the phases $w^{}_{\mathcal{O}}$, in Eq.~\eqref{eq:G-gen}, and $\rho^{}_{\mathcal{O},s}$, in Eqs.~\eqref{eq:T1T2_psg_u1}--\eqref{eq:time_psg_u1}, yielding a total of 50 U(1) PSGs. Note that $w^{}_{T_1}$ and $w^{}_{T_2}$ are always 0. Each vector $\{ \cdots \}$ lists the six values of $\rho^{}_{\mc{O},s}$ for $s=1, \ldots, 6$. The parameters $n_{\ldots}$ are binary variables, which can be either 0 or 1. For example, in the first line, both $n^{}_{\mathcal{T}R}$ and $n^{}_{R \sigma}$ can take the values 0, 1, which together with the two possibilities for $\xi$, lead to eight distinct gauge-inequivalent PSG solutions, as noted in the rightmost column. $\xi^{}_{R\sigma}$, in the last line, denotes a continuous phase variable, which can range from $0 \le \xi^{}_{R\sigma} < 2\pi$, and $\mathbb{Q}$ denotes rational numbers.}
\end{table*}

\section{Symmetries and PSG solutions}
\label{sec:symmetry}
The ruby lattice is composed of six sublattices, which we label by $s = 1,2,\ldots,6$ in Fig.~\ref{fig:fig1}(a). The coordinates of any site can be written as
\begin{align}\label{eq:GCC}
	(x,y;s) & \equiv \vect{r} \equiv x\mathbf{T}_{1} +y\mathbf{T}_{2}+\bm{\lambda}_s,\;
\end{align}
with the translation vectors of the underlying triangular Bravais lattice given by
\begin{alignat}{1}
\label{eq:bravais_vector}
	 \mathbf{T}^{}_{1}&=\frac{1}{2}\left(\sqrt{3}(b+\sqrt{3}a)\mathbf{\hat x}-(b+\sqrt{3}a)\mathbf{\hat y}\right), \; \\
 \mathbf{T}^{}_{2}&=\left(b+\sqrt{3}a\right)\mathbf{\hat y}, \; 
 \end{alignat}
where $a$ and $b$ are the lengths of the hexagonal and triangular sides, respectively ($a \ge b$), and $\bm{\lambda}_s$ is a vector specifying the position of the sublattice $s$ within the unit cell. We choose a unit cell such that its center coincides with the center of the hexagonal plaquettes of the ruby lattice, as depicted in Fig.~\ref{fig:fig1}(a). The position of every site inside the unit cell is specified as $\bm{\lambda}_s = x_s \mathbf{a}_1 + y_s \mathbf{a}_2$, where $\mathbf{a}_1 = a(1,\sqrt{3})/2$ and $\mathbf{a}_2 = a(1,0)$.

The space (wallpaper) group of the ruby lattice is $p6m$, which is generated by two translations ($T_1$ and $T_2$) along $\bf{T}_1$ and $\bf{T}_2$, a $C_6$ rotation around an axis passing through the origin---the center of a hexagon---and perpendicular to the lattice plane ($R$), and a reflection about the $x$ axis ($\sigma$). Under these symmetries, illustrated in Fig.~\ref{fig:fig1}(a), the lattice coordinates transform as 
\begin{equation}
 \label{eq:generators}
	\left.\begin{aligned}
		T^{}_{1}&:(x,y;s)\rightarrow (x+1,y;s),\\
		T^{}_{2}&:(x,y;s)\rightarrow (x,y+1;s),\\
		R&:(x,y;s)\rightarrow (x-y,x;R(s)),\\
		\sigma&:(x,y;s)\rightarrow (x,x-y;\sigma(s)),\\
	\end{aligned}\right.
\end{equation}
where $R(\{1,2,3,4,5,6\})=\{2,3,4,5,6,1\}$ and $\sigma(\{$1,2,3, 4,5,6$\}) = \{3,2,1,6,5,4\}$. In addition to the spatial symmetries, a fully symmetric QSL solution also requires the inclusion of time-reversal symmetry $\mathcal{T}$, which naturally leaves the lattice coordinates invariant. The mutual relations between the different symmetry operations in Eq.~\eqref{eq:generators} lead to the following set of algebraic conditions:
\begin{align}
 T^{}_1 T^{}_2 &= T^{}_2 T^{}_1, \label{eq:T1T2} \\
R T^{}_2 R^{-1} T^{}_1 &= 1, \label{eq:R_T1}\\
 R T^{-1}_2 T^{-1}_1R^{-1}T^{}_2&=1, \label{eq:R_T2}\\
 R^6 &= 1, \label{eq:R}\\
 \sigma T^{-1}_1 T^{-1}_2\sigma^{-1} T^{}_1 &= 1, \label{eq:sig_T1}\\
 \sigma T^{}_2\sigma^{-1}T^{}_2&=1, \label{eq:sig_T2}\\
 \mathcal{\sigma}^2 &= 1, \label{eq:sigma}\\
 R\sigma R \sigma &= 1, \label{eq:sigma_R}\\
 \mathcal{T}^2 &= 1, \label{eq:time}\\
 \mathcal{T} \mathcal{O} \mathcal{T}^{-1} \mathcal{O}^{-1} &= 1, \;\mathcal{O}\in\{T_1,T_2,R,\sigma\} \label{eq:time_O}.
\end{align}

Having specified the symmetries $\mc{O} \in \{T_1,T_2,R, \sigma,\mc{T}\}$, we can now enumerate the allowed PSG solutions for the ruby lattice by associating a gauge transformation $G_\mc{O} \in$ SU(2) for each $\mc{O}$ and using the algebraic relations in Eqs.~\eqref{eq:T1T2}--\eqref{eq:time_O}. Distinct choices for the gauge transformations $G_\mc{O}$ lead to distinct QSL states, which differ in the projective action of symmetry operations. Here, we construct all possible gauge-inequivalent representations $G_{\mathcal{O}}(x,y;s)$ for $\mathcal{G}\simeq$ U(1), $\mathbb{Z}_2$.

\subsection{U(1) solutions}

\label{sec:psg_sol}
Due to the local SU(2) gauge redundancy, an \textit{Ansatz} may not always appear in its canonical form, meaning that the IGG may not be explicitly visible. In particular, for a U(1) IGG, the canonical form dictates that any \textit{Ansatz} can be expressed solely in terms of hopping amplitudes. However, an arbitrary choice of gauge can distort this canonical form, introducing pairing terms. Note that the presence of such terms does not imply that the IGG is $\mathbb{Z}_2$, as a suitable local SU(2) gauge transformation can always restore the canonical form of the U(1) \textit{Ansatz}. Therefore, it is crucial to adopt a judiciously chosen generic gauge form that preserves the canonical structure of the \textit{Ansatz}.
The generic form of a U(1) PSG solution for a symmetry operator $\mathcal{O}$ can be written, in the canonical gauge, as
\begin{equation}
\label{eq:G-gen}
 G^{}_\mathcal{O}(x,y;s)=\exp\left(\dot{\iota}\, \phi^{}_\mathcal{O}(x,y;s) \,\tau^3\right) \left(\dot{\iota}\tau^1\right)^{w^{}_\mathcal{O}}, 
\end{equation}
where $\phi^{}_\mathcal{O}(x,y;s)$ is a real-valued, position- and sublattice-dependent phase function specific to the symmetry $\mathcal{O}$ (its explicit form is given below), and $w^{}_\mathcal{O}\in\{0,1\}$ encodes whether $\mathcal{O}$ is accompanied, in the canonical gauge, by an additional factor of $\dot{\iota}\tau^{1}$. One can easily verify that this PSG structure preserves the canonical form of a U(1) \textit{Ansatz}, meaning that after a gauge transformation of this type, the U(1) \textit{Ansatz} still consists solely of hopping terms. Even though this describes the most general U($1$) gauge transformation, not all possibilities of the form of Eq.~\eqref{eq:G-gen} are allowed because some of them do not satisfy the gauge-enriched algebraic relations extending Eqs.~\eqref{eq:T1T2}--\eqref{eq:time_O}, derived in Appendix~\ref{sec:genric_gauge_con}. In particular, note that there exist no solutions with $w^{}_\mathcal{O}=1$ for $\mathcal{O}\in\{T_1,T_2\}$ (refer to Appendix~\ref{app:u1_psg_derivation} for details). 

The possible solutions for $\phi^{}_\mathcal{O}(x,y;s)$ are conveniently parametrized as 
\begin{align}
&\phi^{}_{T_1}(x,y;s)=y\,\xi,\;\phi^{}_{T_2}(x,y;s)=0, \label{eq:T1T2_psg_u1} \\
&\phi^{}_R(x,y;s)=y\left[x-\frac{1}{2}(y+1)\right]\xi+\rho^{}_{R,s}, \label{eq:R_psg_u1} \\
&\phi^{}_\sigma(x,y;s)=\frac{1}{2}x(x+1)\xi+\rho^{}_{\sigma,s}, \label{eq:sig_psg_u1}\\
&\phi^{}_\mathcal{T}(x,y;s)=\rho^{}_{\mathcal{T},s}. \label{eq:time_psg_u1} 
\end{align}
Here $\{\xi, \rho^{}_{R,s},$ $\rho^{}_{\sigma,s},\rho^{}_{\mc{T},s} \}$ are U(1) phases which can take values in $[0,2\pi)$. In order to define a particular PSG, we therefore need to specify the three numbers $\{w^{}_R, w^{}_\sigma, w^{}_\mc{T} \}$, which feed into Eq.~\eqref{eq:G-gen}, together with the corresponding set $\{\xi, \rho^{}_{R,s},$ $\rho^{}_{\sigma,s},\rho^{}_{\mc{T},s} \}$.

The different gauge-inequivalent choices of $\rho^{}_{\mathcal{O},s}$ (which determine $\phi^{}_{\mc{O}}$) as well as $w^{}_{\mc{O}}$ are listed in Table~\ref{table:u1_psg}; in this way, we obtain a total of 50 U(1) PSG solutions.

\subsection{$\mathbb{Z}_2$ solutions}
\label{sec:psg_sol_z2}

Similarly, for a $\mathbb{Z}_2$ IGG, the PSG solutions (for details, see Appendix~\ref{app:z2_psg_derivation}) are given by
\begin{align}
&G^{}_{T_1}(x,y;s)=\eta^{y}\tau^0,\;G^{}_{T_2}(x,y;s)=\tau^0, \label{eq:T1T2_psg} \\
&G^{}_R(x,y;s)=\eta^{xy+\frac{y}{2}(y+1)}\,g^{}_{R,s}, \label{eq:R_psg} \\
&G^{}_\sigma(x,y;s)=\eta^{\frac{x}{2}(x+1)}\,g^{}_{\sigma,s}, \label{eq:sig_psg}\\
&G^{}_\mathcal{T}(x,y;s)=\eta^s_{\mathcal{T}R}\,g^{}_{\mathcal{T}}, \label{eq:time_psg} 
\end{align}
where
\begin{align}
g^{}_{R,s}&=\{1,1,1,1,1,\eta^{}_R\}\tau^0, \\
g^{}_{\sigma,s}&=\{1,\eta^{}_{R\sigma},1,\eta^{}_{R\sigma},\eta^{}_R,\eta^{}_{R\sigma}\}g^{}_\sigma,\;g^2_\sigma=\eta^{}_\sigma\tau^0,
\end{align}
$g^{}_{\sigma}$ and $g_\mathcal{T}$ are $2\times2$ SU(2) matrices, and all the parameters $\eta_{\ldots}$ take values $\pm1$. All possible gauge-inequivalent choices of these matrices are summarized in Table~\ref{table:z2_psg}.
In this case, we obtain a total of $64$ projective
extensions of lattice and time-reversal symmetries defining fully symmetric $\mathbb{Z}_2$ \textit{Ans\"atze}.

\begin{table}[h]
	\begin{ruledtabular}
		\begin{tabular}{ccccc}
		$\eta^{}_\sigma$&$g^{}_{\sigma}$&$g^{}_{\mathcal{T}}$ & Set of $\eta$ parameters& \# of PSGs\\
			\hline
$+1$&$\tau^0$ & $\dot\iota\tau^2$ & $\{\eta^{}_{\mathcal{T}R},\eta,\eta^{}_R,\eta^{}_{R\sigma}\}$& $2^4$\\
$+1$&$\tau^0$ & $\tau^0$ & $\{\eta^{}_{\mathcal{T}R}=-1,\eta,\eta^{}_R,\eta^{}_{R\sigma}\}$& $2^3$\\
$-1$&$\dot\iota\tau^3$ & $\dot\iota\tau^2$ & $\{\eta^{}_{\mathcal{T}R},\eta,\eta^{}_R,\eta^{}_{R\sigma}\}$& $2^4$\\
$-1$&$\dot\iota\tau^3$ & $\dot\iota\tau^3$ & $\{\eta^{}_{\mathcal{T}R},\eta,\eta^{}_R,\eta^{}_{R\sigma}\}$& $2^4$\\
$-1$&$\dot\iota\tau^3$ & $\tau^0$ & $\{\eta^{}_{\mathcal{T}R}=-1,\eta,\eta^{}_R,\eta^{}_{R\sigma}\}$& $2^3$\\
		\end{tabular}
	\end{ruledtabular}
	\caption{\label{table:z2_psg}The possible gauge-inequivalent choices of $\eta_\sigma$ and the matrices $g^{}_{\mathcal{O}}$, defining a total of $64$ $\mathbb{Z}_2$ PSG solutions.}
\end{table}

\begin{table}[t]
	\begin{ruledtabular}
		\begin{tabular}{ccccccccc}
Label & $\phi^{}_h$ & $\phi^{}_s$ & $\phi^{}_t$ & $\phi^{}_{t'}$ & $\phi^{}_{p}$ & $\phi^{}_{w}$ & $\phi^{}_{w'}$ & $\phi^{}_{b}$\\
\hline
U1$000$ & $0$ & $0$ & $0$ & $0$ & $0$ & $0$ & $0$ & $0$\\
U1$\pi00$ & $0$ & $\pi$ & $0$ & $\star$ & $\star$ & $\star$ & $\star$ & $\star$\\
U1$0\pi\pi$ & $\pi$ & $\pi$ & $0$ & $\star$ & $\star$ & $\star$ & $\star$ & $\star$\\
U1$\pi\pi\pi$ & $\pi$ & $0$ & $0$ & $0$ & $\pi$ & $0$ & $0$ & $0$\\
U1$00\pi$ & $\star$ & $\star$ & $0$ & $\star$ & $0$ & $\star$ & $\star$ & $\pi$\\
U1$\pi\pi0$ & $\star$ & $\star$ & $0$ & $\star$ & $\pi$ & $\star$ & $\star$ & $\pi$\\
U3$0\pi\pi$ & $0$ & $\pi$ & $\pi$ & $\pi$ & $\pi$ & $\pi$ & $\pi$ & $\pi$ \\
U3$\pi0\pi$ & $\pi$ & $\pi$ & $\pi$ & $\pi$ & $0$ & $0$ & $0$ & $\pi$\\
U3$0\pi0$ & $\star$ & $\star$ & $\pi$ & $\star$ & $\pi$ & $\star$ & $\star$ & $0$ \\
U3$\pi00$ & $\star$ & $\star$ & $0$ & $\star$ & $0$ & $\star$ & $\star$ & $0$ \\
U3$00\pi$ & $\pi$ & $0$ & $0$ & $\star$ & $\star$ & $\star$ & $\star$ & $0$ \\
U3$\pi\pi\pi$ & $0$ & $0$ & $\pi$ & $\star$ & $\star$ & $\star$ & $\star$ & $\star$ \\
U7$0\pi\pi$ & $0$ & $\star$ & $\star$ & $\star$ & $\star$ & $\phi$ & $-\phi$ & $\star$ \\
U7$\pi0\pi$ & $\pi$ & $\star$ & $\star$ & $\star$ & $\star$ & $\phi$ & $-\phi$ & $\star$ \\
U8$0\pi\pi$ & $0$ & $\star$ & $\star$ & $\star$ & $\star$ & $\phi$ & $\phi$ & $\star$ \\
U8$\pi0\pi$ & $\pi$ & $\star$ & $\star$ & $\star$ & $\star$ & $\phi$ & $\phi+\pi$ & $\star$ \\
U8$0\pi0$ & $0$ & $\star$ & $\star$ & $\star$ & $\star$ & $\phi$ & $\phi+\pi$ & $\star$ \\
U8$\pi00$ & $\pi$ & $\star$ & $\star$ & $\star$ & $\star$ & $\phi$ & $\phi$ & $\star$ \\
		\end{tabular}
	\end{ruledtabular}	
 	\caption{\label{table:u1_flux_structures}Flux structures of all U(1) \textit{Ans\"atze}, defined by the fluxes threading the loops illustrated in Fig.~\ref{fig:loops}. A $\star$ indicates that the flux is not defined due to vanishing (due to symmetry) mean-field amplitudes on some bonds of that particular loop.}
\end{table}

\begin{table}
\begin{ruledtabular}
\begin{tabular}{cccccc}
 \multirow{2}{*}{Label} & 1NN & 2NN & %
 3NN & \multirow{2}{*}{Onsite} & \multirow{2}{*}{Parent U(1)}\\
\cline{2-4}
 & $u^{}_{1}$ & $u^{}_{1'}$ & $u^{}_{2}$ & & \\
\hline
\textcolor{teal}{Z10000} & $\tau^{\textcolor{blue}{1},3}$ & $\tau^{3}$ & $\tau^{1,3}$ & $\tau^{\textcolor{blue}{1},3}$ & U1$000$, U3$0\pi\pi$, U8$0\pi\pi$, U7$0\pi\pi$ \\
\textcolor{teal}{Z11100} & $\tau^{\textcolor{blue}{1},3}$ & $\tau^{3}$ & $\tau^{1,3}$ & $\tau^{\textcolor{blue}{1},3}$ & U3$\pi0\pi$,U8$\pi0\pi$,U7$\pi0\pi$ \\
\textcolor{teal}{Z10100} & $\tau^{\textcolor{blue}{1},3}$ & $\tau^{3}$ & $0$ & $\tau^{\textcolor{blue}{1},3}$ & U3$00\pi$\\
\textcolor{teal}{Z11000} & $\tau^{\textcolor{blue}{1},3}$ & $\tau^{3}$ & $0$ & $\tau^{\textcolor{blue}{1},3}$ & U1$\pi00$, U3$\pi\pi\pi$\\
\textcolor{teal}{Z10010} & $0$ & $\tau^{3}$ & $\tau^{1,3}$ & $\tau^{\textcolor{blue}{1},3}$ & U3$0\pi0$, U8$0\pi0$ \\
\textcolor{teal}{Z11110} & $0$ & $\tau^{3}$ & $\tau^{1,3}$ & $\tau^{\textcolor{blue}{1},3}$ & U3$\pi00$, U8$\pi00$ \\
Z30000 & $\tau^{3}$ & $\tau^{3}$ & $\tau^{\textcolor{blue}{1},3}$ & $\tau^{3}$ & U1$000$, U3$0\pi0$ \\
Z31100 & $\tau^{3}$ & $\tau^{3}$ & $\tau^{\textcolor{blue}{1},3}$ & $\tau^{3}$ & U3$\pi00$ \\
Z30010 & $\tau^1$ & $\tau^{3}$ & $\tau^{\textcolor{blue}{1},3}$ & $\tau^{3}$ & U3$0\pi\pi$ \\
Z31110 & $\tau^1$ & $\tau^{3}$ & $\tau^{\textcolor{blue}{1},3}$ & $\tau^{3}$ & U3$\pi0\pi$\\
\hline
\textcolor{teal}{Z10001} & $\tau^{2}$ & $\tau^{3}$ & $\tau^{2}$ & $\tau^{\textcolor{blue}{1},3}$ & U3$0\pi\pi$ \\
\textcolor{teal}{Z11101} & $\tau^{2}$ & $\tau^{3}$ & $\tau^{2}$ & $\tau^{\textcolor{blue}{1},3}$ & U3$\pi0\pi$ \\
\textcolor{teal}{Z10101} & $\tau^{2}$ & $\tau^{3}$ & $\tau^0$ & $\tau^{\textcolor{blue}{1},3}$ & U3$00\pi$ \\
\textcolor{teal}{Z11001} & $\tau^{2}$ & $\tau^{3}$ & $\tau^0$ & $\tau^{\textcolor{blue}{1},3}$ & U3$\pi\pi\pi$ \\
\textcolor{teal}{Z10011} & $\tau^{0}$ & $\tau^{3}$ & $\tau^{2}$ & $\tau^{\textcolor{blue}{1},3}$ & U3$0\pi0$ \\
\textcolor{teal}{Z11111} & $\tau^{0}$ & $\tau^{3}$ & $\tau^{2}$ & $\tau^{\textcolor{blue}{1},3}$ & U3$\pi00$ \\
\textcolor{teal}{Z10111} & $\tau^{0}$ & $\tau^{3}$ & $\tau^{0}$ & $\tau^{\textcolor{blue}{1},3}$ & U1$000$ \\
\textcolor{teal}{Z11011} & $\tau^{0}$ & $\tau^{3}$ & $\tau^{0}$ & $\tau^{\textcolor{blue}{1},3}$ & U1$\pi\pi\pi$\\
\textcolor{teal}{Z30011} & $\tau^{0,\textcolor{blue}{2}}$ & $\tau^{3}$ & $\tau^{2}$ & $\tau^{3}$ & U3$0\pi\pi$, U1$0\pi\pi$ \\
\textcolor{teal}{Z31111} & $\tau^{0,\textcolor{blue}{2}}$ & $\tau^{3}$ & $\tau^{2}$ & $\tau^{3}$ & U3$\pi0\pi$ \\
\textcolor{teal}{Z30111} & $\tau^{0,\textcolor{blue}{2}}$ & $\tau^{3}$ & $\tau^0$ & $\tau^{3}$ & U3$00\pi$ \\
\textcolor{teal}{Z31011} & $\tau^{0,\textcolor{blue}{2}}$ & $\tau^{3}$ & $\tau^0$ & $\tau^{3}$ & U3$\pi\pi\pi$ \\
\end{tabular}
\end{ruledtabular}
\caption{\label{table:ansatze_z2}Symmetric $\mathbb{Z}_2$ mean-field {\it Ans\"atze}. The \textit{Ans\"atze} with teal labels can be realized with a $\mathbb{Z}_2$ IGG even with only first- and second-neighbor amplitudes $u^{}_{1}$ and $u^{}_{1'}$, i.e., when the mean-field amplitudes are nonvanishing on both the blue $(J^{}_{1})$ \emph{and} the green $(J^{\prime}_{1})$ bonds in Fig.~\ref{fig:fig1} simultaneously; third-neighbor amplitudes $u^{}_{2}$ are not required to break the IGG from U(1) down to $\mathbb{Z}_{2}$. The blue indices indicate the first terms responsible for breaking the U(1) IGG down to $\mathbb{Z}_2$. Note that, although the $\tau^{1}$ structure on the blue $u_{1}$ bonds may seem to reduce the IGG from U(1) to $\mathbb{Z}_{2}$, an unbroken U(1) (in a non-canonical gauge) can be masked as $\mathbb{Z}_{2}$ due to local SU(2) gauge redundancy. The presence of the pairing terms in the true $\mathbb{Z}_{2}$ \textit{Ansätze} is unavoidable, whereas disguised U(1) \textit{Ansätze} can be gauge-transformed into purely hopping forms.}
\end{table}

\begin{figure*}	\includegraphics[width=0.9\linewidth]{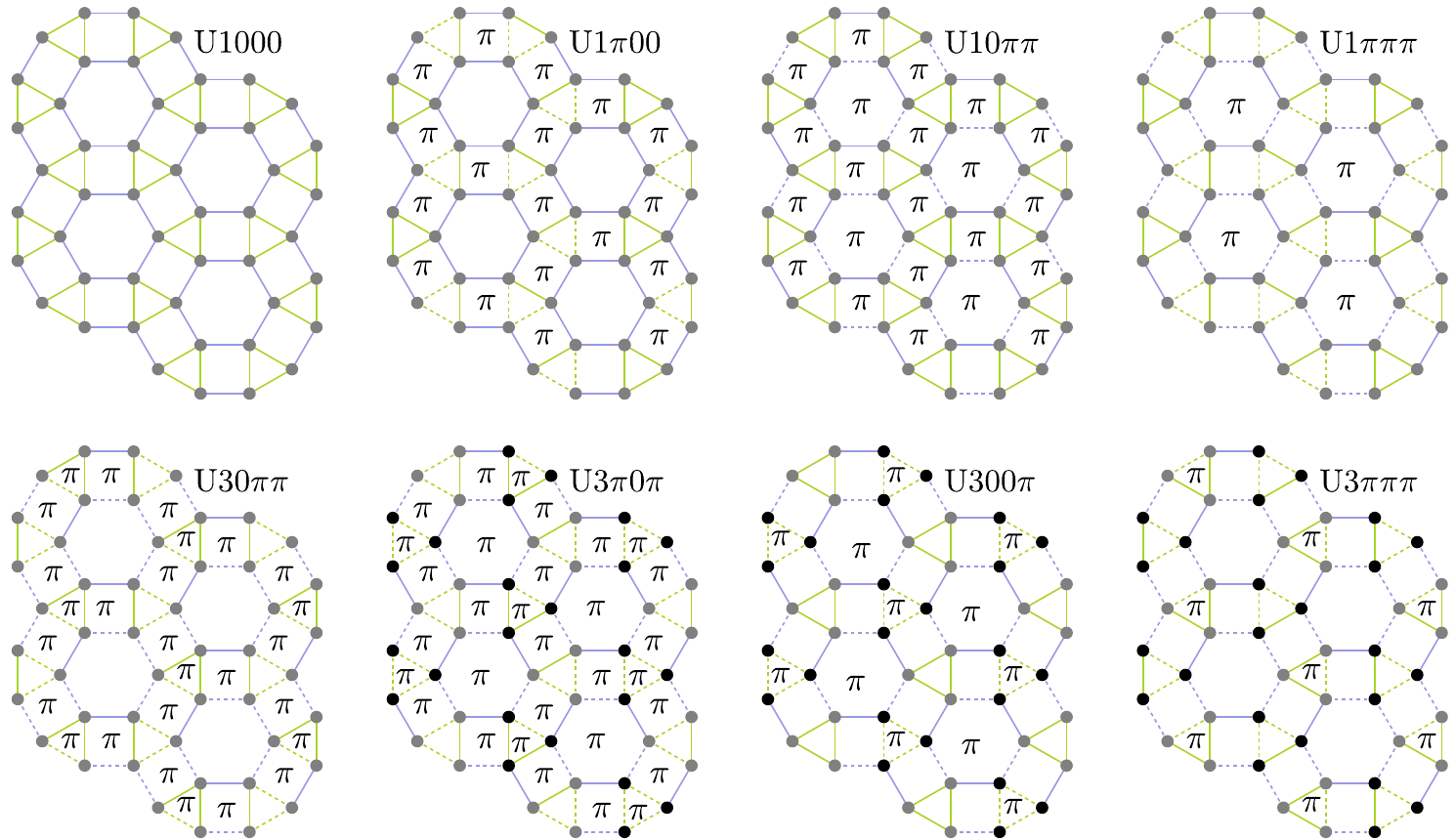}
\caption{Graphical representation of all U(1) \textit{Ans\"atze} that can be realized with mean-field Hamiltonians that include up to 2NN amplitudes. The gray (black) points denote positive (negative) on-site hoppings (chemical potential), i.e., $+$ ($-$) $a^{}_3\tau^3$. The solid (dashed) blue and green links represent real 1NN and 2NN hoppings which are positive (negative), respectively.} 
\label{fig:1nn_u1_ansatze}
\end{figure*}

\section{Short-ranged mean-field Ans\"atze}
Having enumerated the different PSG solutions, we can now write down all possible mean-field \textit{Ans\"atze} with U(1) or $\mathbb{Z}_2$ IGGs. The number of U(1) and $\mathbb{Z}_2$ PSG solutions represents the maximum possible number of Ansätze (with hopping and pairing amplitudes on geometrical distances which could, in principle, extend up to infinity) that can be realized with IGG U(1) and $\mathbb{Z}_2$, respectively. However, breaking the SU(2) IGG down to one of its subgroups, U(1) or $\mathbb{Z}_2$, across all PSGs requires considering a specific range of bonds with nonvanishing mean-field amplitudes. For instance, if one considers a U(1) PSG and constructs Ansätze up to the $p$-th neighbor bonds, the real hopping terms—responsible for breaking SU(2) IGG down to U(1)—may emerge only beyond the $p$-th neighbor bonds. This implies that under a $p$-th neighbor restriction/cutoff, a U(1) Ansatz associated with a particular U(1) PSG may not be realizable, and effectively still have an SU(2) IGG. Consequently, upon restricting the mean-field amplitudes to only short-range bonds (of relevance as ground states of spin models with short-range interactions), not all U(1) PSGs can give rise to U(1) Ansätze. As a result, the number of realizable U(1) Ansätze (with mean-field amplitudes on bonds up to a finite p-th neighbor distance) may be smaller than the total number of U(1) PSGs. The same principle applies to realizing QSLs with $\mathbb{Z}_2$ IGG as well.
 Here, we restrict our analysis to a subset of those PSGs which realize distinct phases for mean-field \textit{Ans\"atze} comprising hopping and pairing terms between up to third-nearest-neighboring (3NN) sites.
\subsection{U(1) QSL states}
\label{sec:u1_ansatze}
To label the individual states, we adopt the notation 
\begin{equation} 
\text{U}\;\text{PSG}_\text{row}\,\xi \,\xi^{}_{\mathcal{T}R}\,\xi^{}_{R\sigma},
\end{equation}
where ``$\text{PSG}_\text{row}$'' refers to the row number in Table~\ref{table:u1_psg}. Besides the variables $\xi$ and $\xi^{}_{R\sigma}$ introduced above, our nomenclature also uses the index $\xi^{}_{\mathcal{T}R}$ (see Appendix~\ref{app:u1_psg_derivation}), which is defined as
\begin{alignat}{1}
\xi^{}_{\mathcal{T}R} =
\begin{cases}
n^{}_{\mathcal{T}R}\pi \quad & \mbox{for rows 1, 2}\,\\
n^{}_{R}\pi \quad & \mbox{for rows 3, 4, 7, 8}\,\\
0 \quad & \mbox{for rows 5, 6}\,
\end{cases}.
\end{alignat}

We find a total of 8 U(1) \textit{Ans\"atze} which can be realized up to 2NN and are depicted in Fig.~\ref{fig:1nn_u1_ansatze}. With the inclusion of 3NN coupling terms, a total of 18 U(1) \textit{Ans\"atze} can be realized and these are graphically illustrated in Fig.~\ref{fig:3nn_u1_ansatze} of Appendix~\ref{app:3NN}.

\begin{figure}[b]	\includegraphics[width=0.7\linewidth]{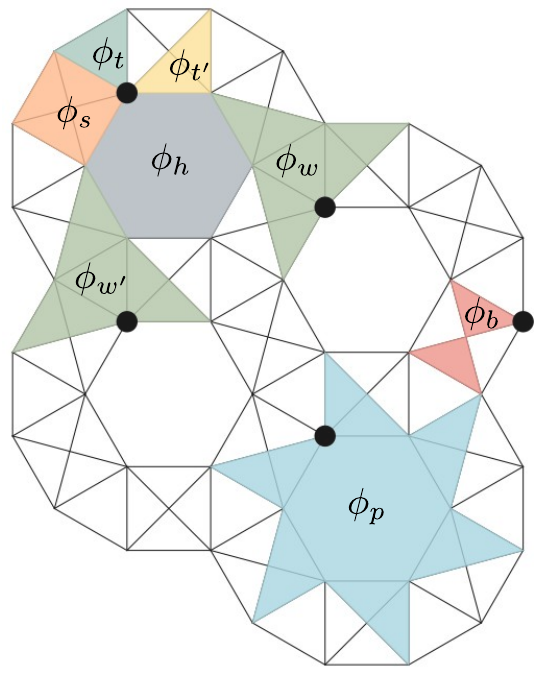}
\caption{Definition of the fluxes characterizing the U(1) \textit{Ans\"atze}. The black dots mark the sites belonging to sublattice 1 in the unit cell (per the ordering in Fig.~\ref{fig:fig1}), which are taken as the base sites for the respective loops.} 
\label{fig:loops}
\end{figure}

In addition to their PSG description, all the \textit{Ans\"atze} can equivalently be characterized by fluxes $\phi^{}_{i}$ associated with different loop operators $P_{\mathcal{C}_{i}}(\phi^{}_{i})$, defined for $q$-sided loops $\mathcal{C}_{i}$ with a ``base site'' $i$ as
\begin{align} 
P^{\pdagger}_{\mathcal{C}_{i}}(\phi^{\pdagger}_{i})&=u^{\pdagger}_{ij}u^{\pdagger}_{jk}\cdots u^{\pdagger}_{j'i}\propto g^{\pdagger}_ie^{\dot{\iota}\phi^{}_{i}\tau^3}(\tau^3)^qg^\dagger_i,
\end{align}
where $g_i\in$ SU(2).
Here, $\phi^{}_i$ can be interpreted as a gauge magnetic flux threading the loop $\mathcal{C}_{i}$. In Fig.~\ref{fig:loops}, denoting the base sites by black points, we define eight loops: two triangular ones, one each about the square and hexagonal plaquettes of the ruby lattice, two windmill-shaped loops, one pinwheel, and finally, one bowtie. The fluxes threading these loops then characterize all the U(1) states, as classified in Table~\ref{table:u1_flux_structures}. 

\begin{figure}[t]
\centering
	\includegraphics[width=0.55\linewidth]{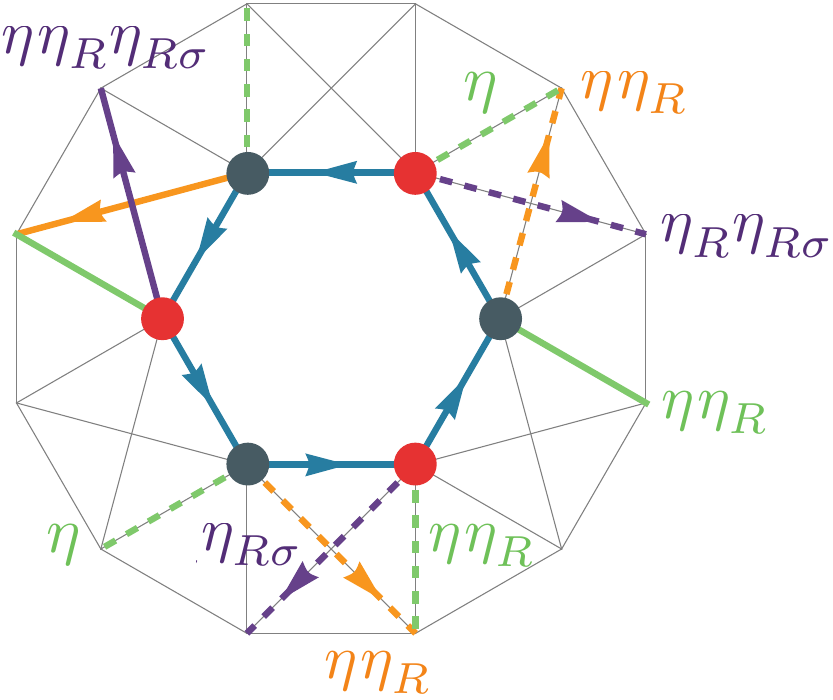}
\caption{Representation of mean-field amplitudes for all $\mathbb{Z}_2$ \textit{Ans\"atze}. The mean-field parameters on the blue, green, and orange bonds should be read off as $u^{}_1$, $u^{}_{1'}$, and $u^{}_2$, respectively, as specified in Table~\ref{table:ansatze_z2}. The parameters on all other bonds can be obtained by translations. In addition to the matrix structures of $u^{}_1$, $u^{}_{1'}$, $u^{}_2$, and $u'_2$, each $\mathbb{Z}_2$ \textit{Ansatz} is characterized by a specific sign of the binary variables $\eta_{\ldots} \in \pm 1$. For $\eta = -1$, the dashed bonds alternate in sign under translations along $\mathbf{T}^{}_{1}$. The parameters on the purple bonds are given by $g^\dagger_\sigma u^{\pdagger}_2g^{\pdagger}_\sigma$, where $g^{\pdagger}_\sigma=\tau^0$ ($=i\tau^3$) for states labeled by Z1$\cdots$ (Z3$\cdots$), respectively. The dark gray dots represent the on-site terms $a_\mu$, while the red sites correspond to $g^\dagger_\sigma a^{\pdagger}_\mu g^{\pdagger}_\sigma$. }
\label{fig:Z2MF}
\end{figure}

\subsection{$\mathbb{Z}_2$ QSL states}
For a $\mathbb{Z}_2$ IGG, we find that of the 64 projective extensions of the lattice and time-reversal symmetries, only 22 $\mathbb{Z}_2$ states are realizable with mean-field \textit{Ans\"atze} up to 3NNs. Let us denote the link field [see Eq.~\eqref{eq:link_singlet}] on a directed intra- (inter-) unit-cell bond by $u_{s_1s_2}$ $(u'_{s_1s_2})$, where $s_1, s_2 = 1, \ldots, 6$ represent the sublattice indices as per the numbering convention in Fig.~\ref{fig:fig1}(a). Then, the mean-field parameters of the $\mathbb{Z}_2$ \textit{Ans\"atze} can be written as
\begin{align}
u^{}_{12}&=u^{}_{23}=u^{}_{34}=u^{}_{45}=\eta^{}_Ru^{}_{56}=u^{}_{61} \equiv u^{}_1,\\
u^{}_{13}&=u^{}_{24}=\eta u^{}_{35}=\eta\eta^{}_Ru^{}_{46}=\eta\eta^{}_Ru^{}_{51}=\eta u^{}_{62} \equiv u^{}_{1'},\\
u^{}_{14}&=\eta\eta^{}_R u^{}_{36}=\eta\eta^{}_Ru^{}_{52} \equiv u^{}_{2}, \\
u'_{25}&=\eta\eta^{}_Ru'_{41}=\eta u'_{63}=\eta\eta^{}_R\eta^{}_{R\sigma}g^\dagger_\sigma u^{}_2g^{}_\sigma,
\end{align}
where the subscript $1$, $1'$, or $2$ signifies the $J^{}_1$, $J'_1$, or $J^{}_2$ bond type, respectively, and we have temporarily used the notation $u_{s_1s_2}$ $(s_1,s_2 =1,\ldots,6)$ to denote the directed link \textit{from} a site on sublattice $s_1$ \textit{to} a site on sublattice $s_2$.
This sign structure of the terms is schematically illustrated in Fig.~\ref{fig:Z2MF}.

The associated matrix structures of the mean-field amplitudes are listed in Table~\ref{table:ansatze_z2} for the various \textit{Ans\"atze}. Akin to the U(1) case,
the taxonomy of these states follows the notation
\begin{equation} 
\text{Z}\;\text{PSG}_\text{row}\left(\frac{1-\eta}{2}\right)\left(\frac{1-\eta^{}_R}{2}\right)\left(\frac{1-\eta^{}_{R\sigma}}{2}\right)\left(\frac{1-\eta^{}_{\mathcal{T}R}}{2}\right),
\end{equation}
where $\text{PSG}_\text{row}$ indicates the corresponding row in Table~\ref{table:z2_psg}, and the fractions in the brackets simply map the variables $\eta_{\ldots} \in \pm 1$ to $\{0, 1\}$.

\subsection{Phase diagram}

\begin{figure}[t]
\centering
	\includegraphics[width=0.8\linewidth]{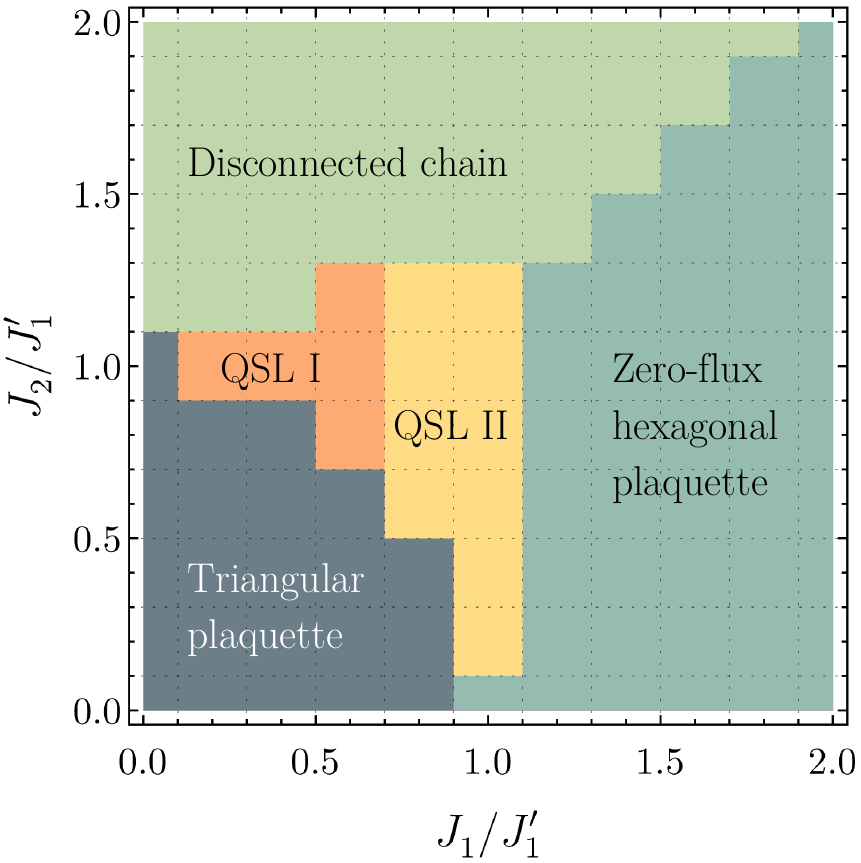}
\caption{Phase diagram obtained from a self-consistent mean-field treatment of the U(1) and $\mathbb{Z}_2$ \textit{Ans\"atze}. The ``QSL I'' phase corresponds to the U(1) state U3$\pi$0$\pi$ (though its $\mathbb{Z}_2$ descendants labeled by Z11100, Z11101, Z31110, and Z31111 have energies that are slightly lower, by $\mathcal{O}(10^{-5})J_{1}'$, or similar). ``QSL II'' represents the U(1) state U30$\pi\pi$ (its $\mathbb{Z}_2$ descendants, namely, Z10000, Z10001, Z30010, and Z30011, too have marginally lower or comparable energies). Thus, one cannot unambiguously distill the energetically favored mean-field ground state, as this would require a variational treatment after Gutzwiller projection. 
} 
\label{fig:PD}
\end{figure}

\begin{figure*}[tb]	\includegraphics[width=1.0\linewidth]{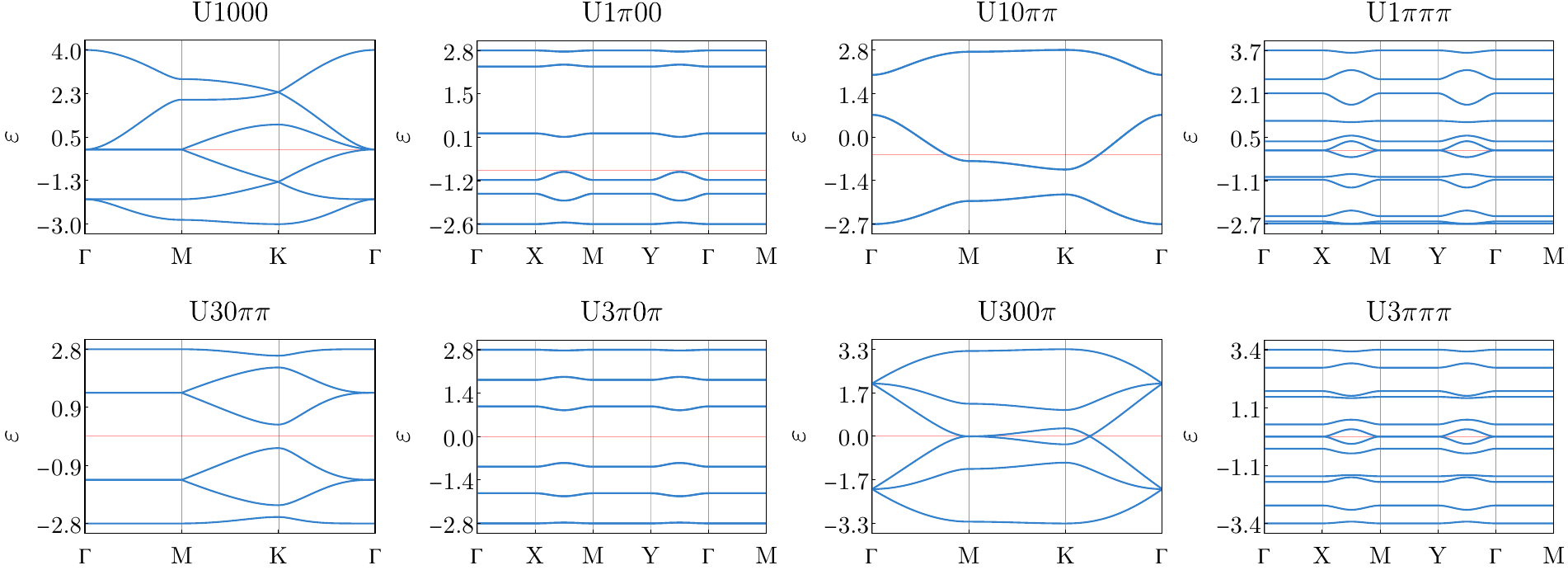}
	\caption{Spinon band structures for the eight different U(1) \textit{Ans\"atze} realizable with first- and second-nearest-neighbor mean-field amplitudes. The magnitude of the symmetry-allowed hoppings is set to one for $\chi^{}_1=\chi_{1}'$, and zero for further-neighbor bonds. The red line marks the Fermi level.\\}
	\label{fig:U12NNSBS}
\end{figure*}

Equipped with the list of possible U(1) and $\mathbb{Z}_2$ \textit{Ans\"atze} derived above, we now systematically construct a mean-field phase diagram of the Hamiltonian~\eqref{eq:mod-ham} and its candidate ground-state spin-liquid phases. To do so, we self-consistently determine the mean-field parameters for all the \textit{Ans\"atze} as a function of $J_1/J'_1$ and $J_2/J'_1$ together with their corresponding energies. Then, for each point in the space of Heisenberg couplings, we identify the lowest-energy disordered states. By a ``disordered'' state here, we mean a state that preserves SU(2) spin-rotation symmetry, i.e., one without any conventional dipolar (or higher multipolar) magnetic order. Such an analysis is restricted, by construction, to spin-rotation-invariant phases: the Abrikosov-fermion mean-field framework adopted in this work cannot diagnose instabilities towards magnetically ordered states (which would require, e.g., a Schwinger-boson treatment or a Gutzwiller-projected variational Monte Carlo scheme augmented by a Zeeman-like term and Jastrow factors), nor towards other symmetry-broken phases such as valence-bond crystals, lattice nematics, or chiral spin liquids. We address possible instabilities of the candidate U(1) Dirac-like phases towards such symmetry-broken states in Sec.~\ref{sec:end}. Within these caveats, the present analysis leads to the phase diagram presented in Fig.~\ref{fig:PD}.

To build some intuition, let us first discuss the three limiting cases. For small $J^{}_1/J'_1$ and $J^{}_2/J'_1$, all states with nonvanishing mean-field parameters on the $J'_1$ bonds effectively reduce to a triangular plaquette phase, i.e., the mean-field parameters on the other bonds (such as, e.g., along the sides of the hexagons) turn out to be negligibly small or zero. In Fig.~\ref{fig:PD}, this phase is represented by the gray region in the bottom-left corner. When $J^{}_1/J'_1$ is increased to much larger values (the dark green region in the phase diagram), the system belongs to a zero-flux hexagonal plaquette singlet phase, as expected by virtue of the bonds along the perimeter of the hexagons being the strongest. All QSL \textit{Ans\"atze} with $\phi^{}_h=0$ collapse to this zero-flux state in this regime. On the other hand, for large $J^{}_2/J'_1$, the mean-field parameters evaluate to be negligibly small on the $J^{}_1$ and $J'^{}_1$ bonds and the system in this phase is comprised of disconnected singlet chains. 

Interestingly, in the intermediate region between these limits, two kinds of spin liquids (labeled QSL I and QSL II) are seen to emerge in Fig.~\ref{fig:PD}. The \textit{Ans\"atze} denoted U3$\pi0\pi$ and U3$0\pi\pi$ yield the QSL I and QSL II phases, respectively; both correspond to gapped U(1) states. While we do allow for the possibility of $\mathbb{Z}_2$ QSL states in the phase diagram, in our mean-field calculations, we find the self-consistently determined amplitudes of the terms that are responsible for breaking the U(1) IGG to $\mathbb{Z}_2$ to be very small. Strictly speaking, the energies of the $\mathbb{Z}_{2}$ states are lower than those of their optimal parent U(1) states by $\sim \mathcal{O}(10^{-5})J_{1}'$. Hence, these $\mathbb{Z}_2$ \textit{Ans\"atze} collapse to their lowest-energy parent U(1) states (though the IGG is still broken down, even if weakly, from U(1) to $\mathbb{Z}_2$). For instance, in the QSL I region of the phase diagram, the $\mathbb{Z}_2$ states labeled by Z11100, Z11101, Z31110, and Z31111 effectively behave as the parent U(1) state U3$\pi$0$\pi$. A similar argument holds for the parent U(1) state U30$\pi\pi$ and its $\mathbb{Z}_2$ descendants---Z10000, Z10001, Z30010, and Z30011---in the QSL II region. Further resolution of this delicate energetic competition is intimately tied to the fate of the gapped U(1) parent QSL, i.e., the instability to which it flows once gauge fluctuations beyond mean-field are accounted for~\cite{Polyakov-1977}, e.g., upon Gutzwiller projection.

\section{Characterization of Ans\"atze}

\label{sec:U12NN}

After identifying the candidate QSL states on the ruby lattice, we now examine the properties of their spinon excitations in more detail. Given that U(1) solutions appear to be energetically indistinguishable from the $\mathbb{Z}_2$ ones in our self-consistent mean-field phase diagram above, here, we choose to focus only on the various U(1) \textit{Ans\"atze}. Moreover, motivated by the results of \citet{Schmoll-2024} suggesting a QSL ground state of the \textit{isotropic} $S$\,$=$\,$1/2$ Heisenberg antiferromagnet on the ruby lattice, with $J^{}_1$\,$=$\,$J'^{}_1$, $J^{}_2$\,$=$\,$0$, we restrict our analysis to mean-field Hamiltonians which include up to 2NN terms only. The properties of the realizable mean-field U(1) \textit{Ans\"atze} when allowing for up to 3NN hoppings are elaborated on in Appendix~\ref{app:3NN}.

The properties of any \textit{Ansatz} and its excitations depend solely on the mean-field parameters, which---in principle---should be computed self-consistently by optimization with respect to a given model (as we indeed did in obtaining Fig.~\ref{fig:PD}). However, to avoid such model dependencies, it is more convenient to discuss the general properties of the \textit{Ansatz} itself, without reference to any underlying microscopic Hamiltonian. Therefore, in the following, we provide a summary of the generic spinon excitation spectra and dynamical structure factors of the different U(1) \textit{Ans\"atze}, adopting the same gauge choices as in Sec.~\ref{sec:u1_ansatze}.

\label{sec:spectrum}

First, in Fig.~\ref{fig:U12NNSBS}, we present the spinon spectra obtained by fixing the magnitude of the symmetry-allowed first- and second-nearest-neighbor hoppings to unity, and setting all further-neighbor hoppings to zero. We plot the energy along a high-symmetry path $\Gamma\rightarrow $\,M\,$\rightarrow$\,K\,$\rightarrow\Gamma$ in the first Brillouin zone [light gray hexagon in Fig.~\ref{fig:fig1}(b)] for the \textit{Ans\"atze} that are realizable in a single unit cell, and along $\Gamma\rightarrow$\,X\,$\rightarrow$\,M\,$\rightarrow$\, Y\,$\rightarrow\Gamma\rightarrow$\,M of the reduced Brillouin zone [dark gray rectangle in Fig.~\ref{fig:fig1}(b)] for those \textit{Ans\"atze} which are realizable only with a doubled unit cell. 

\begin{figure*}[t]	\includegraphics[width=\linewidth]{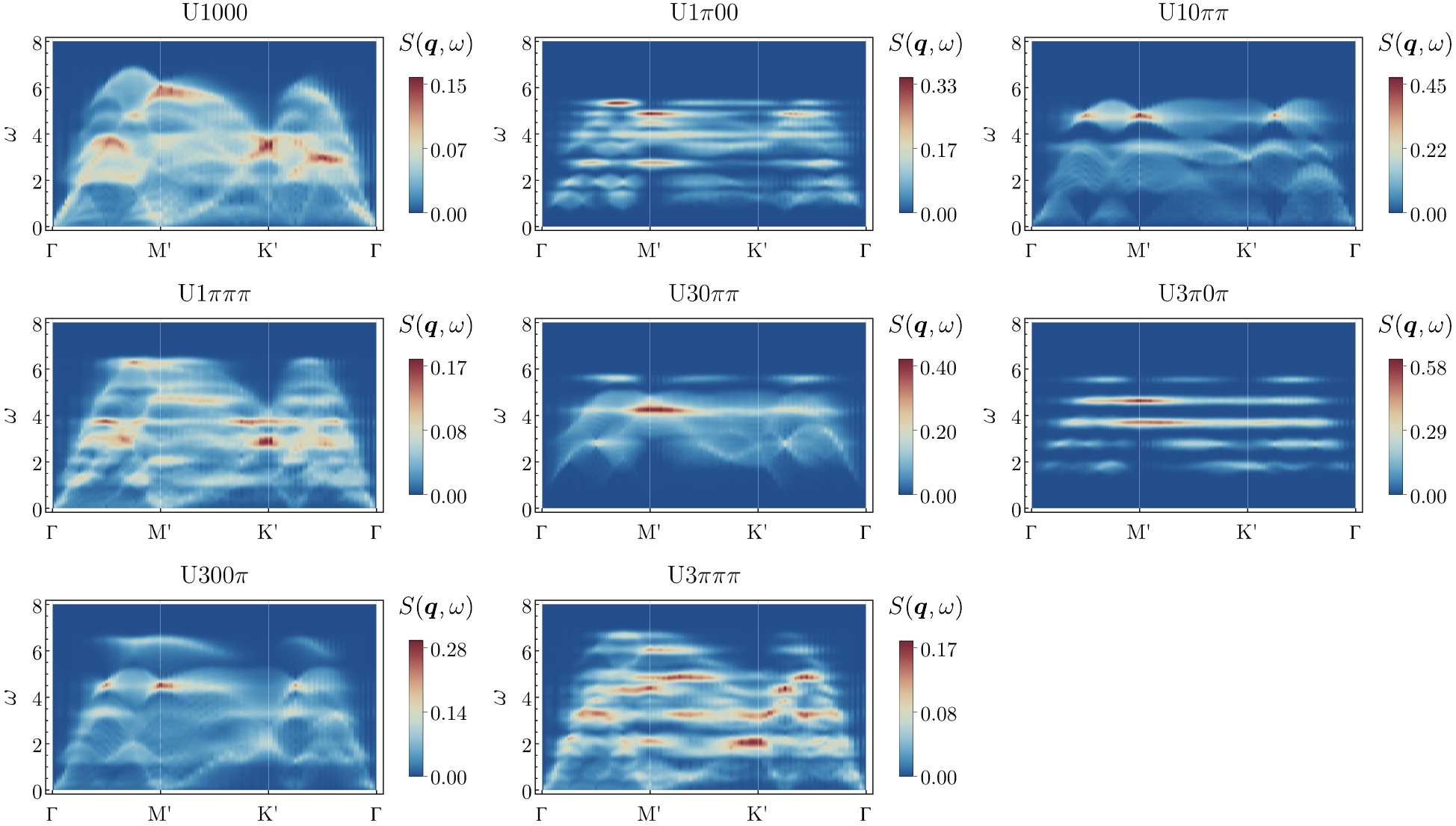}
\caption{Dynamical structure factors of the eight different U(1) \textit{Ans\"atze} realizable with first- and second-nearest-neighbor mean-field amplitudes. The DSF is evaluated along a high-symmetry path in the extended Brillouin zone [see Fig.~\ref{fig:fig1}(b)] for a system size of $14\times14\times6$ sites.} 
\label{fig:U12NNDSF}
\end{figure*}

To further characterize these states, we calculate the dynamical structure factor (DSF), defined by the two-time momentum-resolved spin-spin correlation function
\begin{equation}
 \mathcal{S}^{\mu\nu}(\mathbf{q},\omega)=\int \frac{d\tau}{2\pi N} e^{\dot\iota\omega\tau}\sum_{\langle i,j\rangle}e^{\dot\iota\mathbf{q}\cdot(\mathbf{r}_i-\mathbf{r}_j)}\left\langle \hat{\mathbf{S}}^\mu_i(\tau)\hat{\mathbf{S}}^\nu_j(0)\right\rangle,
\end{equation}
where $\mu,\nu=x,y,z$. Owing to the spin-rotational symmetry of our problem, it suffices to compute only the longitudinal component, i.e., $\mu$\,$=$\,$\nu$\,$=$\,$z$. Figure~\ref{fig:U12NNDSF} illustrates the DSFs for the U(1) \textit{Ans\"atze} realizable up to 2NN---plotted along the high-symmetry lines $\Gamma\rightarrow \mathrm{M}'\rightarrow \mathrm{K}'\rightarrow\Gamma$ of the extended Brillouin zone [see Fig.~\ref{fig:fig1}(b)]---for a system with $14\times14\times6$ sites and parameters identical to those chosen for the spinon dispersion plots. These structure factors provide a reference for direct comparison to neutron-scattering measurements. 

Based on the information in Figs.~\ref{fig:U12NNSBS} and \ref{fig:U12NNDSF}, we can note some salient features of the different \textit{Ans\"atze}:
\begin{itemize}
\item U1000: This \textit{Ansatz} corresponds to the uniform RVB state where all symmetries act nonprojectively. For the reference parameters chosen, the spectrum consists of a nodal line along $\overline{\Gamma \mathrm{M}}$ at the Fermi level. However, this nodal Fermi line is an artifact of the particular choice of parameters, i.e., $\chi^{}_1$\,$=$\,$\chi'_{1}$. In general, this state exhibits a gapped excitation spectrum for $\chi^{}_{1}$\,$>$\,$\chi'_{1}$ and is gapless (at the $\Gamma$ point) for $\chi^{}_{1}$\,$<$\,$\chi'_{1}$. In the DSF in Fig.~\ref{fig:U12NNDSF}, we observe finite intensity down to $\omega=0$ around the $\Gamma$ and $\mathrm{M}'$ points, which is a consequence of scattering near the Fermi surface.
 
\item U1$\pi$00: In general, the excitations of this \textit{Ansatz} are gapped and the bands are quasiflat, as also reflected in its DSF in Fig.~\ref{fig:U12NNDSF}. Note that there are three major regions in which the spectral weight is concentrated. The first occurs at $\omega$\,$\approx$\,$2.7$, as a result of excitations from the lowest filled band to the lowest empty band in Fig.~\ref{fig:U12NNSBS}. The spectral weight within $5.0<\omega<5.5$ arises due to scattering from the lowest filled band to the two highest nearly flat bands.
 
\item U10$\pi\pi$: As noted in Table~\ref{maple} of Appendix~\ref{app:3NN}, this \textit{Ansatz} is also realized as a U(1) QSL state (labeled as UC01) on the maple-leaf lattice \cite{Sonnenschein-2024}. On both the ruby and maple-leaf lattices, the spectrum clearly shows the presence of a Fermi surface for generic parameters. Correspondingly, the DSF also exhibits finite intensity down to zero energy.
 
\item U1$\pi\pi\pi$: The spectrum of this \textit{Ansatz} exhibits nodal structures (as can be seen from the band touchings along the segments $\overline{\Gamma \mathrm{X}}$, $\overline{\Gamma \mathrm{M}}$, and $\overline{\mathrm{MY}}$) at the Fermi level for $\chi^{}_{1}=\chi'_{1}$ whereas for $\chi'_{1}>1.2\chi^{}_{1}$, it manifests as a Dirac spin liquid. Once again, we find finite intensity down to zero energy in the DSF due to the presence of a Fermi surface.
 
\item U30$\pi\pi$: For this \textit{Ansatz}, the spectrum is generically gapped. The DSF, notably, features high intensity around the $\mathrm{M}'$ point at $\omega\approx4.3$. 
 
\item U3$\pi$0$\pi$: As for the previous case, we find that the spinon dispersions of this \textit{Ansatz} too are generically gapped. The flat bands in the spectrum are reflected in the DSF as well, as seen in Fig.~\ref{fig:U12NNDSF}. 
 
\item U300$\pi$: This \textit{Ansatz} exhibits a nodal Fermi surface for $\chi^{}_{1}\geq 1.2\chi'_{1}$ and is gapped otherwise. Similar to the U1000 and U1$\pi\pi\pi$ cases above, the DSF for this \textit{Ansatz} shows finite intensities down to zero energy around the $\Gamma$ and $\mathrm{M}'$ points due to the presence of a nodal Fermi surface. 
 
\item U3$\pi\pi\pi$: Lastly, this \textit{Ansatz} hosts spinon excitations with Fermi nodal structures passing through the $\mathrm{M}$ point for $\chi^{}_{1}=\chi'_{1}$ and is gapped otherwise. The DSF again reveals spectral weight down to zero energies but additionally, there are flat high-intensity regions at relatively high energies originating from the quasiflatness of the excitation spectrum. 
\end{itemize}

\section{Discussion and outlook}
\label{sec:end}

In this work, we have classified and constructed quadratic spinon Hamiltonians on the ruby (or bounce) lattice with U(1) and $\mathbb{Z}_{2}$ low-energy gauge groups. This is achieved using the projective symmetry group framework for fermionic spinons and respecting {\it all} symmetries of the spin-$1/2$ lattice model, i.e., space-group, time-reversal, and spin-rotation, thus realizing fully symmetric QSLs. We obtain 50 U(1) and 64 $\mathbb{Z}_{2}$ algebraic PSGs, leading to as many distinct QSL phases at the mean-field level. The realization of these PSGs via {\it singlet} mean-field {\it Ansätze} restricted to short-range (up to second-neighbor) amplitudes, of relevance to the models concerned, leads to only 8 U(1) and 18 $\mathbb{Z}_{2}$ distinct QSL phases. 

In light of a recent tensor-network study~\cite{Schmoll-2024} lending evidence for a symmetric gapless QSL ground state of the $J_{1}=J_{1}'$ Heisenberg antiferromagnet on the ruby lattice, our classification sets the stage for future works aimed at characterizing its precise microscopic nature. The Gutzwiller-projected static and dynamical spin structure factors for the different variational states could be compared to those obtained from unconstrained numerical approaches to narrow down and pinpoint promising candidate ground states. For a precise identification of the nature of the spin liquid, it would be worthwhile to perform a variational Monte Carlo study towards optimizing the corresponding Gutzwiller-projected wave functions and evaluating the energetic competitiveness of the gapless U(1) and $\mathbb{Z}_2$ states for Hamiltonian parameter regimes displaying QSL ground states. This would also enable one to address the intriguing issue of how bond depletion induces a zero-temperature quantum phase transition from the gapped QSL on the maple-leaf lattice to a gapless QSL on the ruby lattice~\cite{Schmoll-2024}, since the latter lattice is a bond-diluted version of the former~\cite{Richter2004}. An alternate treatment of these {\it Ansätze} would be their assessment within the pseudofermion functional renormalization group framework~\cite{Mueller-2024} by using the effective low-energy vertex functions (instead of the bare couplings) within a self-consistent Fock-like mean-field scheme to compute low-energy theories for emergent spinon excitations~\cite{Hering-2019,Hering-2022}. 

Since our classification features gapless states with a rich variety of Fermi surfaces, consisting of either extended surfaces, lines, or Dirac points, and given that the ruby lattice has the same coordination number as the kagome, it would be useful to understand the impact of quantum fluctuations. In particular, determining the stability of these states towards $\mathbb{Z}_{2}$ QSLs~\cite{Iqbal-2011b}, chiral QSLs~\cite{Hu-2015}, valence-bond crystals~\cite{Iqbal-2011a,Iqbal-2012}, and magnetic orders~\cite{Chauhan-2023} possibly driven by monopole condensation~\cite{Song-2019,Song-2020,Kiese-2023,Budaraju-2023} and fermion bilinear terms~\cite{Hermele-2008,Iqbal-2011a,Iqbal-2012}, will constitute fruitful future endeavors. Of particular significance would be identifying perturbations to the Hamiltonian (e.g., multispin exchanges) that preserve the gapless nature of the QSL. Lastly, it would also be interesting to extend the current analysis to classify chiral spin liquids since, typically, the inclusion of longer-range Heisenberg couplings stabilizes noncoplanar orders in corresponding classical models. For $S$\,$=$\,$1/2$ spins, such orders could melt via quantum fluctuations and potentially give rise to chiral spin liquids, as recently investigated on the related maple-leaf lattice~\cite{Gembe-2024}.

\begin{acknowledgements}
We thank R. Thomale for useful discussions. This work was initiated and completed at the Aspen Center for Physics, which is supported by National Science Foundation Grant No.~PHY-2210452 and a grant from the Simons Foundation (1161654, Troyer). A.M. is supported by the Deutsche Forschungsgemeinschaft (DFG, German Research Foundation) through Project-ID 258499086 - SFB 1170 and through the W\"urzburg-Dresden Cluster of Excellence on Complexity, Topology and Dynamics in Quantum Matter - ctd.qmat Project-ID 390858490 - EXC 2147. R.S. is supported by the Princeton Quantum Initiative Fellowship. This research was also supported in part by Grant No.~NSF PHY-2309135 to the Kavli Institute for Theoretical Physics (KITP) during the ``A New Spin on Quantum Magnets'' program in summer 2023. Y.I.\ acknowledges support from the ICTP through the Associates Programme, from the Simons Foundation through Grant No.~284558FY19, and from IIT Madras through the Institute of Eminence (IoE) program for establishing QuCenDiEM (Project No.~SP22231244CPETWOQCDHOC). Y.I.\ also acknowledges the use of the computing resources at HPCE, IIT Madras.
\end{acknowledgements}

\newpage

\pagebreak
\appendix
\section{Gauge-enriched symmetry relations}
\label{sec:genric_gauge_con}
Since the PSGs $G_\mathcal{O}(x,y;s)$ are the projective generalizations of the symmetry group with elements $\mathcal{O}\in\{$lattice space group, time reversal$\}$, they necessarily follow the same algebraic relations that all the symmetry elements do, i.e., Eqs.~\eqref{eq:T1T2}--\eqref{eq:time_O}. On the right-hand side of all these relations is the identity element. For the U(1) and $\mathbb{Z}_2$ IGGs that we consider, the projective action of the identity is determined either up to a phase $\xi$, where $0 \le \xi < 2\pi$, or a sign parameter $\eta = \pm 1$, respectively. This leads to the following set of algebraic conditions for the PSGs:
\begin{widetext}
\begin{align}
&G^{}_{T_1}(x,y;s) G^{}_{T_{2}}(x-1,y;s) G^{-1}_{T_1}(x,y-1;s)G^{-1}_{T_{2}}(x,y;s) =e^{\dot\iota\xi\tau^3}/\eta\tau^0 \label{eq:gauge_T1T2} \\
&G^{}_R(x,y;s)G^{}_{T_2}(y,y-x;R^{-1}(s))G^{-1}_R(x+1,y;s) G^{}_{T_1}(x+1,y;s) =e^{\dot\iota\xi^{}_{RT_1}\tau^3}/\eta^{}_{RT_1}\tau^0 \label{eq:gauge_R_T1}\\
&G^{}_R(x,y;s)G^{-1}_{T_2}(y,y-x+1;R^{-1}(s)) G^{-1}_{T_1}(y+1,y-x+1;R^{-1}(s))G^{-1}_R(x,y+1;s) G^{}_{T_2}(x,y+1;s) =e^{\dot\iota\xi^{}_{RT_2}\tau^3}/\eta^{}_{RT_2}\tau^0 \label{eq:gauge_R_T2}\\
&G^{}_R(x,y;s)G^{}_R(y,y-x;R^5(s))G^{}_R(y-x,-x;R^4(s)) \notag\\ 
&G^{}_R(-x,-y;R^3(s))G^{}_R(-y,x-y;R^2(s)) G^{}_R(x-y,x;R(s)) =e^{\dot\iota\xi^{}_{R}\tau^3}/\eta^{}_{R}\tau^0 \label{eq:gauge_R}\\
&G^{}_\sigma(x,y;s)G^{-1}_{T_2}(x,x-y+1;\sigma^{-1}(s)) G^{-1}_{T_1}(x+1,x-y+1;\sigma^{-1}(s))G^{-1}_\sigma(x+1,y;s) G^{}_{T_1}(x+1,y;s) =e^{\dot\iota\xi^{}_{\sigma T_1}\tau^3}/\eta^{}_{\sigma T_1}\tau^0 \label{eq:gauge_sig_T1}\\
&G^{}_\sigma(x,y;s)G^{}_{T_2}(x,x-y;\sigma^{-1}(s))G^{-1}_\sigma(x,y+1;s) G^{}_{T_2}(x,y+1;s) =e^{\dot\iota\xi^{}_{\sigma T_2}\tau^3}/\eta^{}_{\sigma T_2}\tau^0 \label{eq:gauge_sig_T2}\\
&G^{}_\sigma(x,y;s)G^{}_\sigma(x,x-y;\sigma(s)) =e^{\dot\iota\xi^{}_{\sigma}\tau^3}/\eta^{}_\sigma\tau^0 \label{eq:gauge_sig}\\
&G^{}_R(x,y;s)G^{}_\sigma(y,y-x;R^{-1}(s))G^{}_R(y,x;R\sigma(s)) G^{}_\sigma(x,x-y;\sigma(s)) =e^{\dot\iota\xi^{}_{R\sigma}\tau^3}/\eta^{}_{R\sigma}\tau^0\label{eq:gauge_R_sig}\\
&G^{}_\mathcal{T}(x,y;s)G^{}_\mathcal{O}(x,y;s)G^{-1}_\mathcal{T}(\mathcal{O}^{-1}(x,y;s)) G^{-1}_\mathcal{O}(x,y;s) =e^{\dot\iota\xi^{}_{\mathcal{T}\mathcal{O}}\tau^3}/\eta^{}_{\mathcal{T}\mathcal{O}}\tau^0\label{eq:gauge_time_O}\\
&\left[G^{}_\mathcal{T}(x,y;s)\right]^2 =e^{\dot\iota\xi^{}_{\mathcal{T}}\tau^3}/\eta^{}_{\mathcal{T}}\tau^0 \label{eq:gauge_time}.
\end{align}
\end{widetext}
The solutions obtained from the relations above define the algebraic PSGs for a given symmetry group. 

Here, we consider a choice of gauge such that the IGG defines a global gauge group, i.e., the IGG elements take the form of a global U(1) phase parameter $\xi$ (for a U(1) IGG) or a global sign parameter $\eta$ (for a $\mathbb{Z}_2$ IGG). Consequently, this gauge choice has the advantage that the link fields $u_{ij}$, i.e., the \textit{Ans\"atze}, manifestly reflect their U(1) or $\mathbb{Z}_2$ nature. In the following, we sequentially treat the U(1) and $\mathbb{Z}_2$ IGGs.
\section{$\mathbf{U(1)}$ PSG classification}
\label{app:u1_psg_derivation}
In the canonical form, a U(1) \textit{Ansatz} includes only real and imaginary hopping parameters, i.e., $u_{ij}=\dot{\iota}\chi^0_{ij} \tau^0 + \chi^3_{ij} \tau^3$, and the structure of the gauge transformations that keep this canonical form intact is given by
\begin{equation}
 \label{eq:canonical_u1_gauge_structure}
 G^{}_\mathcal{O}(x,y;s)=g^{}_3\left( \phi^{}_\mathcal{O}(x,y;s)\right)\left(\dot{\iota}\tau^1\right)^{w^{}_\mathcal{O}},
\end{equation}
where $w^{}_\mathcal{O}$ can take values 0,1, $\mathcal{O}\in\{T_1,T_2,R,\sigma,\mathcal{T}\}$, and $g^{}_3 (\lambda) \equiv \exp(\dot{\iota}\lambda \tau^3)$.

\subsection{Lattice symmetries}
For $\mathcal{O}\in\{T_1,T_2\}$, there are three cases to consider for $w^{}_\mathcal{O}$: (i) $(w^{}_{T_1},w^{}_{T_2})=(0,0)$, (ii) $(w^{}_{T_1},w^{}_{T_2})=(1,0)$, and (iii) $(w^{}_{T_1},w^{}_{T_2})=(1,1)$. As cases (ii) and (iii) do not satisfy Eqs.~\eqref{eq:gauge_R_T1} and \eqref{eq:gauge_R_T2}, we restrict ourselves to case (i) only, i.e., $w^{}_{T_1}=w^{}_{T_2}=0$. Now, using an appropriate local gauge transformation, one can always fix
\begin{equation}
 \phi^{}_{T_1}(x,0;s)=\phi^{}_{T_2}(x,y;s)=0.
\end{equation}
Together with Eq.~\eqref{eq:gauge_T1T2}, this yields
\begin{equation}\label{eq:tran^{}_1_u1}
 G^{}_{T_1}(x,y;s)=g^{}_3(y\,\xi),\; G^{}_{T_2}(x,y;s)=\tau^0.
\end{equation}

Next, to find the PSG solutions for point-group symmetries, we first define $\Delta_i\phi^{}_{\mathcal{O}}(x,y;s) \equiv \phi^{}_{\mathcal{O}}(x,y;s)-\phi^{}_{\mathcal{O}}[T^{-1}_i(x,y;s)]$. Then, using Eq.~\eqref{eq:tran^{}_1_u1}, we can recast Eqs.~\eqref{eq:gauge_R_T1} and~\eqref{eq:gauge_R_T2} as
\begin{equation}\label{eq:R_tran_1}
	\left.\begin{aligned}
&\Delta^{}_1\phi^{}_R(x,y;s)=-\xi^{}_{RT_1}+y\,\xi,\\
&\Delta^{}_2\phi^{}_R(x,y;s)=-\xi^{}_{RT_2}+(-1)^{w^{}_R}(x-y)\xi,\\
\end{aligned}\right.
\end{equation}
which must obey the consistency relation
\begin{alignat}{1}
\label{eq:consistent}
\Delta^{}_1\phi^{}_\mathcal{O}(x,y;s)&+\Delta^{}_{2}\phi^{}_{\mathcal{O}}\left[T^{-1}_1(x,y;s)\right]\\
\nonumber
&=\Delta^{}_{2}\phi^{}_\mathcal{O}(x,y;s)+\Delta^{}_1\phi^{}_{\mathcal{O}}\left[T^{-1}_{2}(x,y;s)\right],
\end{alignat}
thus yielding
\begin{equation}\label{eq:R_consistent}
 (1-(-1)^{w^{}_{R}})\xi=0.
\end{equation}
This implies that for $w^{}_R$\,$=$\,$1$, $2\xi$\,$=$\,$0$, while there are no constraints on $\xi$ for $w^{}_R$\,$=$\,$0$. Consequently, from Eq.~\eqref{eq:R_tran_1}, we find the solution
\begin{equation}\label{eq:R_sol_u_1}
\left.\begin{aligned}
 \phi^{}_R(x,y;s)&=y\left[x-\frac{(y+1)}{2}\right]\xi -\left(x\xi^{}_{RT_1}+y\xi^{}_{RT_2}\right)+\rho^{}_{R,s},\\
\end{aligned}\right.
\end{equation}
where $\rho^{}_{R,s}$ is the sublattice-dependent part of the U(1) phase $\phi^{}_R(x,y;s)$, i.e., $\rho^{}_{R,s}\equiv\phi^{}_R(0,0;s)$. 

Similarly, Eqs.~\eqref{eq:gauge_sig_T1} and \eqref{eq:gauge_sig_T2} lead to
\begin{align}\label{eq:sig_tran_1}
&\Delta^{}_1\phi^{}_\sigma(x,y;s)=-\xi^{}_{\sigma T_1}-(-1)^{w^{}_\sigma}x\xi+(1+(-1)^{w^{}_\sigma})y\xi,\notag\\
&\Delta^{}_2\phi^{}_\sigma(x,y;s)=-\xi^{}_{\sigma T_2}.
\end{align}
The consistency condition~\eqref{eq:consistent} with $\mathcal{O}=\sigma$ requires
\begin{equation}\label{eq:sigma_consistent}
 (1+(-1)^{w^{}_{\sigma}})\xi=0.
\end{equation}
This shows that for $w^{}_\sigma$\,$=$\,$0$, $2\xi$\,$=$\,$0$ but there are no constraints on $\xi$ for $w^{}_\sigma$\,$=$\,$1$. Consequently, from Eq.~\eqref{eq:sig_tran_1}, we obtain the solution 
\begin{equation}\label{eq:sig_sol_u_1}
\left.\begin{aligned}
&\phi^{}_\sigma(x,y;s)=\frac{1}{2}x(x+1)\xi-(x\xi^{}_{\sigma T_1}+y\xi^{}_{\sigma T_2})+\rho^{}_{\sigma,s},\\
\end{aligned}\right.
\end{equation}
where $\rho^{}_{\sigma,s}$ is defined similarly to $\rho^{}_{R,s}$. Furthermore, the cyclic condition for $R$ specified by Eq.~\eqref{eq:gauge_R} gives
\begin{align}\label{eq:R_cyclic_u}
\xi^{}_R&=\rho^{}_{R,s}+\rho^{}_{R,R^2(s)}+\rho^{}_{R,R^{4}(s)}\notag\\
&+(-1)^{w^{}_R}\left(\rho^{}_{R,R(s)}+\rho^{}_{R,R^3(s)}+\rho^{}_{R,R^{5}(s)}\right).
\end{align}
while the one for $\sigma$ in Eq.~\eqref{eq:gauge_sig} gives
\begin{equation}\label{eq:sig_cyclic_u}
\rho^{}_{\sigma,s}+(-1)^{w^{}_\sigma}\rho^{}_{\sigma,\sigma(s)}=\xi^{}_\sigma,
\end{equation}
and
\begin{align}
w^{}_\sigma=0:\;&2\xi^{}_{\sigma T_1}+\xi^{}_{\sigma T_2}=0,\label{eq:sig_cyclic_u_1}\\
w^{}_\sigma=1:\;&\xi^{}_{\sigma T_2}=0\label{eq:sig_cyclic_u_2}.
\end{align}

We know that under a local gauge transformation $W(x,y;s)$, an element of the projective symmetry group $G_\mathcal{O}$ transforms as $G_\mathcal{O}(x,y;s)\rightarrow W^\dagger(x,y;s)$ $G_\mathcal{O}(x,y;s) W[\mathcal{O}^{-1}(x,y;s)]$. Thus, a local gauge transformation of the form
\begin{equation}
\label{eq:local}
 W(x,y;s)=g^{}_3(x\theta^{}_x+y\theta^{}_y)
\end{equation}
does not affect the $G_{T_i}$ up to a global phase that has no consequence for the \textit{Ans\"atze} and can be safely ignored. However, the phases $\xi^{}_{\mathcal{O}T_i}$ do get modified locally, so one can choose an appropriate $\theta^{}_{x/y}$ to set
\begin{equation}
 \xi^{}_{RT_1}=\xi^{}_{RT_2}=0.
\end{equation}

Also, note that we have the condition in Eq.~\eqref{eq:gauge_R_sig} yielding
\begin{align}
(w^{}_R,w^{}_\sigma)=(0,0):\;&\xi^{}_{\sigma T_i}=0,\label{eq:sig_R_u_00}\\
(w^{}_R,w^{}_\sigma)=(0,1):\;&\xi^{}_{\sigma T_i}=0,\label{eq:sig_R_u_01}\\
(w^{}_R,w^{}_\sigma)=(1,0):\;&\xi^{}_{\sigma T_1}=\xi^{}_{\sigma T_2}=\frac{2\pi p_{\sigma {T_1}}}{3},\label{eq:sig_R_u_10}\\
(w^{}_R,w^{}_\sigma)=(1,1):\;&\xi^{}_{\sigma T_i}=0,\label{eq:sig_R_u_11}
\end{align}
and
\begin{align}\label{eq:R_sig_u}
\rho^{}_{R,s}&+(-1)^{w^{}_R+w^{}_\sigma}\rho^{}_{R,R\sigma(s)}\notag\\
&+(-1)^{w^{}_R}\rho^{}_{\sigma,R^{-1}(s)}+(-1)^{w^{}_\sigma}\rho^{}_{\sigma,\sigma(s)}=\xi^{}_{R\sigma}.
\end{align}
So, all phases $\xi^{}_{\mathcal{O}T_i}$ are zero except in the case when $(w^{}_R,w^{}_\sigma)$\,$=$\,$(1,0)$, as can be seen in Eq.~\eqref{eq:sig_R_u_10}. However, these can also be set to zero with a gauge transformation of the form $W(x,y;s)=g^{}_3((x+y)\xi^{}_{\sigma T_1})=g^{}_3((x+y)\, 2\pi p_{\sigma {T_1}}/3)$. Thus, our solutions simplify to the forms:
\begin{equation}\label{eq:R_sig_sol_u}
\left.\begin{aligned}
 \phi^{}_R(x,y;s)&=y\left[x-\frac{1}{2}(y+1)\right]\xi+\rho^{}_{R,s},\\
 \phi^{}_\sigma(x,y;s)&=\frac{1}{2}x(x+1)\xi+\rho^{}_{\sigma,s}.\\
\end{aligned}\right.
\end{equation}

We are still left though with a sublattice-dependent gauge degree of freedom. Under a gauge transformation of the form $W(x,y;s)=g^{}_3(\theta^{}_s)$, the phases $\rho^{}_{\mathcal{O},s}$ transform as
\begin{equation}
\left.\begin{aligned}
&\rho^{}_{R,s}\rightarrow\rho^{}_{R,s}-\theta^{}_{s}+(-1)^{w^{}_R}\theta^{}_{R^{-1}(s)},\\
&\rho^{}_{\sigma,s}\rightarrow\rho^{}_{\sigma,s}-\theta^{}_{s}+(-1)^{w^{}_\sigma}\theta^{}_{\sigma^{-1}(s)}.\\
\end{aligned}\right.
\end{equation}
With appropriate choices of the parameters $\theta$ and using Eq.~\eqref{eq:R_cyclic_u}, one can fix
\begin{align}
w^{}_R=0:\;&\rho^{}_{R,s}=\xi^{}_R/6=0\label{eq:rho_g_fix_wr_0},\\
w^{}_R=1:\;&\rho^{}_{R,s}=n^{}_{R}\pi\delta_{s,6}\label{eq:rho_g_fix_wr_1},\\
w^{}_\sigma=1:\;&\rho^{}_{\sigma,1}=0\label{eq:rho_g_fix_ws_1}.
\end{align}
In Eq.~\eqref{eq:rho_g_fix_wr_0}, we have set the phase to zero using the fact that a global phase has no consequence on the U(1) \textit{Ans\"atze}. Furthermore, one can use Eqs.~\eqref{eq:sig_cyclic_u} and \eqref{eq:R_sig_u} to fix the following: 
\begingroup
\allowdisplaybreaks
\begin{align}
(w^{}_R,w^{}_\sigma)&=(0,0):\notag\\
\rho^{}_{\sigma,s}&=\{0,n^{}_{R\sigma}\pi,0,n^{}_{R\sigma}\pi,0,n^{}_{R\sigma}\pi\}\label{eq:sig_fix_u_00},\\
(w^{}_R,w^{}_\sigma)&=(0,1):\notag\\
\rho^{}_{\sigma,s}&=\{0,n^{}_{R\sigma}\pi,0,n^{}_{R\sigma}\pi,0,n^{}_{R\sigma}\pi\}\label{eq:sig_fix_u_01},\\
(w^{}_R,w^{}_\sigma)&=(1,0):\notag\\
\rho^{}_{\sigma,s}&=\{0,n^{}_{R\sigma}\pi,0,n^{}_{R\sigma}\pi,n^{}_R\pi,n^{}_{R\sigma}\pi\}\label{eq:sig_fix_u_10},\\
(w^{}_R,w^{}_\sigma)&=(1,1):\notag\\
\rho^{}_{\sigma,s}&=\{0,\xi^{}_{R\sigma},0,\xi^{}_{R\sigma},n^{}_R\pi,\xi^{}_{R\sigma}\}\label{eq:sig_fix_u_11}.
\end{align}
\subsection{Time-reversal symmetry}
Now, we proceed to find the PSG solutions for time-reversal symmetry. Using Eq.~\eqref{eq:gauge_time_O} for $\mathcal{O}\in T_1,T_2$, we have
\begin{equation}
\label{eq:time_u1_sol_1}
\left.\begin{aligned}
&\Delta_1\phi^{}_{\mathcal{T}}(x,y;s)=\xi^{}_{\mathcal{T}T_1}+[1-(-1)^{w^{}_{\mathcal{T}}}]y\,\xi,\\
&\Delta_2\phi^{}_{\mathcal{T}}(x,y;s)=\xi^{}_{\mathcal{T}T_2}.\\
\end{aligned}\right.
\end{equation}
The consistency condition~\eqref{eq:consistent} for $\mathcal{O}\in\mathcal{T}$ posits 
\begin{equation}
 [1-(-1)^{w^{}_{\mathcal{T}}}]\xi=0,
\end{equation}
implying that for $w^{}_{\mathcal{T}}=1$, $2\xi=0$. The solution for $G_\mathcal{T}$ can be obtained from Eq.~\eqref{eq:time_u1_sol_1} as
\begin{equation}\label{eq:time_u1_sol_2}
 \phi^{}_\mathcal{T}(x,y;s)=x\xi^{}_{\mathcal{T}T_1}+y\xi^{}_{\mathcal{T}T_2}+\rho^{}_{\mathcal{T},s}\, .
\end{equation}
Now, let us consider the remaining conditions separately for $w^{}_\mathcal{T}=0$ and $w^{}_\mathcal{T}=1$.

\subsubsection{$w^{}_\mathcal{T}=0$}
In this case, Eq.~\eqref{eq:gauge_time} yields
\begin{equation}\label{eq:utime_fixing_1} 2\xi^{}_{\mathcal{T}T_i}=0,\;\rho^{}_{\mathcal{T},s}=\frac{\xi^{}_{\mathcal{T}}}{2}+n^{}_{\mathcal{T},s}\pi,\;\mbox{ for } n^{}_{\mathcal{T},s}=0,1,
\end{equation}
while, from Eq.~\eqref{eq:gauge_time_O} with $\mathcal{O}\in R$ and using Eq.~\eqref{eq:utime_fixing_1}, we obtain
\begin{align}
 \xi^{}_{\mathcal{T}T_1}=\xi^{}_{\mathcal{T}T_2}&=0 \label{eq:utime_fixing_2a}, \\
 \rho^{}_{\mathcal{T},s}-(-1)^{w^{}_R}\rho^{}_{\mathcal{T},R^{-1}(s)}&=\xi^{}_{\mathcal{T}R}. \label{eq:utime_fixing_2b}
\end{align}
Furthermore, Eq.~\eqref{eq:gauge_time_O} with $\mathcal{O}\in\sigma$ results in
\begin{equation}\label{eq:utime_fixing_4} 
\rho^{}_{\mathcal{T},s}-(-1)^{w^{}_\sigma}\rho^{}_{\mathcal{T},\sigma(s)}=\xi^{}_{\mathcal{T}\sigma}.
\end{equation}
Finally, using Eqs.~\eqref{eq:utime_fixing_1}, \eqref{eq:utime_fixing_2b}, and \eqref{eq:utime_fixing_4}, we can fix $\rho^{}_{\mathcal{T},s}$ as 
\begin{equation}\label{eq:utime_fixing_5} \rho^{}_{\mathcal{T},1}=\rho^{}_{\mathcal{T},3}=\rho^{}_{\mathcal{T},5}=0,\;\rho^{}_{\mathcal{T},2}=\rho^{}_{\mathcal{T},4}=\rho^{}_{\mathcal{T},6}=\pi . 
\end{equation}

\subsubsection{$w^{}_\mathcal{T}=1$}
Similar to the $w^{}_\mathcal{T}=0$ case above, here also, one can fix $\xi^{}_{\mathcal{T}T_1}=\xi^{}_{\mathcal{T}T_2}=0$, i.e.,
\begin{equation}\label{eq:time_u1_wt1_1}
 \phi^{}_\mathcal{T}(x,y;s)=\rho^{}_{\mathcal{T},s}\, .
\end{equation}
First, consider the case when $(w^{}_{R},w^{}_{\sigma}) = (0,0)$. Using Eqs.~\eqref{eq:R_sig_sol_u}, \eqref{eq:rho_g_fix_wr_0}, and the fact that $2\xi=0$ for $w^{}_\sigma=0$, one can write Eq.~\eqref{eq:gauge_time_O} with $\mathcal{O}\in R$ as
\begin{equation}\label{eq:time_wt1_2}
\rho^{}_{\mathcal{T},s}-\rho^{}_{\mathcal{T},R^{-1}(s)}=\xi^{}_{\mathcal{T}R}. 
\end{equation}
Consequently, we find
\begin{equation}\label{eq:time_wt1_3}
\rho^{}_{\mathcal{T},s}=\rho^{}_{\mathcal{T},1}+(s-1)\xi^{}_{\mathcal{T}R},\;\text{with}\; 6\,\xi^{}_{\mathcal{T}R}=0. 
\end{equation}
With the help of the gauge freedom of the IGG, one can set the global phase $\rho^{}_{\mathcal{T},1}$ to zero. Likewise, one can exploit Eq.~\eqref{eq:gauge_time_O} with $\mathcal{O}\in \sigma$ to find $\rho^{}_{\mathcal{T},3}=\rho^{}_{\mathcal{T},1}$ and $\rho^{}_{\mathcal{T},4}=\rho^{}_{\mathcal{T},6}$, which require 
\begin{equation}\label{eq:time_wt1_4}
2\xi^{}_{\mathcal{T}R}=0\Rightarrow \xi^{}_{\mathcal{T}R}=n^{}_{\mathcal{T}R}\pi,\;\text{with}\; n^{}_{\mathcal{T}R}=0,1. 
\end{equation}
Therefore, Eq.~\eqref{eq:time_wt1_3} takes the form
\begin{equation}\label{eq:time_wt1_5}
\rho^{}_{\mathcal{T},s}=n^{}_{\mathcal{T}R}\pi\delta_{s,2/4/6}. 
\end{equation}
Furthermore, one can choose a gauge to fix $\rho^{}_{\mathcal{T},s}=0$. The associated gauge transformation is given by
\begin{equation}\label{eq:time_u1_wt1_gauge}
 W(x,y;s)=g^{}_3\left(\frac{\rho^{}_{\mathcal{T},s}}{2}\right)\, .
\end{equation}
For the case of $(w^{}_{R},w^{}_{\sigma})=(0,0)$, $\rho^{}_{\sigma,s}$ remains unaffected by this gauge transformation while $\rho^{}_{R,s}$ takes the form
\begin{equation}\label{eq:time_u1_wt1_6}
\rho^{}_{R,s}=(-1)^{s+1}n^{}_{\mathcal{T}R}\pi/2, 
\end{equation}
which, after a global phase shift, reduces to $\rho^{}_{R,s}=n^{}_{\mathcal{T}R}\pi\delta_{s,2/4/6}$. Accordingly, for $(w^{}_{R},w^{}_{\sigma})$\,$=$\,$(0,0)$, one has $\rho^{}_{\mathcal{T},s}=0$ with
\begin{align}
&\rho^{}_{R,s}=\{0,n^{}_{\mathcal{T}R}\pi,0,n^{}_{\mathcal{T}R}\pi,0,n^{}_{\mathcal{T}R}\pi\},\notag\\
&\rho^{}_{\sigma,s}=\{0,n^{}_{R\sigma}\pi,0,n^{}_{R\sigma}\pi,0,n^{}_{R\sigma}\pi\}\label{eq:R_sig_fix_u_001}.
\end{align}
The advantage of the choice $\rho^{}_{\mathcal{T},s}=0$, obtained by a gauge transformation of the form~\eqref{eq:time_u1_wt1_gauge}, is that the resultant mean-field parameters on all the bonds include only real hopping terms. This holds true for all other choices of $(w^{}_{R},w^{}_{\sigma})$ as well; however, the phases $\rho^{}_{R,s}$ and $\rho^{}_{\sigma,s}$ may differ from those given in Eqs.~\eqref{eq:sig_fix_u_01}, \eqref{eq:sig_fix_u_10}, and Eq.~\eqref{eq:sig_fix_u_11}.
The new choices for the remaining cases are as follows:
\begin{widetext}
\begin{align}
&(w^{}_{R},w^{}_{\sigma})=(0,1): \rho^{}_{R,s}=0, \,
\rho^{}_{\sigma,s}=\{0,n^{}_{R\sigma}\pi,0,(n^{}_{R\sigma}+n^{}_{\mathcal{T}R})\pi,n^{}_{\mathcal{T}R}\pi,(n^{}_{R\sigma}+n^{}_{\mathcal{T}R})\pi\}, 
\label{eq:R_sig_fix_u_011},\\
&(w^{}_{R},w^{}_{\sigma})=(1,0): \rho^{}_{R,s}=\{0,0,0,0,0,n^{}_{R}\pi\},\, \rho^{}_{\sigma,s}=\{0,n^{}_{R\sigma}\pi,0,n^{}_{R\sigma}\pi,n^{}_R\pi,n^{}_{R\sigma}\pi\}\label{eq:R_sig_fix_u_101},\\
&(w^{}_{R},w^{}_{\sigma})=(1,1): \rho^{}_{R,s}=\{0,0,0,0,0,n^{}_{R}\pi\},\, \rho^{}_{\sigma,s}=\{0,n^{}_{R\sigma}\pi,0,n^{}_{R\sigma}\pi,n^{}_R\pi,n^{}_{R\sigma}\pi\}\label{eq:R_sig_fix_u_111}.
\end{align}
\end{widetext}

\begin{figure*}[t]	\includegraphics[width=0.9\linewidth]{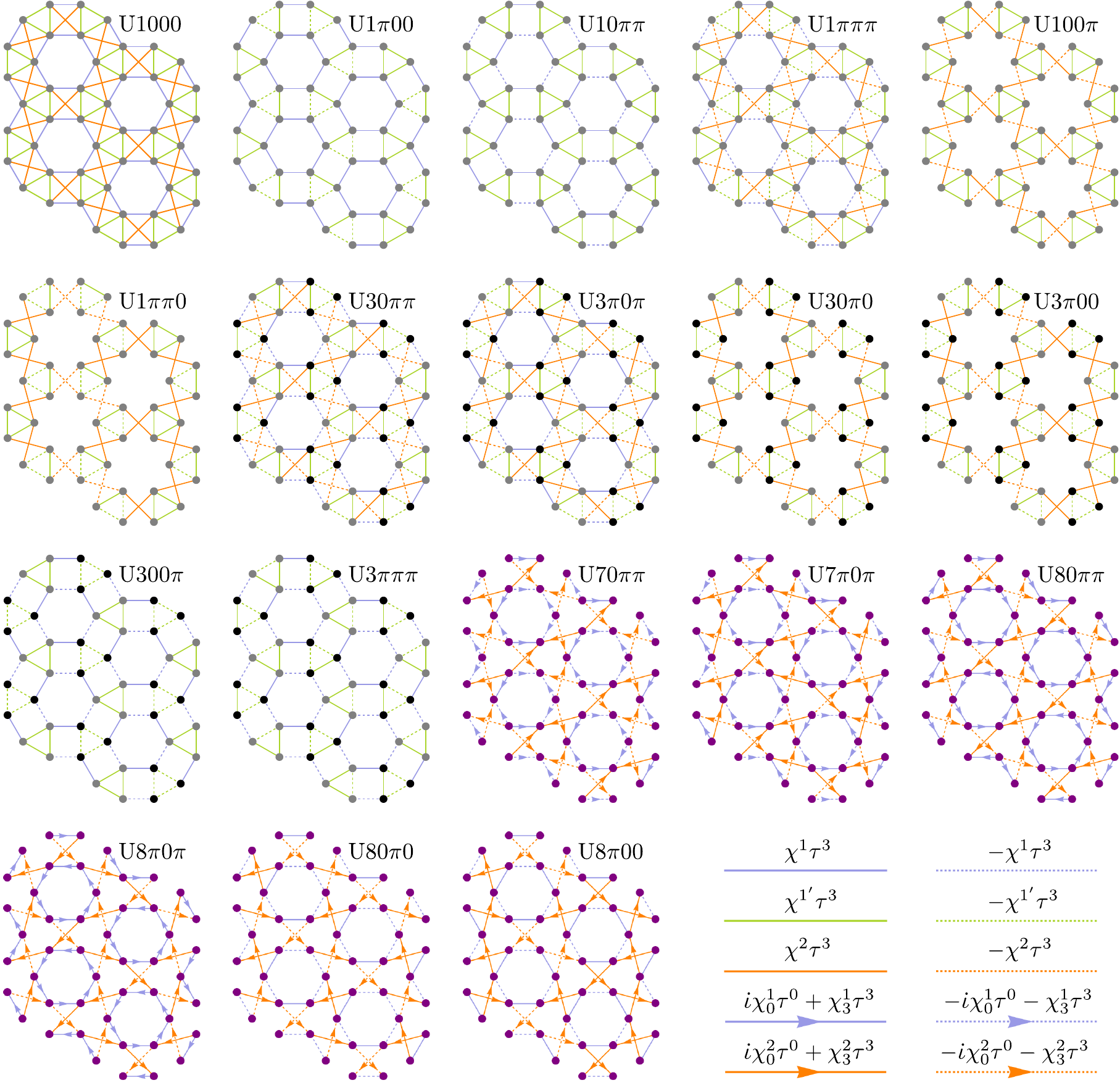}
\caption{Graphical illustrations of all the U(1) \textit{Ans\"atze} which can be realized up to 3NN. The gray (black) points denote positive (negative) on-site hoppings and should be read as $+$ ($-$) $a^{}_3\tau^3$. The purple dots denote a vanishing on-site hopping amplitude. The mean-field parameters on the dashed lines or arrows are associated with a negative sign. All other notations employed here are defined in the legend in the bottom-right corner. For the \textit{Ans\"atze} U80$\pi\pi$ and U8$\pi$0$\pi$, $\chi^1_3=0$, i.e., there are only imaginary hoppings on the blue bonds.} 
\label{fig:3nn_u1_ansatze}
\end{figure*}

\section{$\mathbf{\mathbb{Z}_2}$ PSG classification}
\label{app:z2_psg_derivation}
\subsection{Lattice symmetries}
Unlike U(1) \textit{Ans\"atze} which allow only real and imaginary hopping parameters in the canonical gauge,
for $\mathbb{Z}_2$ \textit{Ans\"atze}, all the terms in Eq.~\eqref{eq:link_singlet} are permitted. So, the projective gauges can generically have an SU(2) structure.
Making use of the local gauge redundancy similar to the U(1) case, one can set $G_{T_1}(x,0;s)=G_{T_2}(x,y;s)=\tau^0$. Consequently, the relation~\eqref{eq:gauge_T1T2} yields the projective solution for $\mathcal{O}\in\{T_1,T_2\}$ as 
\begin{equation}
	\label{eq:tran^{}_1_z2}
	\left.\begin{aligned}
		&G^{}_{T_1}(x,y;s)=\eta^{y}\tau^0,\;G^{}_{T_2}(x,y;s)=\tau^0.\\
	\end{aligned}\right.
\end{equation}
Additionally, Eqs.~\eqref{eq:gauge_R_T1} and~\eqref{eq:gauge_R_T2} give
\begin{equation}
	\label{eq:R_sol_1}		G^{}_R(x,y;s)=\eta^x_{RT_1}\eta^y_{RT_2}\eta^{xy+\frac{y}{2}(y+1)}g^{}_{R,s}.
\end{equation}
Furthermore, the cyclic property of $R$ in Eq.~\eqref{eq:gauge_R} shows that
\begin{equation}\label{eq:R_cyclic} \prod_{s}g^{}_{R,s}=\eta^{}_R\tau^0.
\end{equation}
From Eqs.~\eqref{eq:gauge_sig_T1} and \eqref{eq:gauge_sig_T2}, the projective solution for $\sigma$ reads as
\begin{equation}
	\label{eq:sig_sol_1}	G^{}_\sigma(x,y;s)=\eta^x_{\sigma T_1}\eta^y_{\sigma T_2}\eta^{\frac{x}{2}(x+1)}g^{}_{\sigma,s},
\end{equation}
while the cyclic property of $\sigma$ in Eq.~\eqref{eq:gauge_sig} imposes the constraints
\begin{equation}\label{eq:sig_cyclic}		
\eta^{}_{\sigma T_2}=1,\text{ and } g^{}_{\sigma,s}g^{}_{\sigma,\sigma(s)}=\eta^{}_\sigma\tau^0.
\end{equation}
Furthermore, we have another lattice-symmetry constraint arising from Eq.~\eqref{eq:gauge_R_sig}, which results in
\begin{align}
\eta^{}_{\sigma T_1}&=\eta^{}_{RT_1}\eta^{}_{RT_2} \label{eq:R_sig_eta}, \\
g^{}_{R,s}g^{}_{\sigma,R^{-1}(s)}g^{}_{R,R\sigma(s)}g^{}_{\sigma,\sigma(s)}&=\eta^{}_{R\sigma}\tau^0 \label{eq:R_sig} .
\end{align}
At this point, we are left with three $\eta$ parameters ($\eta$, $\eta^{}_{RT_1}$, and $\eta^{}_{RT_2}$). However, not all possible choices of these three are gauge independent. Indeed, they can be further fixed by employing a local gauge transformation of the form $W(x,y;s)=\
\eta^x_x\eta^y_y\tau^0$. Under this transformation, $G_{T_1}$ and $G_{T_2}$ remain unchanged up to an unimportant global sign which has no consequences on the \textit{Ans\"atze}. However, $G_{R}$ and $G_{\sigma}$ are modified and take the forms:
\begin{align}
&G^{}_R(x,y;s)\rightarrow(\eta^{}_{RT_1}\eta^{}_x\eta^{}_y)^x(\eta^{}_{RT_2}\eta^{}_x)^y\eta^{xy+\frac{y}{2}(y+1)}g^{}_{R,s}, \label{eq:R_local} \\
&G^{}_\sigma(x,y;s)\rightarrow(\eta^{}_y\eta^{}_{RT_1}\eta^{}_{RT_2})^x\eta^{\frac{x}{2}(x+1)}g^{}_{\sigma,s} \label{eq:sig_local} .
\end{align}
For $\eta^{}_x=\eta^{}_{RT_2}$ and $\eta^{}_y=\eta^{}_{RT_1}\eta^{}_{RT_2}$, one can also set $\eta^{}_{RT_1} $ $=\eta^{}_{RT_2}=1$. In the new gauge, the solutions described above are given by
\begin{align}
&G^{}_R(x,y;s)=\eta^{xy+\frac{y}{2}(y+1)}g^{}_{R,s} \label{eq:R_sol_2} \\
&G^{}_\sigma(x,y;s)=\eta^{\frac{x}{2}(x+1)}g^{}_{\sigma,s} \label{eq:sig_sol2} .
\end{align}
In a similar fashion, one can fix the $g$ matrices by employing a sublattice-dependent gauge transformation of the form $W(x,y;s)=w^{}_s$: 
\begin{alignat}{3}
&g^{\pdagger}_{R,1}\rightarrow w^{\dagger}_{1}g^{\pdagger}_{R,1}w^{\pdagger}_{6},\,\,&&g^{\pdagger}_{R,2}\rightarrow w^{\dagger}_{2}g^{\pdagger}_{R,2}w^{\pdagger}_{1}, \,\,&&g^{\pdagger}_{R,3}\rightarrow w^{\dagger}_{3}g^{\pdagger}_{R,3}w^{\pdagger}_{2}, \notag \\
&g^{\pdagger}_{R,4}\rightarrow w^{\dagger}_{4}g^{\pdagger}_{R,4}w^{\pdagger}_{3}, \,\,&&g^{\pdagger}_{R,5}\rightarrow w^{\dagger}_{5}g^{\pdagger}_{R,5}w^{\pdagger}_{4}, \,\,&&g^{\pdagger}_{R,6}\rightarrow w^{\dagger}_{6}g^{\pdagger}_{R,6}w^{\pdagger}_{5},\notag\\
&g^{\pdagger}_{\sigma,1}\rightarrow w^{\dagger}_{1}g^{\pdagger}_{\sigma,1}w^{\pdagger}_{3}, \,\,&&g^{\pdagger}_{\sigma,2}\rightarrow w^{\dagger}_{2}g^{\pdagger}_{\sigma,2}w^{\pdagger}_{2}, \,\,&&g^{\pdagger}_{\sigma,3}\rightarrow w^{\dagger}_{3}g^{\pdagger}_{\sigma,3}w^{\pdagger}_{1},\notag \\
&g^{\pdagger}_{\sigma,4}\rightarrow w^{\dagger}_{4}g^{\pdagger}_{\sigma,4}w^{\pdagger}_{6}, \,\,&&g^{\pdagger}_{\sigma,5}\rightarrow w^{\dagger}_{5}g^{\pdagger}_{\sigma,5}w^{\pdagger}_{5}, \,\,&&g^{\pdagger}_{\sigma,6}\rightarrow w^{\dagger}_{6}g^{\pdagger}_{\sigma,6}w^{\pdagger}_{4}\notag .
\end{alignat}
Now, with the choices $w^{}_1=g^{}_{R,1}w^{}_6$, $w^{}_2=g^{}_{R,2}g^{}_{R,1}w^{}_6$, $w^{}_3=g^{}_{R,3}g^{}_{R,2}g^{}_{R,1}w^{}_6$, $w^{}_4=g^{}_{R,4}g^{}_{R,3}g^{}_{R,2}g^{}_{R,1}w^{}_6$, and $w^{}_5=g^{}_{R,5}g^{}_{R,4}g^{}_{R,3}g^{}_{R,2}g^{}_{R,1}w^{}_6$, we can set
\begin{equation}\label{eq:R_fixing_1}
g^{}_{R,1}=g^{}_{R,2}=g^{}_{R,3}=g^{}_{R,4}=g^{}_{R,5}=\tau^0.
\end{equation}
Inserting this in Eq.~\eqref{eq:R_cyclic}, all the $g^{}_{R,s}$ can be fixed as
\begin{equation}\label{eq:R_fixing}
g^{}_{R,1}=g^{}_{R,2}=g^{}_{R,3}=g^{}_{R,4}=g^{}_{R,5}=\eta^{}_Rg^{}_{R,6}=\tau^0.
\end{equation}
Now, in the new gauge $g^{}_{\sigma,1}\rightarrow w^\dagger_6g^\dagger_{R,1}g^{}_{\sigma,1}g^{}_{R,3}g^{}_{R,2}g^{}_{R,1}w^{}_6$, and we are still left with the freedom to choose $w^{}_6$. With an appropriate choice, one can set \begin{equation}\label{eq:sig_fixing_1} g^{}_{\sigma,1}=e^{\dot\iota\phi\tau^3},
 \end{equation}
whereupon, using Eqs.~\eqref{eq:sig_cyclic}, \eqref{eq:R_sig} and defining $\eta^{}_{\sigma}\eta^{}_{R\sigma}\rightarrow\eta^{}_{R\sigma}$, one can fix $g^{}_{\sigma,s}$ as 
\begin{equation}\label{eq:sig_fixing} g^{}_{\sigma,s}=\{1,\eta^{}_{R\sigma},1,\eta^{}_{R\sigma},\eta^{}_R,\eta^{}_{R\sigma}\}g^{}_{\sigma,1}
 \end{equation}
with $g^2_{\sigma,1}=\eta^{}_{\sigma}\tau^0$. Lastly, using Eq.~\eqref{eq:sig_fixing_1}, we fix
 \begin{align}\label{eq:g_sig_1_fixing}
&\text{for }\eta^{}_\sigma=+1,\;g^{}_{\sigma,1}=\tau^0\\
&\text{for }\eta^{}_\sigma=-1,\;g^{}_{\sigma,1}=\dot\iota\tau^3 .
\end{align}
Noting that there are eventually four gauge-independent $\eta$ parameters ($\eta$, $\eta^{}_\sigma$, $\eta^{}_R$, and $\eta^{}_{R\sigma}$), we conclude there are a total of $2^4=16$ projective extensions of full lattice space-group symmetries.

\begin{figure*}[tb]	\includegraphics[width=1.0\linewidth]{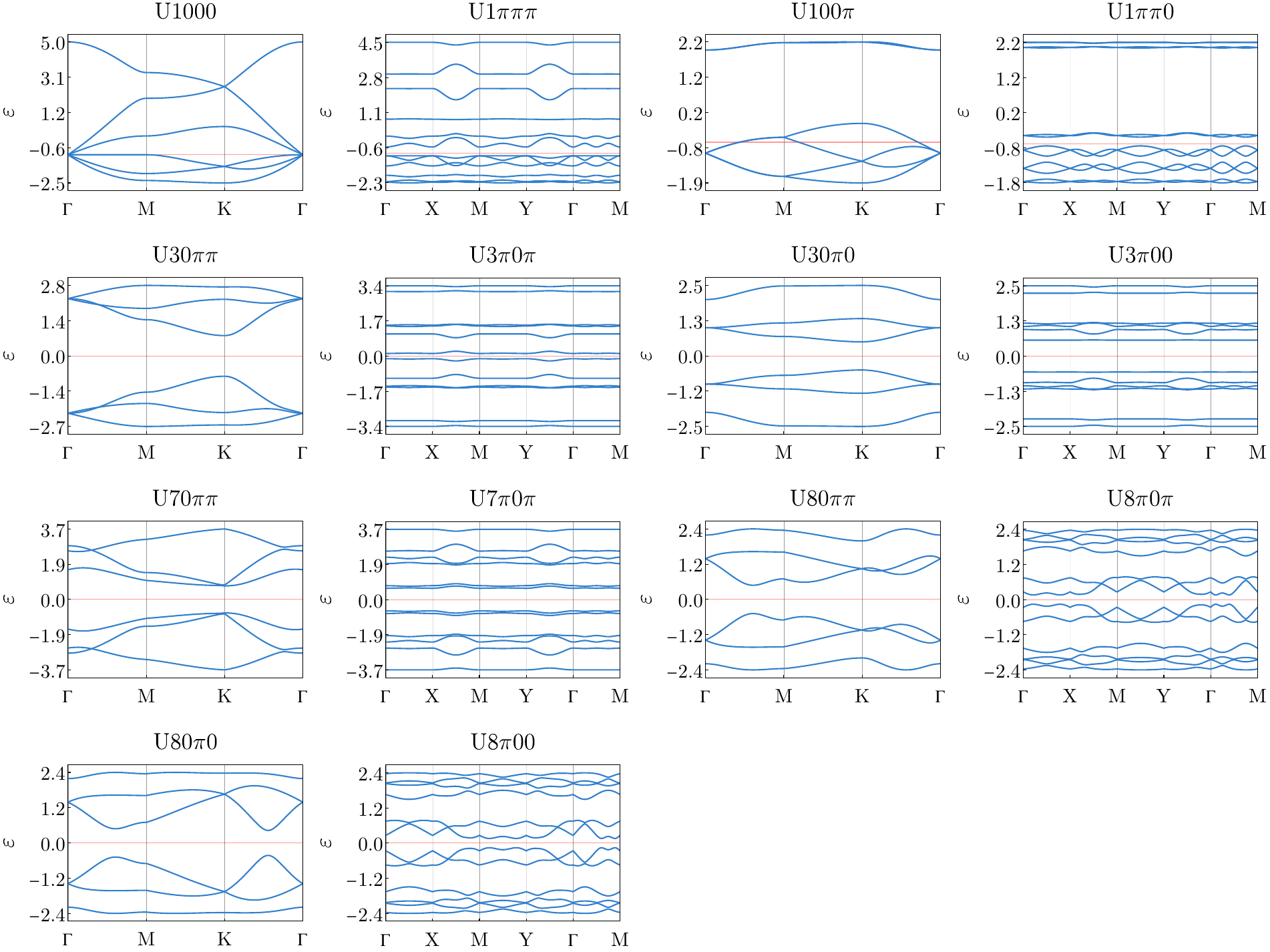}
	\caption{Spinon band structures for the U(1) \textit{Ans\"atze} realizable up to 3NN. The magnitude of the symmetry-allowed hoppings is set to one for 1NN and 2NN bonds, $0.5$ for 3NN bonds, and $0$ for further-neighbor bonds. The red line indicates the Fermi level. The U7 and U8 class \textit{Ans\"atze} cannot be realized when including mean-field amplitudes up to only 2NN.}
	\label{fig:U13NNSBS}
\end{figure*}

\subsection{Time-reversal symmetry}
Let us now derive the projective solutions for time-reversal symmetry. Using Eq.~\eqref{eq:gauge_time_O} for $\mathcal{O}\in T_i$, the solution for $G_{\mathcal{T}}(x,y;s)$ can be written as
\begin{equation}\label{eq:T_sol_1}
	G^{}_{\mathcal{T}}(x,y;s)=\eta^{x}_{\mathcal{T}T_1}\eta^{y}_{\mathcal{T}T_2}g^{}_{\mathcal{T},s},
\end{equation} 
whereas Eq.~\eqref{eq:gauge_time} leads to the condition
\begin{equation}\label{eq:T_constraint}		
\left(g^{}_{\mathcal{T},s}\right)^2=\eta^{}_\mathcal{T}\tau^0.
\end{equation}

Moreover,
 Eq.~\eqref{eq:gauge_time_O} for $\mathcal{O}\in \{R,\sigma\}$ asserts that
 \begin{align}
\eta^{}_{\mathcal{T}T_1}&=\eta^{}_{\mathcal{T}T_2}=+1,\\
g^{}_{\mathcal{T},s}g^{}_{R,s}&=\eta^{}_{\mathcal{T}R}g^{}_{R,s}g^{}_{\mathcal{T},R^{-1}(s)},\label{eq:time_R}\\
g^{}_{\mathcal{T},s}g^{}_{\sigma,s}&=\eta^{}_{\mathcal{T}\sigma}g^{}_{\sigma,s}g^{}_{\mathcal{T},\sigma^{-1}(s)}\label{eq:time_sig}.
\end{align}
Another condition stems from Eqs.~\eqref{eq:R_fixing} and \eqref{eq:time_R}, which together yield 
\begin{equation}
 g^{}_{\mathcal{T},R^{-1}(s)}=\eta^{}_{\mathcal{T}R}g^{}_{\mathcal{T},s}.
\end{equation}
Moreover, we use Eq.~\eqref{eq:time_sig} and finally obtain all the symmetric PSG solutions detailed in Sec.~\ref{sec:psg_sol_z2}.

\section{Third-nearest-neighbor U(1) \textit{Ans\"atze}}
\label{app:3NN}

\begin{figure*}	\includegraphics[width=\linewidth]{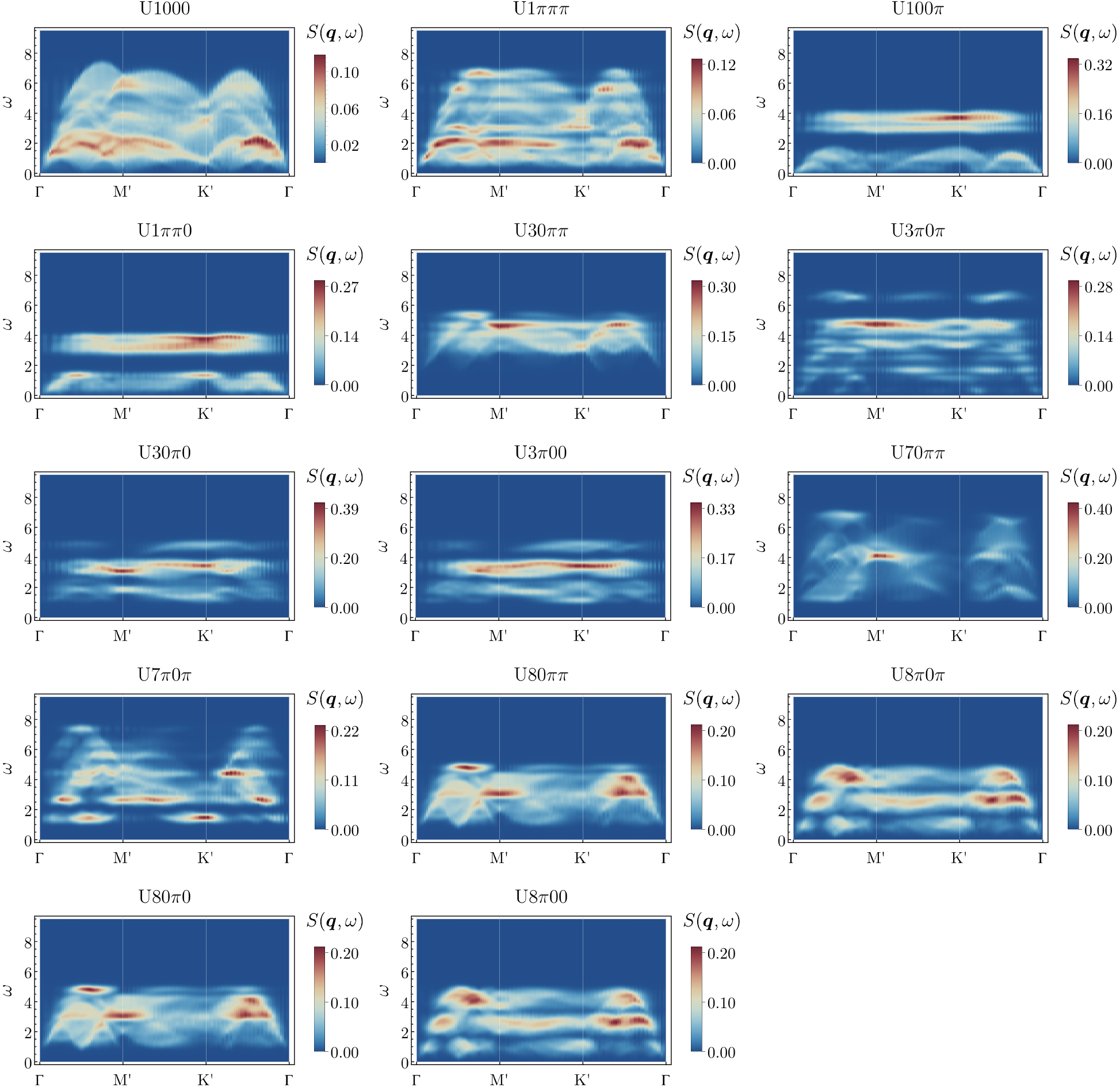}
\caption{Dynamical structure factors of the U(1) \textit{Ans\"atze} realizable up to 3NN, plotted along a high-symmetry path in the extended Brillouin zone for a system size of $14\times14\times6$ sites.} 
\label{fig:U13NNDSF}
\end{figure*}

The different U(1) \textit{Ans\"atze} which can be realized by mean-field Hamiltonians with up to 3NN couplings are illustrated in Fig.~\ref{fig:3nn_u1_ansatze}. This also shows why we need to consider eight distinct loops in Table~\ref{table:u1_flux_structures} to differentiate between the possible U(1) QSL states.

For example, it is essential to define the flux $\phi^{}_{p}$ threading the pinwheel loop to distinguish between the states labeled U3$0\pi0$ and U3$\pi00$, as the mean-field parameters vanish on the blue bonds for both. Although the flux through a triangle, $\phi^{}_t$, equals $\pi$ for U3$0\pi0$ and $0$ for U3$\pi00$, these can be interchanged by a global rotation for an odd-sided loop, so they are insufficient to distinguish the two states.

Similarly, for the last six states, the mean-field parameters vanish on the green bonds, and to tell them apart, in addition to the flux threading the hexagonal plaquette, one must define two fluxes, $\phi^{}_{w}$ and $\phi^{}_{w'}$, threading the windmill loops, illustrated by the green plaquettes in Fig.~\ref{fig:loops}. Additionally, the flux $\phi^{}_{b}$ through the bowtie-shaped loop, illustrated by the pink plaquette in Fig.~\ref{fig:loops}, is required to distinguish between the states labeled U1$00\pi$, U1$\pi\pi0$, U3$0\pi0$, and U3$\pi00$.

Moreover, there exist other classes of \textit{Ansätze}, the realization of which requires enlarging the unit cell by tripling or more along the $\mathbf{T}^{}_{1}$ direction. These correspond to the PSG class listed in the sixth row of Table~\ref{table:u1_psg}, where $\xi$ can take values $m \pi/ n$, with $m$ and $n$ being integers. However, we restrict our discussion to cases realizable with only the original unit cell or a doubling thereof.

Next, we extend our analysis in Sec.~\ref{sec:U12NN} to include the case of mean-field \textit{Ans\"atze} with third-nearest-neighbor (3NN) amplitudes, setting, without loss of generality, the mean-field parameters for the 3NN bonds to 0.5. We begin with the \textit{Ansatz} labeled U1000. For our reference parameters, the spectrum, shown in Fig.~\ref{fig:U13NNSBS}, features a Dirac cone at the zone center. However, the presence of this Dirac point is an artifact of the specific choice of parameters and it can be gapped out with different parameters (unlike in the case with 2NNs alone discussed in Sec.~\ref{sec:U12NN}). The addition of 3NN interactions also opens up a gap for the state U1$\pi\pi\pi$, as shown in Fig.~\ref{fig:U13NNSBS}. The corresponding DSFs for these two U(1) QSLs are displayed in the first two panels of Fig.~\ref{fig:U13NNDSF}.

For the states U1$\pi$00 and U10$\pi\pi$, the properties remain unchanged from the 2NN case because the projective symmetries do not permit mean-field parameters on the 3NN bonds. However, with the inclusion of 3NN amplitudes, two new U(1) \textit{Ansätze}, labeled U100$\pi$ and U1$\pi\pi$0, appear, which cannot be realized in a mean-field Hamiltonian with only 2NNs. Among these, U100$\pi$ consistently possesses a Fermi surface, as reflected in the DSF plot in Fig.~\ref{fig:U13NNDSF}, where a dome-like region of finite intensity extends down to zero energy. On the other hand, the U1$\pi\pi$0 state is generally gapped. Its DSF is shown in Fig.~\ref{fig:U13NNDSF}: the large intensity spread out flatly at high energies is due to excitations from all the filled bands to the topmost empty band.

In the U3 class, six \textit{Ansätze} are realizable up to 3NNs. Among them, U30$\pi\pi$, U3$\pi$0$\pi$, U300$\pi$, and U3$\pi\pi\pi$ can also be realized in the 2NN case. The first two generally exhibit gapped excitations, as previously, with their corresponding DSF plots shown in Fig.~\ref{fig:U13NNDSF}. The last two are unaffected by the 3NN couplings as the mean-field parameters identically vanish on the 3NN bonds.
Furthermore, two new \textit{Ansätze}, labeled U30$\pi$0 and U3$\pi$00, appear after considering the 3NN interactions. Both of these generally exhibit gapped excitations, as shown in Fig.~\ref{fig:U13NNSBS}, with their DSFs presented in the third row of Fig.~\ref{fig:U13NNDSF}.

Finally, the analysis including up to 3NNs also yields six additional U(1) \textit{Ansätze} that belong to the U7 and U8 classes. All of these are generically characterized by gapped excitations, as illustrated in the last six panels of Fig.~\ref{fig:U13NNSBS} and reflected in their DSF profiles in Fig.~\ref{fig:U13NNDSF}.

\begin{table}[t]
	\begin{ruledtabular}
		\begin{tabular}{cc}
 Maple-leaf ($u^{}_{\mathrm{dimer}}=0$)& Ruby ($u^{}_{J_2}=0$)\vspace*{0.1cm}\\
\hline
UC01 & U10$\pi\pi$ \\
UC10 & U1$\pi$00 \\
UD00 & U300$\pi$ \\
UD11 & U3$\pi\pi\pi$ \\
		\end{tabular}
	\end{ruledtabular}	
 	\caption{\label{maple}There are four U(1) \textit{Ans\"atze} on the maple-leaf (ruby) lattice for which the mean-field amplitudes vanish on the dimer ($J_2$) bonds. All of these \textit{Ans\"atze} are therefore the same on the two lattices. Many more such correspondences could be established once we enforce reflection symmetry on the maple-leaf lattice, but such identifications are beyond the scope of our present study.}
\end{table}

\bibliography{references}

\begin{thebibliography}{89}%
\makeatletter
\providecommand \@ifxundefined [1]{%
 \@ifx{#1\undefined}
}%
\providecommand \@ifnum [1]{%
 \ifnum #1\expandafter \@firstoftwo
 \else \expandafter \@secondoftwo
 \fi
}%
\providecommand \@ifx [1]{%
 \ifx #1\expandafter \@firstoftwo
 \else \expandafter \@secondoftwo
 \fi
}%
\providecommand \natexlab [1]{#1}%
\providecommand \enquote  [1]{``#1''}%
\providecommand \bibnamefont  [1]{#1}%
\providecommand \bibfnamefont [1]{#1}%
\providecommand \citenamefont [1]{#1}%
\providecommand \href@noop [0]{\@secondoftwo}%
\providecommand \href [0]{\begingroup \@sanitize@url \@href}%
\providecommand \@href[1]{\@@startlink{#1}\@@href}%
\providecommand \@@href[1]{\endgroup#1\@@endlink}%
\providecommand \@sanitize@url [0]{\catcode `\\12\catcode `\$12\catcode
  `\&12\catcode `\#12\catcode `\^12\catcode `\_12\catcode `\%12\relax}%
\providecommand \@@startlink[1]{}%
\providecommand \@@endlink[0]{}%
\providecommand \url  [0]{\begingroup\@sanitize@url \@url }%
\providecommand \@url [1]{\endgroup\@href {#1}{\urlprefix }}%
\providecommand \urlprefix  [0]{URL }%
\providecommand \Eprint [0]{\href }%
\providecommand \doibase [0]{https://doi.org/}%
\providecommand \selectlanguage [0]{\@gobble}%
\providecommand \bibinfo  [0]{\@secondoftwo}%
\providecommand \bibfield  [0]{\@secondoftwo}%
\providecommand \translation [1]{[#1]}%
\providecommand \BibitemOpen [0]{}%
\providecommand \bibitemStop [0]{}%
\providecommand \bibitemNoStop [0]{.\EOS\space}%
\providecommand \EOS [0]{\spacefactor3000\relax}%
\providecommand \BibitemShut  [1]{\csname bibitem#1\endcsname}%
\let\auto@bib@innerbib\@empty
\bibitem [{\citenamefont {Savary}\ and\ \citenamefont
  {Balents}(2016)}]{Savary-2017}%
  \BibitemOpen
  \bibfield  {author} {\bibinfo {author} {\bibfnamefont {L.}~\bibnamefont
  {Savary}}\ and\ \bibinfo {author} {\bibfnamefont {L.}~\bibnamefont
  {Balents}},\ }\bibfield  {title} {\bibinfo {title} {Quantum spin liquids: a
  review},\ }\href {https://doi.org/10.1088/0034-4885/80/1/016502} {\bibfield
  {journal} {\bibinfo  {journal} {Rep. Prog. Phys.}\ }\textbf {\bibinfo
  {volume} {80}},\ \bibinfo {pages} {016502} (\bibinfo {year}
  {2016})}\BibitemShut {NoStop}%
\bibitem [{\citenamefont {Iqbal}\ \emph {et~al.}(2013)\citenamefont {Iqbal},
  \citenamefont {Becca}, \citenamefont {Sorella},\ and\ \citenamefont
  {Poilblanc}}]{Iqbal-2013}%
  \BibitemOpen
  \bibfield  {author} {\bibinfo {author} {\bibfnamefont {Y.}~\bibnamefont
  {Iqbal}}, \bibinfo {author} {\bibfnamefont {F.}~\bibnamefont {Becca}},
  \bibinfo {author} {\bibfnamefont {S.}~\bibnamefont {Sorella}},\ and\ \bibinfo
  {author} {\bibfnamefont {D.}~\bibnamefont {Poilblanc}},\ }\bibfield  {title}
  {\bibinfo {title} {{Gapless spin-liquid phase in the kagome
  spin-$\frac{1}{2}$ Heisenberg antiferromagnet}},\ }\href
  {https://doi.org/10.1103/PhysRevB.87.060405} {\bibfield  {journal} {\bibinfo
  {journal} {Phys. Rev. B}\ }\textbf {\bibinfo {volume} {87}},\ \bibinfo
  {pages} {060405} (\bibinfo {year} {2013})}\BibitemShut {NoStop}%
\bibitem [{\citenamefont {He}\ \emph {et~al.}(2017)\citenamefont {He},
  \citenamefont {Zaletel}, \citenamefont {Oshikawa},\ and\ \citenamefont
  {Pollmann}}]{He-2017}%
  \BibitemOpen
  \bibfield  {author} {\bibinfo {author} {\bibfnamefont {Y.-C.}\ \bibnamefont
  {He}}, \bibinfo {author} {\bibfnamefont {M.~P.}\ \bibnamefont {Zaletel}},
  \bibinfo {author} {\bibfnamefont {M.}~\bibnamefont {Oshikawa}},\ and\
  \bibinfo {author} {\bibfnamefont {F.}~\bibnamefont {Pollmann}},\ }\bibfield
  {title} {\bibinfo {title} {{Signatures of Dirac Cones in a DMRG Study of the
  Kagome Heisenberg Model}},\ }\href
  {https://doi.org/10.1103/PhysRevX.7.031020} {\bibfield  {journal} {\bibinfo
  {journal} {Phys. Rev. X}\ }\textbf {\bibinfo {volume} {7}},\ \bibinfo {pages}
  {031020} (\bibinfo {year} {2017})}\BibitemShut {NoStop}%
\bibitem [{\citenamefont {Iqbal}\ \emph {et~al.}(2016)\citenamefont {Iqbal},
  \citenamefont {Hu}, \citenamefont {Thomale}, \citenamefont {Poilblanc},\ and\
  \citenamefont {Becca}}]{Iqbal-2016}%
  \BibitemOpen
  \bibfield  {author} {\bibinfo {author} {\bibfnamefont {Y.}~\bibnamefont
  {Iqbal}}, \bibinfo {author} {\bibfnamefont {W.-J.}\ \bibnamefont {Hu}},
  \bibinfo {author} {\bibfnamefont {R.}~\bibnamefont {Thomale}}, \bibinfo
  {author} {\bibfnamefont {D.}~\bibnamefont {Poilblanc}},\ and\ \bibinfo
  {author} {\bibfnamefont {F.}~\bibnamefont {Becca}},\ }\bibfield  {title}
  {\bibinfo {title} {{Spin liquid nature in the Heisenberg
  ${J}_{1}\ensuremath{-}{J}_{2}$ triangular antiferromagnet}},\ }\href
  {https://doi.org/10.1103/PhysRevB.93.144411} {\bibfield  {journal} {\bibinfo
  {journal} {Phys. Rev. B}\ }\textbf {\bibinfo {volume} {93}},\ \bibinfo
  {pages} {144411} (\bibinfo {year} {2016})}\BibitemShut {NoStop}%
\bibitem [{\citenamefont {Hu}\ \emph {et~al.}(2019)\citenamefont {Hu},
  \citenamefont {Zhu}, \citenamefont {Eggert},\ and\ \citenamefont
  {He}}]{Hu-2019}%
  \BibitemOpen
  \bibfield  {author} {\bibinfo {author} {\bibfnamefont {S.}~\bibnamefont
  {Hu}}, \bibinfo {author} {\bibfnamefont {W.}~\bibnamefont {Zhu}}, \bibinfo
  {author} {\bibfnamefont {S.}~\bibnamefont {Eggert}},\ and\ \bibinfo {author}
  {\bibfnamefont {Y.-C.}\ \bibnamefont {He}},\ }\bibfield  {title} {\bibinfo
  {title} {{Dirac Spin Liquid on the Spin-$1/2$ Triangular Heisenberg
  Antiferromagnet}},\ }\href {https://doi.org/10.1103/PhysRevLett.123.207203}
  {\bibfield  {journal} {\bibinfo  {journal} {Phys. Rev. Lett.}\ }\textbf
  {\bibinfo {volume} {123}},\ \bibinfo {pages} {207203} (\bibinfo {year}
  {2019})}\BibitemShut {NoStop}%
\bibitem [{\citenamefont {Semeghini}\ \emph {et~al.}(2021)\citenamefont
  {Semeghini}, \citenamefont {Levine}, \citenamefont {Keesling}, \citenamefont
  {Ebadi}, \citenamefont {Wang}, \citenamefont {Bluvstein}, \citenamefont
  {Verresen}, \citenamefont {Pichler}, \citenamefont {Kalinowski},
  \citenamefont {Samajdar}, \citenamefont {Omran}, \citenamefont {Sachdev},
  \citenamefont {Vishwanath}, \citenamefont {Greiner}, \citenamefont
  {Vuletić},\ and\ \citenamefont {Lukin}}]{Semeghini-2021}%
  \BibitemOpen
  \bibfield  {author} {\bibinfo {author} {\bibfnamefont {G.}~\bibnamefont
  {Semeghini}}, \bibinfo {author} {\bibfnamefont {H.}~\bibnamefont {Levine}},
  \bibinfo {author} {\bibfnamefont {A.}~\bibnamefont {Keesling}}, \bibinfo
  {author} {\bibfnamefont {S.}~\bibnamefont {Ebadi}}, \bibinfo {author}
  {\bibfnamefont {T.~T.}\ \bibnamefont {Wang}}, \bibinfo {author}
  {\bibfnamefont {D.}~\bibnamefont {Bluvstein}}, \bibinfo {author}
  {\bibfnamefont {R.}~\bibnamefont {Verresen}}, \bibinfo {author}
  {\bibfnamefont {H.}~\bibnamefont {Pichler}}, \bibinfo {author} {\bibfnamefont
  {M.}~\bibnamefont {Kalinowski}}, \bibinfo {author} {\bibfnamefont
  {R.}~\bibnamefont {Samajdar}}, \bibinfo {author} {\bibfnamefont
  {A.}~\bibnamefont {Omran}}, \bibinfo {author} {\bibfnamefont
  {S.}~\bibnamefont {Sachdev}}, \bibinfo {author} {\bibfnamefont
  {A.}~\bibnamefont {Vishwanath}}, \bibinfo {author} {\bibfnamefont
  {M.}~\bibnamefont {Greiner}}, \bibinfo {author} {\bibfnamefont
  {V.}~\bibnamefont {Vuletić}},\ and\ \bibinfo {author} {\bibfnamefont
  {M.~D.}\ \bibnamefont {Lukin}},\ }\bibfield  {title} {\bibinfo {title}
  {{Probing topological spin liquids on a programmable quantum simulator}},\
  }\href {https://doi.org/10.1126/science.abi8794} {\bibfield  {journal}
  {\bibinfo  {journal} {Science}\ }\textbf {\bibinfo {volume} {374}},\ \bibinfo
  {pages} {1242} (\bibinfo {year} {2021})}\BibitemShut {NoStop}%
\bibitem [{\citenamefont {Giudici}\ \emph {et~al.}(2022)\citenamefont
  {Giudici}, \citenamefont {Lukin},\ and\ \citenamefont
  {Pichler}}]{Giudici-2022}%
  \BibitemOpen
  \bibfield  {author} {\bibinfo {author} {\bibfnamefont {G.}~\bibnamefont
  {Giudici}}, \bibinfo {author} {\bibfnamefont {M.~D.}\ \bibnamefont {Lukin}},\
  and\ \bibinfo {author} {\bibfnamefont {H.}~\bibnamefont {Pichler}},\
  }\bibfield  {title} {\bibinfo {title} {{Dynamical preparation of quantum spin
  liquids in Rydberg atom arrays}},\ }\href
  {https://doi.org/10.1103/PhysRevLett.129.090401} {\bibfield  {journal}
  {\bibinfo  {journal} {Phys. Rev. Lett.}\ }\textbf {\bibinfo {volume} {129}},\
  \bibinfo {pages} {090401} (\bibinfo {year} {2022})}\BibitemShut {NoStop}%
\bibitem [{\citenamefont {Samajdar}\ \emph {et~al.}(2023)\citenamefont
  {Samajdar}, \citenamefont {Joshi}, \citenamefont {Teng},\ and\ \citenamefont
  {Sachdev}}]{Samajdar-2023}%
  \BibitemOpen
  \bibfield  {author} {\bibinfo {author} {\bibfnamefont {R.}~\bibnamefont
  {Samajdar}}, \bibinfo {author} {\bibfnamefont {D.~G.}\ \bibnamefont {Joshi}},
  \bibinfo {author} {\bibfnamefont {Y.}~\bibnamefont {Teng}},\ and\ \bibinfo
  {author} {\bibfnamefont {S.}~\bibnamefont {Sachdev}},\ }\bibfield  {title}
  {\bibinfo {title} {{Emergent ${\mathbb{Z}}_{2}$ gauge theories and
  topological excitations in Rydberg atom arrays}},\ }\href
  {https://doi.org/10.1103/PhysRevLett.130.043601} {\bibfield  {journal}
  {\bibinfo  {journal} {Phys. Rev. Lett.}\ }\textbf {\bibinfo {volume} {130}},\
  \bibinfo {pages} {043601} (\bibinfo {year} {2023})}\BibitemShut {NoStop}%
\bibitem [{\citenamefont {Tarabunga}\ \emph {et~al.}(2022)\citenamefont
  {Tarabunga}, \citenamefont {Surace}, \citenamefont {Andreoni}, \citenamefont
  {Angelone},\ and\ \citenamefont {Dalmonte}}]{tarabunga2022gauge}%
  \BibitemOpen
  \bibfield  {author} {\bibinfo {author} {\bibfnamefont {P.~S.}\ \bibnamefont
  {Tarabunga}}, \bibinfo {author} {\bibfnamefont {F.~M.}\ \bibnamefont
  {Surace}}, \bibinfo {author} {\bibfnamefont {R.}~\bibnamefont {Andreoni}},
  \bibinfo {author} {\bibfnamefont {A.}~\bibnamefont {Angelone}},\ and\
  \bibinfo {author} {\bibfnamefont {M.}~\bibnamefont {Dalmonte}},\ }\bibfield
  {title} {\bibinfo {title} {{Gauge-Theoretic Origin of Rydberg Quantum Spin
  Liquids}},\ }\href {https://doi.org/10.1103/PhysRevLett.129.195301}
  {\bibfield  {journal} {\bibinfo  {journal} {Phys. Rev. Lett.}\ }\textbf
  {\bibinfo {volume} {129}},\ \bibinfo {pages} {195301} (\bibinfo {year}
  {2022})}\BibitemShut {NoStop}%
\bibitem [{\citenamefont {Samajdar}\ \emph {et~al.}(2021)\citenamefont
  {Samajdar}, \citenamefont {Ho}, \citenamefont {Pichler}, \citenamefont
  {Lukin},\ and\ \citenamefont {Sachdev}}]{Samajdar.2021}%
  \BibitemOpen
  \bibfield  {author} {\bibinfo {author} {\bibfnamefont {R.}~\bibnamefont
  {Samajdar}}, \bibinfo {author} {\bibfnamefont {W.~W.}\ \bibnamefont {Ho}},
  \bibinfo {author} {\bibfnamefont {H.}~\bibnamefont {Pichler}}, \bibinfo
  {author} {\bibfnamefont {M.~D.}\ \bibnamefont {Lukin}},\ and\ \bibinfo
  {author} {\bibfnamefont {S.}~\bibnamefont {Sachdev}},\ }\bibfield  {title}
  {\bibinfo {title} {{Quantum phases of Rydberg atoms on a kagome lattice}},\
  }\href {https://doi.org/10.1073/pnas.2015785118} {\bibfield  {journal}
  {\bibinfo  {journal} {Proc. Natl. Acad. Sci. U.S.A.}\ }\textbf {\bibinfo
  {volume} {118}},\ \bibinfo {pages} {e2015785118} (\bibinfo {year}
  {2021})}\BibitemShut {NoStop}%
\bibitem [{\citenamefont {Verresen}\ \emph {et~al.}(2021)\citenamefont
  {Verresen}, \citenamefont {Lukin},\ and\ \citenamefont
  {Vishwanath}}]{Veressen-2021}%
  \BibitemOpen
  \bibfield  {author} {\bibinfo {author} {\bibfnamefont {R.}~\bibnamefont
  {Verresen}}, \bibinfo {author} {\bibfnamefont {M.~D.}\ \bibnamefont
  {Lukin}},\ and\ \bibinfo {author} {\bibfnamefont {A.}~\bibnamefont
  {Vishwanath}},\ }\bibfield  {title} {\bibinfo {title} {{Prediction of toric
  code topological order from Rydberg blockade}},\ }\href
  {https://doi.org/10.1103/PhysRevX.11.031005} {\bibfield  {journal} {\bibinfo
  {journal} {Phys. Rev. X}\ }\textbf {\bibinfo {volume} {11}},\ \bibinfo
  {pages} {031005} (\bibinfo {year} {2021})}\BibitemShut {NoStop}%
\bibitem [{\citenamefont {Verresen}\ and\ \citenamefont
  {Vishwanath}(2022)}]{Veressen-2022}%
  \BibitemOpen
  \bibfield  {author} {\bibinfo {author} {\bibfnamefont {R.}~\bibnamefont
  {Verresen}}\ and\ \bibinfo {author} {\bibfnamefont {A.}~\bibnamefont
  {Vishwanath}},\ }\bibfield  {title} {\bibinfo {title} {{Unifying Kitaev
  magnets, kagom\'e dimer models, and ruby Rydberg spin liquids}},\ }\href
  {https://doi.org/10.1103/PhysRevX.12.041029} {\bibfield  {journal} {\bibinfo
  {journal} {Phys. Rev. X}\ }\textbf {\bibinfo {volume} {12}},\ \bibinfo
  {pages} {041029} (\bibinfo {year} {2022})}\BibitemShut {NoStop}%
\bibitem [{\citenamefont {Jahromi}\ \emph {et~al.}(2016)\citenamefont
  {Jahromi}, \citenamefont {Kargarian}, \citenamefont {Masoudi},\ and\
  \citenamefont {Langari}}]{Jahromi-2016}%
  \BibitemOpen
  \bibfield  {author} {\bibinfo {author} {\bibfnamefont {S.~S.}\ \bibnamefont
  {Jahromi}}, \bibinfo {author} {\bibfnamefont {M.}~\bibnamefont {Kargarian}},
  \bibinfo {author} {\bibfnamefont {S.~F.}\ \bibnamefont {Masoudi}},\ and\
  \bibinfo {author} {\bibfnamefont {A.}~\bibnamefont {Langari}},\ }\bibfield
  {title} {\bibinfo {title} {{Topological spin liquids in the ruby lattice with
  anisotropic Kitaev interactions}},\ }\href
  {https://doi.org/10.1103/PhysRevB.94.125145} {\bibfield  {journal} {\bibinfo
  {journal} {Phys. Rev. B}\ }\textbf {\bibinfo {volume} {94}},\ \bibinfo
  {pages} {125145} (\bibinfo {year} {2016})}\BibitemShut {NoStop}%
\bibitem [{\citenamefont {Jahromi}\ \emph {et~al.}(2018)\citenamefont
  {Jahromi}, \citenamefont {Or\'us}, \citenamefont {Kargarian},\ and\
  \citenamefont {Langari}}]{Jahromi-2018}%
  \BibitemOpen
  \bibfield  {author} {\bibinfo {author} {\bibfnamefont {S.~S.}\ \bibnamefont
  {Jahromi}}, \bibinfo {author} {\bibfnamefont {R.}~\bibnamefont {Or\'us}},
  \bibinfo {author} {\bibfnamefont {M.}~\bibnamefont {Kargarian}},\ and\
  \bibinfo {author} {\bibfnamefont {A.}~\bibnamefont {Langari}},\ }\bibfield
  {title} {\bibinfo {title} {{Infinite projected entangled-pair state algorithm
  for ruby and triangle-honeycomb lattices}},\ }\href
  {https://doi.org/10.1103/PhysRevB.97.115161} {\bibfield  {journal} {\bibinfo
  {journal} {Phys. Rev. B}\ }\textbf {\bibinfo {volume} {97}},\ \bibinfo
  {pages} {115161} (\bibinfo {year} {2018})}\BibitemShut {NoStop}%
\bibitem [{\citenamefont {Kargarian}\ \emph {et~al.}(2010)\citenamefont
  {Kargarian}, \citenamefont {Bombin},\ and\ \citenamefont
  {Martin-Delgado}}]{Kargarian-2010}%
  \BibitemOpen
  \bibfield  {author} {\bibinfo {author} {\bibfnamefont {M.}~\bibnamefont
  {Kargarian}}, \bibinfo {author} {\bibfnamefont {H.}~\bibnamefont {Bombin}},\
  and\ \bibinfo {author} {\bibfnamefont {M.~A.}\ \bibnamefont
  {Martin-Delgado}},\ }\bibfield  {title} {\bibinfo {title} {{Topological color
  codes and two-body quantum lattice Hamiltonians}},\ }\href
  {https://doi.org/10.1088/1367-2630/12/2/025018} {\bibfield  {journal}
  {\bibinfo  {journal} {New J. Phys.}\ }\textbf {\bibinfo {volume} {12}},\
  \bibinfo {pages} {025018} (\bibinfo {year} {2010})}\BibitemShut {NoStop}%
\bibitem [{\citenamefont {Bombin}\ \emph {et~al.}(2009)\citenamefont {Bombin},
  \citenamefont {Kargarian},\ and\ \citenamefont
  {Martin-Delgado}}]{Bombin-2009}%
  \BibitemOpen
  \bibfield  {author} {\bibinfo {author} {\bibfnamefont {H.}~\bibnamefont
  {Bombin}}, \bibinfo {author} {\bibfnamefont {M.}~\bibnamefont {Kargarian}},\
  and\ \bibinfo {author} {\bibfnamefont {M.~A.}\ \bibnamefont
  {Martin-Delgado}},\ }\bibfield  {title} {\bibinfo {title} {{Interacting
  anyonic fermions in a two-body color code model}},\ }\href
  {https://doi.org/10.1103/PhysRevB.80.075111} {\bibfield  {journal} {\bibinfo
  {journal} {Phys. Rev. B}\ }\textbf {\bibinfo {volume} {80}},\ \bibinfo
  {pages} {075111} (\bibinfo {year} {2009})}\BibitemShut {NoStop}%
\bibitem [{\citenamefont {Farnell}\ \emph {et~al.}(2011)\citenamefont
  {Farnell}, \citenamefont {Darradi}, \citenamefont {Schmidt},\ and\
  \citenamefont {Richter}}]{Farnell2011}%
  \BibitemOpen
  \bibfield  {author} {\bibinfo {author} {\bibfnamefont {D.~J.~J.}\
  \bibnamefont {Farnell}}, \bibinfo {author} {\bibfnamefont {R.}~\bibnamefont
  {Darradi}}, \bibinfo {author} {\bibfnamefont {R.}~\bibnamefont {Schmidt}},\
  and\ \bibinfo {author} {\bibfnamefont {J.}~\bibnamefont {Richter}},\
  }\bibfield  {title} {\bibinfo {title} {{Spin-half Heisenberg antiferromagnet
  on two archimedian lattices: From the bounce lattice to the maple-leaf
  lattice and beyond}},\ }\href {https://doi.org/10.1103/PhysRevB.84.104406}
  {\bibfield  {journal} {\bibinfo  {journal} {Phys. Rev. B}\ }\textbf {\bibinfo
  {volume} {84}},\ \bibinfo {pages} {104406} (\bibinfo {year}
  {2011})}\BibitemShut {NoStop}%
\bibitem [{\citenamefont {Farnell}\ \emph {et~al.}(2014)\citenamefont
  {Farnell}, \citenamefont {G\"otze}, \citenamefont {Richter}, \citenamefont
  {Bishop},\ and\ \citenamefont {Li}}]{Farnell2014}%
  \BibitemOpen
  \bibfield  {author} {\bibinfo {author} {\bibfnamefont {D.~J.~J.}\
  \bibnamefont {Farnell}}, \bibinfo {author} {\bibfnamefont {O.}~\bibnamefont
  {G\"otze}}, \bibinfo {author} {\bibfnamefont {J.}~\bibnamefont {Richter}},
  \bibinfo {author} {\bibfnamefont {R.~F.}\ \bibnamefont {Bishop}},\ and\
  \bibinfo {author} {\bibfnamefont {P.~H.~Y.}\ \bibnamefont {Li}},\ }\bibfield
  {title} {\bibinfo {title} {{Quantum $s=\frac{1}{2}$ antiferromagnets on
  Archimedean lattices: The route from semiclassical magnetic order to
  nonmagnetic quantum states}},\ }\href
  {https://doi.org/10.1103/PhysRevB.89.184407} {\bibfield  {journal} {\bibinfo
  {journal} {Phys. Rev. B}\ }\textbf {\bibinfo {volume} {89}},\ \bibinfo
  {pages} {184407} (\bibinfo {year} {2014})}\BibitemShut {NoStop}%
\bibitem [{\citenamefont {Richter}\ \emph {et~al.}(2004)\citenamefont
  {Richter}, \citenamefont {Schulenburg},\ and\ \citenamefont
  {Honecker}}]{Richter2004}%
  \BibitemOpen
  \bibfield  {author} {\bibinfo {author} {\bibfnamefont {J.}~\bibnamefont
  {Richter}}, \bibinfo {author} {\bibfnamefont {J.}~\bibnamefont
  {Schulenburg}},\ and\ \bibinfo {author} {\bibfnamefont {A.}~\bibnamefont
  {Honecker}},\ }\bibinfo {title} {{Quantum magnetism in two dimensions: From
  semi-classical N{\'e}el order to magnetic disorder}},\ in\ \href
  {https://doi.org/10.1007/BFb0119592} {\emph {\bibinfo {booktitle} {Quantum
  Magnetism}}},\ \bibinfo {editor} {edited by\ \bibinfo {editor} {\bibfnamefont
  {U.}~\bibnamefont {Schollw{\"o}ck}}, \bibinfo {editor} {\bibfnamefont
  {J.}~\bibnamefont {Richter}}, \bibinfo {editor} {\bibfnamefont {D.~J.~J.}\
  \bibnamefont {Farnell}},\ and\ \bibinfo {editor} {\bibfnamefont {R.~F.}\
  \bibnamefont {Bishop}}}\ (\bibinfo  {publisher} {Springer},\ \bibinfo
  {address} {Berlin},\ \bibinfo {year} {2004})\ pp.\ \bibinfo {pages}
  {85--153}\BibitemShut {NoStop}%
\bibitem [{\citenamefont {Jahromi}\ and\ \citenamefont
  {Or\'us}(2020)}]{Jahromi-2020}%
  \BibitemOpen
  \bibfield  {author} {\bibinfo {author} {\bibfnamefont {S.~S.}\ \bibnamefont
  {Jahromi}}\ and\ \bibinfo {author} {\bibfnamefont {R.}~\bibnamefont
  {Or\'us}},\ }\bibfield  {title} {\bibinfo {title} {{Topological
  ${\mathbb{Z}}_{2}$ resonating-valence-bond quantum spin liquid on the ruby
  lattice}},\ }\href {https://doi.org/10.1103/PhysRevB.101.115114} {\bibfield
  {journal} {\bibinfo  {journal} {Phys. Rev. B}\ }\textbf {\bibinfo {volume}
  {101}},\ \bibinfo {pages} {115114} (\bibinfo {year} {2020})}\BibitemShut
  {NoStop}%
\bibitem [{\citenamefont {Nakano}\ and\ \citenamefont
  {Sakai}(2026)}]{Nakano2026}%
  \BibitemOpen
  \bibfield  {author} {\bibinfo {author} {\bibfnamefont {H.}~\bibnamefont
  {Nakano}}\ and\ \bibinfo {author} {\bibfnamefont {T.}~\bibnamefont {Sakai}},\
  }\bibfield  {title} {\bibinfo {title} {{Spin excitation of the Heisenberg
  antiferromagnet with frustration: from the bounce-lattice antiferromagnet
  through the maple-leaf-lattice antiferromagnet to the exact-dimer system}},\
  }\href {https://doi.org/doi:10.1515/zna-2025-0389} {\bibfield  {journal}
  {\bibinfo  {journal} {Z. Naturforsch. A}\ }\textbf {\bibinfo {volume} {81}},\
  \bibinfo {pages} {461} (\bibinfo {year} {2026})}\BibitemShut {NoStop}%
\bibitem [{\citenamefont {Schmoll}\ \emph {et~al.}(2026)\citenamefont
  {Schmoll}, \citenamefont {Naumann}, \citenamefont {Weerda}, \citenamefont
  {Eisert},\ and\ \citenamefont {Iqbal}}]{Schmoll-2024}%
  \BibitemOpen
  \bibfield  {author} {\bibinfo {author} {\bibfnamefont {P.}~\bibnamefont
  {Schmoll}}, \bibinfo {author} {\bibfnamefont {J.}~\bibnamefont {Naumann}},
  \bibinfo {author} {\bibfnamefont {E.~L.}\ \bibnamefont {Weerda}}, \bibinfo
  {author} {\bibfnamefont {J.}~\bibnamefont {Eisert}},\ and\ \bibinfo {author}
  {\bibfnamefont {Y.}~\bibnamefont {Iqbal}},\ }\bibfield  {title} {\bibinfo
  {title} {{An extended non-magnetic phase in the spin-1/2 Heisenberg
  antiferromagnet from the ruby to the maple-leaf lattice}},\ }\href
  {https://arxiv.org/abs/2407.07145} {\  (\bibinfo {year} {2026})},\ \Eprint
  {https://arxiv.org/abs/2407.07145} {arXiv:2407.07145 [cond-mat.str-el]}
  \BibitemShut {NoStop}%
\bibitem [{\citenamefont {Ghosh}\ and\ \citenamefont {Mila}(2025)}]{Ghosh2025}%
  \BibitemOpen
  \bibfield  {author} {\bibinfo {author} {\bibfnamefont {P.}~\bibnamefont
  {Ghosh}}\ and\ \bibinfo {author} {\bibfnamefont {F.}~\bibnamefont {Mila}},\
  }\bibfield  {title} {\bibinfo {title} {{Simplex Crystal Ground State and
  Magnetization Plateaus in the Spin-$1/2$ Heisenberg Model on the Ruby
  Lattice}},\ }\href {https://arxiv.org/abs/2512.14173} {\  (\bibinfo {year}
  {2025})},\ \Eprint {https://arxiv.org/abs/2512.14173} {arXiv:2512.14173
  [cond-mat.str-el]} \BibitemShut {NoStop}%
\bibitem [{\citenamefont {Mello}\ \emph {et~al.}(2026)\citenamefont {Mello},
  \citenamefont {Stoudenmire},\ and\ \citenamefont {Tindall}}]{Mello2026}%
  \BibitemOpen
  \bibfield  {author} {\bibinfo {author} {\bibfnamefont {A.~F.}\ \bibnamefont
  {Mello}}, \bibinfo {author} {\bibfnamefont {E.~M.}\ \bibnamefont
  {Stoudenmire}},\ and\ \bibinfo {author} {\bibfnamefont {J.}~\bibnamefont
  {Tindall}},\ }\bibfield  {title} {\bibinfo {title} {{Finite-temperature
  formation of magnetic plateaus and simplex liquid states on the frustrated
  ruby lattice}},\ }\href {https://arxiv.org/abs/2606.07609} {\  (\bibinfo
  {year} {2026})},\ \Eprint {https://arxiv.org/abs/2606.07609}
  {arXiv:2606.07609 [cond-mat.str-el]} \BibitemShut {NoStop}%
\bibitem [{\citenamefont {Vu}\ \emph {et~al.}(2025)\citenamefont {Vu},
  \citenamefont {Kufel}, \citenamefont {Kemp}, \citenamefont {Pollet},
  \citenamefont {Laumann},\ and\ \citenamefont {Yao}}]{Vu2025}%
  \BibitemOpen
  \bibfield  {author} {\bibinfo {author} {\bibfnamefont {D.}~\bibnamefont
  {Vu}}, \bibinfo {author} {\bibfnamefont {D.~S.}\ \bibnamefont {Kufel}},
  \bibinfo {author} {\bibfnamefont {J.}~\bibnamefont {Kemp}}, \bibinfo {author}
  {\bibfnamefont {L.}~\bibnamefont {Pollet}}, \bibinfo {author} {\bibfnamefont
  {C.~R.}\ \bibnamefont {Laumann}},\ and\ \bibinfo {author} {\bibfnamefont
  {N.~Y.}\ \bibnamefont {Yao}},\ }\bibfield  {title} {\bibinfo {title}
  {{Optimizing the dynamical preparation of quantum spin lakes on the ruby
  lattice}},\ }\href {https://arxiv.org/abs/2512.09040} {\  (\bibinfo {year}
  {2025})},\ \Eprint {https://arxiv.org/abs/2512.09040} {arXiv:2512.09040
  [quant-ph]} \BibitemShut {NoStop}%
\bibitem [{\citenamefont {Ebert}\ \emph {et~al.}(2026)\citenamefont {Ebert},
  \citenamefont {Iqbal},\ and\ \citenamefont {Wietek}}]{Ebert2026}%
  \BibitemOpen
  \bibfield  {author} {\bibinfo {author} {\bibfnamefont {P.~L.}\ \bibnamefont
  {Ebert}}, \bibinfo {author} {\bibfnamefont {Y.}~\bibnamefont {Iqbal}},\ and\
  \bibinfo {author} {\bibfnamefont {A.}~\bibnamefont {Wietek}},\ }\bibfield
  {title} {\bibinfo {title} {{Incommensurate Spin-Density Waves in a Frustrated
  Maple-Leaf Lattice Ferromagnet}},\ }\href {https://arxiv.org/abs/2605.12592}
  {\  (\bibinfo {year} {2026})},\ \Eprint {https://arxiv.org/abs/2605.12592}
  {arXiv:2605.12592 [cond-mat.str-el]} \BibitemShut {NoStop}%
\bibitem [{\citenamefont {Rasche}\ \emph {et~al.}(2013)\citenamefont {Rasche},
  \citenamefont {Isaeva}, \citenamefont {Ruck}, \citenamefont {Borisenko},
  \citenamefont {Zabolotnyy}, \citenamefont {B{\"u}chner}, \citenamefont
  {Koepernik}, \citenamefont {Ortix}, \citenamefont {Richter},\ and\
  \citenamefont {van~den Brink}}]{Rasche-2013}%
  \BibitemOpen
  \bibfield  {author} {\bibinfo {author} {\bibfnamefont {B.}~\bibnamefont
  {Rasche}}, \bibinfo {author} {\bibfnamefont {A.}~\bibnamefont {Isaeva}},
  \bibinfo {author} {\bibfnamefont {M.}~\bibnamefont {Ruck}}, \bibinfo {author}
  {\bibfnamefont {S.}~\bibnamefont {Borisenko}}, \bibinfo {author}
  {\bibfnamefont {V.}~\bibnamefont {Zabolotnyy}}, \bibinfo {author}
  {\bibfnamefont {B.}~\bibnamefont {B{\"u}chner}}, \bibinfo {author}
  {\bibfnamefont {K.}~\bibnamefont {Koepernik}}, \bibinfo {author}
  {\bibfnamefont {C.}~\bibnamefont {Ortix}}, \bibinfo {author} {\bibfnamefont
  {M.}~\bibnamefont {Richter}},\ and\ \bibinfo {author} {\bibfnamefont
  {J.}~\bibnamefont {van~den Brink}},\ }\bibfield  {title} {\bibinfo {title}
  {Stacked topological insulator built from bismuth-based graphene sheet
  analogues},\ }\href {https://doi.org/10.1038/nmat3570} {\bibfield  {journal}
  {\bibinfo  {journal} {Nat. Mater.}\ }\textbf {\bibinfo {volume} {12}},\
  \bibinfo {pages} {422} (\bibinfo {year} {2013})}\BibitemShut {NoStop}%
\bibitem [{\citenamefont {Pauly}\ \emph {et~al.}(2015)\citenamefont {Pauly},
  \citenamefont {Rasche}, \citenamefont {Koepernik}, \citenamefont {Liebmann},
  \citenamefont {Pratzer}, \citenamefont {Richter}, \citenamefont {Kellner},
  \citenamefont {Eschbach}, \citenamefont {Kaufmann}, \citenamefont
  {Plucinski}, \citenamefont {Schneider}, \citenamefont {Ruck}, \citenamefont
  {van~den Brink},\ and\ \citenamefont {Morgenstern}}]{Pauly-2015}%
  \BibitemOpen
  \bibfield  {author} {\bibinfo {author} {\bibfnamefont {C.}~\bibnamefont
  {Pauly}}, \bibinfo {author} {\bibfnamefont {B.}~\bibnamefont {Rasche}},
  \bibinfo {author} {\bibfnamefont {K.}~\bibnamefont {Koepernik}}, \bibinfo
  {author} {\bibfnamefont {M.}~\bibnamefont {Liebmann}}, \bibinfo {author}
  {\bibfnamefont {M.}~\bibnamefont {Pratzer}}, \bibinfo {author} {\bibfnamefont
  {M.}~\bibnamefont {Richter}}, \bibinfo {author} {\bibfnamefont
  {J.}~\bibnamefont {Kellner}}, \bibinfo {author} {\bibfnamefont
  {M.}~\bibnamefont {Eschbach}}, \bibinfo {author} {\bibfnamefont
  {B.}~\bibnamefont {Kaufmann}}, \bibinfo {author} {\bibfnamefont
  {L.}~\bibnamefont {Plucinski}}, \bibinfo {author} {\bibfnamefont {C.~M.}\
  \bibnamefont {Schneider}}, \bibinfo {author} {\bibfnamefont {M.}~\bibnamefont
  {Ruck}}, \bibinfo {author} {\bibfnamefont {J.}~\bibnamefont {van~den
  Brink}},\ and\ \bibinfo {author} {\bibfnamefont {M.}~\bibnamefont
  {Morgenstern}},\ }\bibfield  {title} {\bibinfo {title} {Subnanometre-wide
  electron channels protected by topology},\ }\href
  {https://doi.org/10.1038/nphys3264} {\bibfield  {journal} {\bibinfo
  {journal} {Nat. Phys.}\ }\textbf {\bibinfo {volume} {11}},\ \bibinfo {pages}
  {338} (\bibinfo {year} {2015})}\BibitemShut {NoStop}%
\bibitem [{\citenamefont {Jeschke}\ \emph {et~al.}(2026)\citenamefont
  {Jeschke}, \citenamefont {Guterding},\ and\ \citenamefont
  {Ghosh}}]{Jeschke2026}%
  \BibitemOpen
  \bibfield  {author} {\bibinfo {author} {\bibfnamefont {H.~O.}\ \bibnamefont
  {Jeschke}}, \bibinfo {author} {\bibfnamefont {D.}~\bibnamefont {Guterding}},\
  and\ \bibinfo {author} {\bibfnamefont {P.}~\bibnamefont {Ghosh}},\ }\bibfield
   {title} {\bibinfo {title} {{Realization of the Ruby Lattice Antiferromagnet
  in Layered Transition-Metal Fluorides}},\ }\href
  {https://arxiv.org/abs/2605.28821} {\  (\bibinfo {year} {2026})},\ \Eprint
  {https://arxiv.org/abs/2605.28821} {arXiv:2605.28821 [cond-mat.str-el]}
  \BibitemShut {NoStop}%
\bibitem [{\citenamefont {Sonnenschein}\ \emph {et~al.}(2024)\citenamefont
  {Sonnenschein}, \citenamefont {Maity}, \citenamefont {Liu}, \citenamefont
  {Thomale}, \citenamefont {Ferrari},\ and\ \citenamefont
  {Iqbal}}]{Sonnenschein-2024}%
  \BibitemOpen
  \bibfield  {author} {\bibinfo {author} {\bibfnamefont {J.}~\bibnamefont
  {Sonnenschein}}, \bibinfo {author} {\bibfnamefont {A.}~\bibnamefont {Maity}},
  \bibinfo {author} {\bibfnamefont {C.}~\bibnamefont {Liu}}, \bibinfo {author}
  {\bibfnamefont {R.}~\bibnamefont {Thomale}}, \bibinfo {author} {\bibfnamefont
  {F.}~\bibnamefont {Ferrari}},\ and\ \bibinfo {author} {\bibfnamefont
  {Y.}~\bibnamefont {Iqbal}},\ }\bibfield  {title} {\bibinfo {title}
  {{Candidate quantum spin liquids on the maple-leaf lattice}},\ }\href
  {https://doi.org/10.1103/PhysRevB.110.014414} {\bibfield  {journal} {\bibinfo
   {journal} {Phys. Rev. B}\ }\textbf {\bibinfo {volume} {110}},\ \bibinfo
  {pages} {014414} (\bibinfo {year} {2024})}\BibitemShut {NoStop}%
\bibitem [{\citenamefont {Wen}(2002)}]{Wen-2002}%
  \BibitemOpen
  \bibfield  {author} {\bibinfo {author} {\bibfnamefont {X.-G.}\ \bibnamefont
  {Wen}},\ }\bibfield  {title} {\bibinfo {title} {Quantum orders and symmetric
  spin liquids},\ }\href {https://doi.org/10.1103/PhysRevB.65.165113}
  {\bibfield  {journal} {\bibinfo  {journal} {Phys. Rev. B}\ }\textbf {\bibinfo
  {volume} {65}},\ \bibinfo {pages} {165113} (\bibinfo {year}
  {2002})}\BibitemShut {NoStop}%
\bibitem [{\citenamefont {Wen}(2007)}]{Wenbook}%
  \BibitemOpen
  \bibfield  {author} {\bibinfo {author} {\bibfnamefont {X.-G.}\ \bibnamefont
  {Wen}},\ }\href {https://doi.org/10.1093/acprof:oso/9780199227259.001.0001}
  {\emph {\bibinfo {title} {{Quantum Field Theory of Many-Body Systems}}}}\
  (\bibinfo  {publisher} {Oxford University Press},\ \bibinfo {address}
  {Oxford},\ \bibinfo {year} {2007})\BibitemShut {NoStop}%
\bibitem [{\citenamefont {Wang}\ and\ \citenamefont
  {Vishwanath}(2006)}]{Wang-2006}%
  \BibitemOpen
  \bibfield  {author} {\bibinfo {author} {\bibfnamefont {F.}~\bibnamefont
  {Wang}}\ and\ \bibinfo {author} {\bibfnamefont {A.}~\bibnamefont
  {Vishwanath}},\ }\bibfield  {title} {\bibinfo {title} {{Spin-liquid states on
  the triangular and Kagom\'e lattices: A projective-symmetry-group analysis of
  Schwinger boson states}},\ }\href
  {https://doi.org/10.1103/PhysRevB.74.174423} {\bibfield  {journal} {\bibinfo
  {journal} {Phys. Rev. B}\ }\textbf {\bibinfo {volume} {74}},\ \bibinfo
  {pages} {174423} (\bibinfo {year} {2006})}\BibitemShut {NoStop}%
\bibitem [{\citenamefont {Lu}(2016)}]{Lu-2016a}%
  \BibitemOpen
  \bibfield  {author} {\bibinfo {author} {\bibfnamefont {Y.-M.}\ \bibnamefont
  {Lu}},\ }\bibfield  {title} {\bibinfo {title} {Symmetric ${Z}_{2}$ spin
  liquids and their neighboring phases on triangular lattice},\ }\href
  {https://doi.org/10.1103/PhysRevB.93.165113} {\bibfield  {journal} {\bibinfo
  {journal} {Phys. Rev. B}\ }\textbf {\bibinfo {volume} {93}},\ \bibinfo
  {pages} {165113} (\bibinfo {year} {2016})}\BibitemShut {NoStop}%
\bibitem [{\citenamefont {Lu}\ \emph {et~al.}(2011)\citenamefont {Lu},
  \citenamefont {Ran},\ and\ \citenamefont {Lee}}]{Lu-2011a}%
  \BibitemOpen
  \bibfield  {author} {\bibinfo {author} {\bibfnamefont {Y.-M.}\ \bibnamefont
  {Lu}}, \bibinfo {author} {\bibfnamefont {Y.}~\bibnamefont {Ran}},\ and\
  \bibinfo {author} {\bibfnamefont {P.~A.}\ \bibnamefont {Lee}},\ }\bibfield
  {title} {\bibinfo {title} {{${\mathbb{Z}}_{2}$ spin liquids in the
  $S=\frac{1}{2}$ Heisenberg model on the kagome lattice: A projective
  symmetry-group study of Schwinger fermion mean-field states}},\ }\href
  {https://doi.org/10.1103/PhysRevB.83.224413} {\bibfield  {journal} {\bibinfo
  {journal} {Phys. Rev. B}\ }\textbf {\bibinfo {volume} {83}},\ \bibinfo
  {pages} {224413} (\bibinfo {year} {2011})}\BibitemShut {NoStop}%
\bibitem [{\citenamefont {Yang}\ and\ \citenamefont {Yao}(2012)}]{Yang-2012}%
  \BibitemOpen
  \bibfield  {author} {\bibinfo {author} {\bibfnamefont {F.}~\bibnamefont
  {Yang}}\ and\ \bibinfo {author} {\bibfnamefont {H.}~\bibnamefont {Yao}},\
  }\bibfield  {title} {\bibinfo {title} {Frustrated resonating valence bond
  states in two dimensions: Classification and short-range correlations},\
  }\href {https://doi.org/10.1103/PhysRevLett.109.147209} {\bibfield  {journal}
  {\bibinfo  {journal} {Phys. Rev. Lett.}\ }\textbf {\bibinfo {volume} {109}},\
  \bibinfo {pages} {147209} (\bibinfo {year} {2012})}\BibitemShut {NoStop}%
\bibitem [{\citenamefont {Messio}\ \emph {et~al.}(2013)\citenamefont {Messio},
  \citenamefont {Lhuillier},\ and\ \citenamefont {Misguich}}]{Messio-2013}%
  \BibitemOpen
  \bibfield  {author} {\bibinfo {author} {\bibfnamefont {L.}~\bibnamefont
  {Messio}}, \bibinfo {author} {\bibfnamefont {C.}~\bibnamefont {Lhuillier}},\
  and\ \bibinfo {author} {\bibfnamefont {G.}~\bibnamefont {Misguich}},\
  }\bibfield  {title} {\bibinfo {title} {{Time reversal symmetry breaking
  chiral spin liquids: Projective symmetry group approach of bosonic mean-field
  theories}},\ }\href {https://doi.org/10.1103/PhysRevB.87.125127} {\bibfield
  {journal} {\bibinfo  {journal} {Phys. Rev. B}\ }\textbf {\bibinfo {volume}
  {87}},\ \bibinfo {pages} {125127} (\bibinfo {year} {2013})}\BibitemShut
  {NoStop}%
\bibitem [{\citenamefont {Bieri}\ \emph {et~al.}(2015)\citenamefont {Bieri},
  \citenamefont {Messio}, \citenamefont {Bernu},\ and\ \citenamefont
  {Lhuillier}}]{Bieri-2015}%
  \BibitemOpen
  \bibfield  {author} {\bibinfo {author} {\bibfnamefont {S.}~\bibnamefont
  {Bieri}}, \bibinfo {author} {\bibfnamefont {L.}~\bibnamefont {Messio}},
  \bibinfo {author} {\bibfnamefont {B.}~\bibnamefont {Bernu}},\ and\ \bibinfo
  {author} {\bibfnamefont {C.}~\bibnamefont {Lhuillier}},\ }\bibfield  {title}
  {\bibinfo {title} {{Gapless chiral spin liquid in a kagome Heisenberg
  model}},\ }\href {https://doi.org/10.1103/PhysRevB.92.060407} {\bibfield
  {journal} {\bibinfo  {journal} {Phys. Rev. B}\ }\textbf {\bibinfo {volume}
  {92}},\ \bibinfo {pages} {060407} (\bibinfo {year} {2015})}\BibitemShut
  {NoStop}%
\bibitem [{\citenamefont {Bieri}\ \emph {et~al.}(2016)\citenamefont {Bieri},
  \citenamefont {Lhuillier},\ and\ \citenamefont {Messio}}]{Bieri-2016}%
  \BibitemOpen
  \bibfield  {author} {\bibinfo {author} {\bibfnamefont {S.}~\bibnamefont
  {Bieri}}, \bibinfo {author} {\bibfnamefont {C.}~\bibnamefont {Lhuillier}},\
  and\ \bibinfo {author} {\bibfnamefont {L.}~\bibnamefont {Messio}},\
  }\bibfield  {title} {\bibinfo {title} {{Projective symmetry group
  classification of chiral spin liquids}},\ }\href
  {https://doi.org/10.1103/PhysRevB.93.094437} {\bibfield  {journal} {\bibinfo
  {journal} {Phys. Rev. B}\ }\textbf {\bibinfo {volume} {93}},\ \bibinfo
  {pages} {094437} (\bibinfo {year} {2016})}\BibitemShut {NoStop}%
\bibitem [{\citenamefont {Lu}\ and\ \citenamefont {Ran}(2011)}]{Lu-2011b}%
  \BibitemOpen
  \bibfield  {author} {\bibinfo {author} {\bibfnamefont {Y.-M.}\ \bibnamefont
  {Lu}}\ and\ \bibinfo {author} {\bibfnamefont {Y.}~\bibnamefont {Ran}},\
  }\bibfield  {title} {\bibinfo {title} {{${\mathbb{Z}}_{2}$ spin liquid and
  chiral antiferromagnetic phase in the Hubbard model on a honeycomb
  lattice}},\ }\href {https://doi.org/10.1103/PhysRevB.84.024420} {\bibfield
  {journal} {\bibinfo  {journal} {Phys. Rev. B}\ }\textbf {\bibinfo {volume}
  {84}},\ \bibinfo {pages} {024420} (\bibinfo {year} {2011})}\BibitemShut
  {NoStop}%
\bibitem [{\citenamefont {Yang}\ and\ \citenamefont {Wang}(2016)}]{Yang-2016}%
  \BibitemOpen
  \bibfield  {author} {\bibinfo {author} {\bibfnamefont {X.}~\bibnamefont
  {Yang}}\ and\ \bibinfo {author} {\bibfnamefont {F.}~\bibnamefont {Wang}},\
  }\bibfield  {title} {\bibinfo {title} {{Schwinger boson spin-liquid states on
  square lattice}},\ }\href {https://doi.org/10.1103/PhysRevB.94.035160}
  {\bibfield  {journal} {\bibinfo  {journal} {Phys. Rev. B}\ }\textbf {\bibinfo
  {volume} {94}},\ \bibinfo {pages} {035160} (\bibinfo {year}
  {2016})}\BibitemShut {NoStop}%
\bibitem [{\citenamefont {Yang}\ \emph {et~al.}(2010)\citenamefont {Yang},
  \citenamefont {Paramekanti},\ and\ \citenamefont {Kim}}]{Yang-2010}%
  \BibitemOpen
  \bibfield  {author} {\bibinfo {author} {\bibfnamefont {B.-J.}\ \bibnamefont
  {Yang}}, \bibinfo {author} {\bibfnamefont {A.}~\bibnamefont {Paramekanti}},\
  and\ \bibinfo {author} {\bibfnamefont {Y.~B.}\ \bibnamefont {Kim}},\
  }\bibfield  {title} {\bibinfo {title} {{Competing quantum paramagnetic ground
  states of the Heisenberg antiferromagnet on the star lattice}},\ }\href
  {https://doi.org/10.1103/PhysRevB.81.134418} {\bibfield  {journal} {\bibinfo
  {journal} {Phys. Rev. B}\ }\textbf {\bibinfo {volume} {81}},\ \bibinfo
  {pages} {134418} (\bibinfo {year} {2010})}\BibitemShut {NoStop}%
\bibitem [{\citenamefont {Choy}\ and\ \citenamefont {Kim}(2009)}]{Choy-2009}%
  \BibitemOpen
  \bibfield  {author} {\bibinfo {author} {\bibfnamefont {T.-P.}\ \bibnamefont
  {Choy}}\ and\ \bibinfo {author} {\bibfnamefont {Y.~B.}\ \bibnamefont {Kim}},\
  }\bibfield  {title} {\bibinfo {title} {{Classification of quantum phases for
  the star-lattice antiferromagnet via a projective symmetry group analysis}},\
  }\href {https://doi.org/10.1103/PhysRevB.80.064404} {\bibfield  {journal}
  {\bibinfo  {journal} {Phys. Rev. B}\ }\textbf {\bibinfo {volume} {80}},\
  \bibinfo {pages} {064404} (\bibinfo {year} {2009})}\BibitemShut {NoStop}%
\bibitem [{\citenamefont {Sahoo}\ and\ \citenamefont
  {Flint}(2020)}]{Sahoo-2020}%
  \BibitemOpen
  \bibfield  {author} {\bibinfo {author} {\bibfnamefont {J.}~\bibnamefont
  {Sahoo}}\ and\ \bibinfo {author} {\bibfnamefont {R.}~\bibnamefont {Flint}},\
  }\bibfield  {title} {\bibinfo {title} {{Symmetric spin liquids on the stuffed
  honeycomb lattice}},\ }\href {https://doi.org/10.1103/PhysRevB.101.115103}
  {\bibfield  {journal} {\bibinfo  {journal} {Phys. Rev. B}\ }\textbf {\bibinfo
  {volume} {101}},\ \bibinfo {pages} {115103} (\bibinfo {year}
  {2020})}\BibitemShut {NoStop}%
\bibitem [{\citenamefont {Maity}\ \emph {et~al.}(2023)\citenamefont {Maity},
  \citenamefont {Ferrari}, \citenamefont {Thomale}, \citenamefont {Mandal},\
  and\ \citenamefont {Iqbal}}]{Maity-2023}%
  \BibitemOpen
  \bibfield  {author} {\bibinfo {author} {\bibfnamefont {A.}~\bibnamefont
  {Maity}}, \bibinfo {author} {\bibfnamefont {F.}~\bibnamefont {Ferrari}},
  \bibinfo {author} {\bibfnamefont {R.}~\bibnamefont {Thomale}}, \bibinfo
  {author} {\bibfnamefont {S.}~\bibnamefont {Mandal}},\ and\ \bibinfo {author}
  {\bibfnamefont {Y.}~\bibnamefont {Iqbal}},\ }\bibfield  {title} {\bibinfo
  {title} {{Projective symmetry group classification of Abrikosov fermion
  mean-field ans\"atze on the square-octagon lattice}},\ }\href
  {https://doi.org/10.1103/PhysRevB.107.134438} {\bibfield  {journal} {\bibinfo
   {journal} {Phys. Rev. B}\ }\textbf {\bibinfo {volume} {107}},\ \bibinfo
  {pages} {134438} (\bibinfo {year} {2023})}\BibitemShut {NoStop}%
\bibitem [{\citenamefont {Chatterjee}\ \emph {et~al.}(2026)\citenamefont
  {Chatterjee}, \citenamefont {Maity}, \citenamefont {Potten}, \citenamefont
  {M\"uller}, \citenamefont {Feuerpfeil}, \citenamefont {Thomale},
  \citenamefont {Penc}, \citenamefont {Jeschke}, \citenamefont {Samajdar},\
  and\ \citenamefont {Iqbal}}]{Chatterjee-2026}%
  \BibitemOpen
  \bibfield  {author} {\bibinfo {author} {\bibfnamefont {S.}~\bibnamefont
  {Chatterjee}}, \bibinfo {author} {\bibfnamefont {A.}~\bibnamefont {Maity}},
  \bibinfo {author} {\bibfnamefont {J.}~\bibnamefont {Potten}}, \bibinfo
  {author} {\bibfnamefont {T.}~\bibnamefont {M\"uller}}, \bibinfo {author}
  {\bibfnamefont {A.}~\bibnamefont {Feuerpfeil}}, \bibinfo {author}
  {\bibfnamefont {R.}~\bibnamefont {Thomale}}, \bibinfo {author} {\bibfnamefont
  {K.}~\bibnamefont {Penc}}, \bibinfo {author} {\bibfnamefont {H.~O.}\
  \bibnamefont {Jeschke}}, \bibinfo {author} {\bibfnamefont {R.}~\bibnamefont
  {Samajdar}},\ and\ \bibinfo {author} {\bibfnamefont {Y.}~\bibnamefont
  {Iqbal}},\ }\bibfield  {title} {\bibinfo {title} {{Semi-Dirac spin liquids
  and frustrated quantum magnetism on the trellis lattice}},\ }\href
  {https://doi.org/10.1103/xmtj-4kj9} {\bibfield  {journal} {\bibinfo
  {journal} {Phys. Rev. Res.}\ }\textbf {\bibinfo {volume} {8}},\ \bibinfo
  {pages} {013191} (\bibinfo {year} {2026})}\BibitemShut {NoStop}%
\bibitem [{\citenamefont {Liu}\ and\ \citenamefont {Wang}(2024)}]{Liu-2024}%
  \BibitemOpen
  \bibfield  {author} {\bibinfo {author} {\bibfnamefont {K.}~\bibnamefont
  {Liu}}\ and\ \bibinfo {author} {\bibfnamefont {F.}~\bibnamefont {Wang}},\
  }\bibfield  {title} {\bibinfo {title} {{Schwinger boson symmetric spin
  liquids of Shastry-Sutherland model}},\ }\href
  {https://doi.org/10.1103/PhysRevB.109.134409} {\bibfield  {journal} {\bibinfo
   {journal} {Phys. Rev. B}\ }\textbf {\bibinfo {volume} {109}},\ \bibinfo
  {pages} {134409} (\bibinfo {year} {2024})}\BibitemShut {NoStop}%
\bibitem [{\citenamefont {Maity}\ \emph
  {et~al.}(2026{\natexlab{a}})\citenamefont {Maity}, \citenamefont {Ferrari},
  \citenamefont {Capponi}, \citenamefont {Penc}, \citenamefont {Potten},
  \citenamefont {Müller}, \citenamefont {Feuerpfeil}, \citenamefont {Thomale},
  \citenamefont {Lee}, \citenamefont {Samajdar},\ and\ \citenamefont
  {Iqbal}}]{maity2026}%
  \BibitemOpen
  \bibfield  {author} {\bibinfo {author} {\bibfnamefont {A.}~\bibnamefont
  {Maity}}, \bibinfo {author} {\bibfnamefont {F.}~\bibnamefont {Ferrari}},
  \bibinfo {author} {\bibfnamefont {S.}~\bibnamefont {Capponi}}, \bibinfo
  {author} {\bibfnamefont {K.}~\bibnamefont {Penc}}, \bibinfo {author}
  {\bibfnamefont {J.}~\bibnamefont {Potten}}, \bibinfo {author} {\bibfnamefont
  {T.}~\bibnamefont {Müller}}, \bibinfo {author} {\bibfnamefont
  {A.}~\bibnamefont {Feuerpfeil}}, \bibinfo {author} {\bibfnamefont
  {R.}~\bibnamefont {Thomale}}, \bibinfo {author} {\bibfnamefont {J.~Y.}\
  \bibnamefont {Lee}}, \bibinfo {author} {\bibfnamefont {R.}~\bibnamefont
  {Samajdar}},\ and\ \bibinfo {author} {\bibfnamefont {Y.}~\bibnamefont
  {Iqbal}},\ }\bibfield  {title} {\bibinfo {title} {{Evidence for a
  $\mathbb{Z}_{2}$ Dirac spin liquid in the generalized Shastry-Sutherland
  model}},\ }\href {https://arxiv.org/abs/2501.00096} {\  (\bibinfo {year}
  {2026}{\natexlab{a}})},\ \Eprint {https://arxiv.org/abs/2501.00096}
  {arXiv:2501.00096 [cond-mat.str-el]} \BibitemShut {NoStop}%
\bibitem [{\citenamefont {Schneider}\ \emph {et~al.}(2022)\citenamefont
  {Schneider}, \citenamefont {Halimeh},\ and\ \citenamefont
  {Punk}}]{Benedikt-2022}%
  \BibitemOpen
  \bibfield  {author} {\bibinfo {author} {\bibfnamefont {B.}~\bibnamefont
  {Schneider}}, \bibinfo {author} {\bibfnamefont {J.~C.}\ \bibnamefont
  {Halimeh}},\ and\ \bibinfo {author} {\bibfnamefont {M.}~\bibnamefont
  {Punk}},\ }\bibfield  {title} {\bibinfo {title} {{Projective symmetry group
  classification of chiral ${\mathbb{Z}}_{2}$ spin liquids on the pyrochlore
  lattice: Application to the spin-$\frac{1}{2}$ XXZ Heisenberg model}},\
  }\href {https://doi.org/10.1103/PhysRevB.105.125122} {\bibfield  {journal}
  {\bibinfo  {journal} {Phys. Rev. B}\ }\textbf {\bibinfo {volume} {105}},\
  \bibinfo {pages} {125122} (\bibinfo {year} {2022})}\BibitemShut {NoStop}%
\bibitem [{\citenamefont {Sonnenschein}\ \emph {et~al.}(2020)\citenamefont
  {Sonnenschein}, \citenamefont {Chauhan}, \citenamefont {Iqbal},\ and\
  \citenamefont {Reuther}}]{Sonnenschein-2020}%
  \BibitemOpen
  \bibfield  {author} {\bibinfo {author} {\bibfnamefont {J.}~\bibnamefont
  {Sonnenschein}}, \bibinfo {author} {\bibfnamefont {A.}~\bibnamefont
  {Chauhan}}, \bibinfo {author} {\bibfnamefont {Y.}~\bibnamefont {Iqbal}},\
  and\ \bibinfo {author} {\bibfnamefont {J.}~\bibnamefont {Reuther}},\
  }\bibfield  {title} {\bibinfo {title} {{Projective symmetry group
  classifications of quantum spin liquids on the simple cubic, body centered
  cubic, and face centered cubic lattices}},\ }\href
  {https://doi.org/10.1103/PhysRevB.102.125140} {\bibfield  {journal} {\bibinfo
   {journal} {Phys. Rev. B}\ }\textbf {\bibinfo {volume} {102}},\ \bibinfo
  {pages} {125140} (\bibinfo {year} {2020})}\BibitemShut {NoStop}%
\bibitem [{\citenamefont {Liu}\ \emph {et~al.}(2019)\citenamefont {Liu},
  \citenamefont {Hal\'asz},\ and\ \citenamefont {Balents}}]{Liu-2019}%
  \BibitemOpen
  \bibfield  {author} {\bibinfo {author} {\bibfnamefont {C.}~\bibnamefont
  {Liu}}, \bibinfo {author} {\bibfnamefont {G.~B.}\ \bibnamefont {Hal\'asz}},\
  and\ \bibinfo {author} {\bibfnamefont {L.}~\bibnamefont {Balents}},\
  }\bibfield  {title} {\bibinfo {title} {{Competing orders in pyrochlore
  magnets from a ${\mathbb{Z}}_{2}$ spin liquid perspective}},\ }\href
  {https://doi.org/10.1103/PhysRevB.100.075125} {\bibfield  {journal} {\bibinfo
   {journal} {Phys. Rev. B}\ }\textbf {\bibinfo {volume} {100}},\ \bibinfo
  {pages} {075125} (\bibinfo {year} {2019})}\BibitemShut {NoStop}%
\bibitem [{\citenamefont {Liu}\ \emph {et~al.}(2021)\citenamefont {Liu},
  \citenamefont {Hal\'asz},\ and\ \citenamefont {Balents}}]{Liu-2021}%
  \BibitemOpen
  \bibfield  {author} {\bibinfo {author} {\bibfnamefont {C.}~\bibnamefont
  {Liu}}, \bibinfo {author} {\bibfnamefont {G.~B.}\ \bibnamefont {Hal\'asz}},\
  and\ \bibinfo {author} {\bibfnamefont {L.}~\bibnamefont {Balents}},\
  }\bibfield  {title} {\bibinfo {title} {{Symmetric U(1) and ${\mathbb{Z}}_{2}$
  spin liquids on the pyrochlore lattice}},\ }\href
  {https://doi.org/10.1103/PhysRevB.104.054401} {\bibfield  {journal} {\bibinfo
   {journal} {Phys. Rev. B}\ }\textbf {\bibinfo {volume} {104}},\ \bibinfo
  {pages} {054401} (\bibinfo {year} {2021})}\BibitemShut {NoStop}%
\bibitem [{\citenamefont {Liu}\ \emph {et~al.}(2026)\citenamefont {Liu},
  \citenamefont {Balents},\ and\ \citenamefont {Iqbal}}]{liu2026}%
  \BibitemOpen
  \bibfield  {author} {\bibinfo {author} {\bibfnamefont {C.}~\bibnamefont
  {Liu}}, \bibinfo {author} {\bibfnamefont {L.}~\bibnamefont {Balents}},\ and\
  \bibinfo {author} {\bibfnamefont {Y.}~\bibnamefont {Iqbal}},\ }\bibfield
  {title} {\bibinfo {title} {Classification and correlation signatures of
  chiral spin liquids on the pyrochlore lattice},\ }\href
  {https://arxiv.org/abs/2604.11880} {\  (\bibinfo {year} {2026})},\ \Eprint
  {https://arxiv.org/abs/2604.11880} {arXiv:2604.11880 [cond-mat.str-el]}
  \BibitemShut {NoStop}%
\bibitem [{\citenamefont {Chern}\ \emph {et~al.}(2022)\citenamefont {Chern},
  \citenamefont {Kim},\ and\ \citenamefont {Castelnovo}}]{Chern-2022}%
  \BibitemOpen
  \bibfield  {author} {\bibinfo {author} {\bibfnamefont {L.~E.}\ \bibnamefont
  {Chern}}, \bibinfo {author} {\bibfnamefont {Y.~B.}\ \bibnamefont {Kim}},\
  and\ \bibinfo {author} {\bibfnamefont {C.}~\bibnamefont {Castelnovo}},\
  }\bibfield  {title} {\bibinfo {title} {{Competing quantum spin liquids, gauge
  fluctuations, and anisotropic interactions in a breathing pyrochlore
  lattice}},\ }\href {https://doi.org/10.1103/PhysRevB.106.134402} {\bibfield
  {journal} {\bibinfo  {journal} {Phys. Rev. B}\ }\textbf {\bibinfo {volume}
  {106}},\ \bibinfo {pages} {134402} (\bibinfo {year} {2022})}\BibitemShut
  {NoStop}%
\bibitem [{\citenamefont {Lawler}\ \emph {et~al.}(2008)\citenamefont {Lawler},
  \citenamefont {Paramekanti}, \citenamefont {Kim},\ and\ \citenamefont
  {Balents}}]{Lawler-2008}%
  \BibitemOpen
  \bibfield  {author} {\bibinfo {author} {\bibfnamefont {M.~J.}\ \bibnamefont
  {Lawler}}, \bibinfo {author} {\bibfnamefont {A.}~\bibnamefont {Paramekanti}},
  \bibinfo {author} {\bibfnamefont {Y.~B.}\ \bibnamefont {Kim}},\ and\ \bibinfo
  {author} {\bibfnamefont {L.}~\bibnamefont {Balents}},\ }\bibfield  {title}
  {\bibinfo {title} {{Gapless Spin Liquids on the Three-Dimensional Hyperkagome
  Lattice of ${\mathrm{Na}}_{4}{\mathrm{Ir}}_{3}{\mathrm{O}}_{8}$}},\ }\href
  {https://doi.org/10.1103/PhysRevLett.101.197202} {\bibfield  {journal}
  {\bibinfo  {journal} {Phys. Rev. Lett.}\ }\textbf {\bibinfo {volume} {101}},\
  \bibinfo {pages} {197202} (\bibinfo {year} {2008})}\BibitemShut {NoStop}%
\bibitem [{\citenamefont {Huang}\ \emph {et~al.}(2017)\citenamefont {Huang},
  \citenamefont {Kim},\ and\ \citenamefont {Lu}}]{Huang-2017}%
  \BibitemOpen
  \bibfield  {author} {\bibinfo {author} {\bibfnamefont {B.}~\bibnamefont
  {Huang}}, \bibinfo {author} {\bibfnamefont {Y.~B.}\ \bibnamefont {Kim}},\
  and\ \bibinfo {author} {\bibfnamefont {Y.-M.}\ \bibnamefont {Lu}},\
  }\bibfield  {title} {\bibinfo {title} {{Interplay of nonsymmorphic symmetry
  and spin-orbit coupling in hyperkagome spin liquids: Applications to
  ${\mathrm{Na}}_{4}{\mathrm{Ir}}_{3}{\mathrm{O}}_{8}$}},\ }\href
  {https://doi.org/10.1103/PhysRevB.95.054404} {\bibfield  {journal} {\bibinfo
  {journal} {Phys. Rev. B}\ }\textbf {\bibinfo {volume} {95}},\ \bibinfo
  {pages} {054404} (\bibinfo {year} {2017})}\BibitemShut {NoStop}%
\bibitem [{\citenamefont {Jin}\ and\ \citenamefont {Zhou}(2020)}]{Jin-2020}%
  \BibitemOpen
  \bibfield  {author} {\bibinfo {author} {\bibfnamefont {H.-K.}\ \bibnamefont
  {Jin}}\ and\ \bibinfo {author} {\bibfnamefont {Y.}~\bibnamefont {Zhou}},\
  }\bibfield  {title} {\bibinfo {title} {{Classical and quantum order in
  hyperkagome antiferromagnets}},\ }\href
  {https://doi.org/10.1103/PhysRevB.101.054408} {\bibfield  {journal} {\bibinfo
   {journal} {Phys. Rev. B}\ }\textbf {\bibinfo {volume} {101}},\ \bibinfo
  {pages} {054408} (\bibinfo {year} {2020})}\BibitemShut {NoStop}%
\bibitem [{\citenamefont {Huang}\ \emph {et~al.}(2018)\citenamefont {Huang},
  \citenamefont {Choi}, \citenamefont {Kim},\ and\ \citenamefont
  {Lu}}]{Huang-2018}%
  \BibitemOpen
  \bibfield  {author} {\bibinfo {author} {\bibfnamefont {B.}~\bibnamefont
  {Huang}}, \bibinfo {author} {\bibfnamefont {W.}~\bibnamefont {Choi}},
  \bibinfo {author} {\bibfnamefont {Y.~B.}\ \bibnamefont {Kim}},\ and\ \bibinfo
  {author} {\bibfnamefont {Y.-M.}\ \bibnamefont {Lu}},\ }\bibfield  {title}
  {\bibinfo {title} {{Classification and properties of quantum spin liquids on
  the hyperhoneycomb lattice}},\ }\href
  {https://doi.org/10.1103/PhysRevB.97.195141} {\bibfield  {journal} {\bibinfo
  {journal} {Phys. Rev. B}\ }\textbf {\bibinfo {volume} {97}},\ \bibinfo
  {pages} {195141} (\bibinfo {year} {2018})}\BibitemShut {NoStop}%
\bibitem [{\citenamefont {Chern}\ and\ \citenamefont {Kim}(2021)}]{Chern-2021}%
  \BibitemOpen
  \bibfield  {author} {\bibinfo {author} {\bibfnamefont {L.~E.}\ \bibnamefont
  {Chern}}\ and\ \bibinfo {author} {\bibfnamefont {Y.~B.}\ \bibnamefont
  {Kim}},\ }\bibfield  {title} {\bibinfo {title} {{Theoretical study of quantum
  spin liquids in $S=\frac{1}{2}$ hyper-hyperkagome magnets: Classification,
  heat capacity, and dynamical spin structure factor}},\ }\href
  {https://doi.org/10.1103/PhysRevB.104.094413} {\bibfield  {journal} {\bibinfo
   {journal} {Phys. Rev. B}\ }\textbf {\bibinfo {volume} {104}},\ \bibinfo
  {pages} {094413} (\bibinfo {year} {2021})}\BibitemShut {NoStop}%
\bibitem [{\citenamefont {Chauhan}\ \emph {et~al.}(2023)\citenamefont
  {Chauhan}, \citenamefont {Maity}, \citenamefont {Liu}, \citenamefont
  {Sonnenschein}, \citenamefont {Ferrari},\ and\ \citenamefont
  {Iqbal}}]{Chauhan-2023}%
  \BibitemOpen
  \bibfield  {author} {\bibinfo {author} {\bibfnamefont {A.}~\bibnamefont
  {Chauhan}}, \bibinfo {author} {\bibfnamefont {A.}~\bibnamefont {Maity}},
  \bibinfo {author} {\bibfnamefont {C.}~\bibnamefont {Liu}}, \bibinfo {author}
  {\bibfnamefont {J.}~\bibnamefont {Sonnenschein}}, \bibinfo {author}
  {\bibfnamefont {F.}~\bibnamefont {Ferrari}},\ and\ \bibinfo {author}
  {\bibfnamefont {Y.}~\bibnamefont {Iqbal}},\ }\bibfield  {title} {\bibinfo
  {title} {{Quantum spin liquids on the diamond lattice}},\ }\href
  {https://doi.org/10.1103/PhysRevB.108.134424} {\bibfield  {journal} {\bibinfo
   {journal} {Phys. Rev. B}\ }\textbf {\bibinfo {volume} {108}},\ \bibinfo
  {pages} {134424} (\bibinfo {year} {2023})}\BibitemShut {NoStop}%
\bibitem [{\citenamefont {Li}\ \emph {et~al.}(2025)\citenamefont {Li},
  \citenamefont {Biswas},\ and\ \citenamefont {Parameswaran}}]{Li-2024}%
  \BibitemOpen
  \bibfield  {author} {\bibinfo {author} {\bibfnamefont {M.-H.}\ \bibnamefont
  {Li}}, \bibinfo {author} {\bibfnamefont {S.}~\bibnamefont {Biswas}},\ and\
  \bibinfo {author} {\bibfnamefont {S.~A.}\ \bibnamefont {Parameswaran}},\
  }\bibfield  {title} {\bibinfo {title} {{Classification of spin-$\frac{1}{2}$
  fermionic quantum spin liquids on the trillium lattice}},\ }\href
  {https://doi.org/10.1103/3nmp-1vt2} {\bibfield  {journal} {\bibinfo
  {journal} {Phys. Rev. B}\ }\textbf {\bibinfo {volume} {112}},\ \bibinfo
  {pages} {104429} (\bibinfo {year} {2025})}\BibitemShut {NoStop}%
\bibitem [{\citenamefont {Lu}(2018)}]{Lu-2018}%
  \BibitemOpen
  \bibfield  {author} {\bibinfo {author} {\bibfnamefont {Y.-M.}\ \bibnamefont
  {Lu}},\ }\bibfield  {title} {\bibinfo {title} {{Symmetry-protected gapless
  ${\mathbb{Z}}_{2}$ spin liquids}},\ }\href
  {https://doi.org/10.1103/PhysRevB.97.094422} {\bibfield  {journal} {\bibinfo
  {journal} {Phys. Rev. B}\ }\textbf {\bibinfo {volume} {97}},\ \bibinfo
  {pages} {094422} (\bibinfo {year} {2018})}\BibitemShut {NoStop}%
\bibitem [{\citenamefont {Iqbal}\ \emph
  {et~al.}(2011{\natexlab{a}})\citenamefont {Iqbal}, \citenamefont {Becca},\
  and\ \citenamefont {Poilblanc}}]{Iqbal-2011a}%
  \BibitemOpen
  \bibfield  {author} {\bibinfo {author} {\bibfnamefont {Y.}~\bibnamefont
  {Iqbal}}, \bibinfo {author} {\bibfnamefont {F.}~\bibnamefont {Becca}},\ and\
  \bibinfo {author} {\bibfnamefont {D.}~\bibnamefont {Poilblanc}},\ }\bibfield
  {title} {\bibinfo {title} {{Valence-bond crystal in the extended kagome
  spin-$\frac{1}{2}$ quantum Heisenberg antiferromagnet: A variational Monte
  Carlo approach}},\ }\href {https://doi.org/10.1103/PhysRevB.83.100404}
  {\bibfield  {journal} {\bibinfo  {journal} {Phys. Rev. B}\ }\textbf {\bibinfo
  {volume} {83}},\ \bibinfo {pages} {100404} (\bibinfo {year}
  {2011}{\natexlab{a}})}\BibitemShut {NoStop}%
\bibitem [{\citenamefont {Iqbal}\ \emph
  {et~al.}(2011{\natexlab{b}})\citenamefont {Iqbal}, \citenamefont {Becca},\
  and\ \citenamefont {Poilblanc}}]{Iqbal-2011b}%
  \BibitemOpen
  \bibfield  {author} {\bibinfo {author} {\bibfnamefont {Y.}~\bibnamefont
  {Iqbal}}, \bibinfo {author} {\bibfnamefont {F.}~\bibnamefont {Becca}},\ and\
  \bibinfo {author} {\bibfnamefont {D.}~\bibnamefont {Poilblanc}},\ }\bibfield
  {title} {\bibinfo {title} {{Projected wave function study of
  ${\mathbb{Z}}_{2}$ spin liquids on the kagome lattice for the
  spin-$\frac{1}{2}$ quantum Heisenberg antiferromagnet}},\ }\href
  {https://doi.org/10.1103/PhysRevB.84.020407} {\bibfield  {journal} {\bibinfo
  {journal} {Phys. Rev. B}\ }\textbf {\bibinfo {volume} {84}},\ \bibinfo
  {pages} {020407} (\bibinfo {year} {2011}{\natexlab{b}})}\BibitemShut
  {NoStop}%
\bibitem [{\citenamefont {Iqbal}\ \emph {et~al.}(2012)\citenamefont {Iqbal},
  \citenamefont {Becca},\ and\ \citenamefont {Poilblanc}}]{Iqbal-2012}%
  \BibitemOpen
  \bibfield  {author} {\bibinfo {author} {\bibfnamefont {Y.}~\bibnamefont
  {Iqbal}}, \bibinfo {author} {\bibfnamefont {F.}~\bibnamefont {Becca}},\ and\
  \bibinfo {author} {\bibfnamefont {D.}~\bibnamefont {Poilblanc}},\ }\bibfield
  {title} {\bibinfo {title} {{Valence-bond crystals in the kagom{\'e} spin-1/2
  Heisenberg antiferromagnet: a symmetry classification and projected wave
  function study}},\ }\href {https://doi.org/10.1088/1367-2630/14/11/115031}
  {\bibfield  {journal} {\bibinfo  {journal} {New J. Phys.}\ }\textbf {\bibinfo
  {volume} {14}},\ \bibinfo {pages} {115031} (\bibinfo {year}
  {2012})}\BibitemShut {NoStop}%
\bibitem [{\citenamefont {Iqbal}\ \emph {et~al.}(2014)\citenamefont {Iqbal},
  \citenamefont {Poilblanc},\ and\ \citenamefont {Becca}}]{Iqbal-2014}%
  \BibitemOpen
  \bibfield  {author} {\bibinfo {author} {\bibfnamefont {Y.}~\bibnamefont
  {Iqbal}}, \bibinfo {author} {\bibfnamefont {D.}~\bibnamefont {Poilblanc}},\
  and\ \bibinfo {author} {\bibfnamefont {F.}~\bibnamefont {Becca}},\ }\bibfield
   {title} {\bibinfo {title} {{Vanishing spin gap in a competing spin-liquid
  phase in the kagome Heisenberg antiferromagnet}},\ }\href
  {https://doi.org/10.1103/PhysRevB.89.020407} {\bibfield  {journal} {\bibinfo
  {journal} {Phys. Rev. B}\ }\textbf {\bibinfo {volume} {89}},\ \bibinfo
  {pages} {020407} (\bibinfo {year} {2014})}\BibitemShut {NoStop}%
\bibitem [{\citenamefont {Iqbal}\ \emph {et~al.}(2018)\citenamefont {Iqbal},
  \citenamefont {Poilblanc}, \citenamefont {Thomale},\ and\ \citenamefont
  {Becca}}]{Iqbal-2018_bk}%
  \BibitemOpen
  \bibfield  {author} {\bibinfo {author} {\bibfnamefont {Y.}~\bibnamefont
  {Iqbal}}, \bibinfo {author} {\bibfnamefont {D.}~\bibnamefont {Poilblanc}},
  \bibinfo {author} {\bibfnamefont {R.}~\bibnamefont {Thomale}},\ and\ \bibinfo
  {author} {\bibfnamefont {F.}~\bibnamefont {Becca}},\ }\bibfield  {title}
  {\bibinfo {title} {{Persistence of the gapless spin liquid in the breathing
  kagome Heisenberg antiferromagnet}},\ }\href
  {https://doi.org/10.1103/PhysRevB.97.115127} {\bibfield  {journal} {\bibinfo
  {journal} {Phys. Rev. B}\ }\textbf {\bibinfo {volume} {97}},\ \bibinfo
  {pages} {115127} (\bibinfo {year} {2018})}\BibitemShut {NoStop}%
\bibitem [{\citenamefont {Kiese}\ \emph {et~al.}(2023)\citenamefont {Kiese},
  \citenamefont {Ferrari}, \citenamefont {Astrakhantsev}, \citenamefont
  {Niggemann}, \citenamefont {Ghosh}, \citenamefont {M\"uller}, \citenamefont
  {Thomale}, \citenamefont {Neupert}, \citenamefont {Reuther}, \citenamefont
  {Gingras}, \citenamefont {Trebst},\ and\ \citenamefont {Iqbal}}]{Kiese-2023}%
  \BibitemOpen
  \bibfield  {author} {\bibinfo {author} {\bibfnamefont {D.}~\bibnamefont
  {Kiese}}, \bibinfo {author} {\bibfnamefont {F.}~\bibnamefont {Ferrari}},
  \bibinfo {author} {\bibfnamefont {N.}~\bibnamefont {Astrakhantsev}}, \bibinfo
  {author} {\bibfnamefont {N.}~\bibnamefont {Niggemann}}, \bibinfo {author}
  {\bibfnamefont {P.}~\bibnamefont {Ghosh}}, \bibinfo {author} {\bibfnamefont
  {T.}~\bibnamefont {M\"uller}}, \bibinfo {author} {\bibfnamefont
  {R.}~\bibnamefont {Thomale}}, \bibinfo {author} {\bibfnamefont
  {T.}~\bibnamefont {Neupert}}, \bibinfo {author} {\bibfnamefont
  {J.}~\bibnamefont {Reuther}}, \bibinfo {author} {\bibfnamefont {M.~J.~P.}\
  \bibnamefont {Gingras}}, \bibinfo {author} {\bibfnamefont {S.}~\bibnamefont
  {Trebst}},\ and\ \bibinfo {author} {\bibfnamefont {Y.}~\bibnamefont
  {Iqbal}},\ }\bibfield  {title} {\bibinfo {title} {{Pinch-points to half-moons
  and up in the stars: The kagome skymap}},\ }\href
  {https://doi.org/10.1103/PhysRevResearch.5.L012025} {\bibfield  {journal}
  {\bibinfo  {journal} {Phys. Rev. Res.}\ }\textbf {\bibinfo {volume} {5}},\
  \bibinfo {pages} {L012025} (\bibinfo {year} {2023})}\BibitemShut {NoStop}%
\bibitem [{\citenamefont {Iqbal}\ \emph {et~al.}(2021)\citenamefont {Iqbal},
  \citenamefont {Ferrari}, \citenamefont {Chauhan}, \citenamefont {Parola},
  \citenamefont {Poilblanc},\ and\ \citenamefont {Becca}}]{Iqbal-2021}%
  \BibitemOpen
  \bibfield  {author} {\bibinfo {author} {\bibfnamefont {Y.}~\bibnamefont
  {Iqbal}}, \bibinfo {author} {\bibfnamefont {F.}~\bibnamefont {Ferrari}},
  \bibinfo {author} {\bibfnamefont {A.}~\bibnamefont {Chauhan}}, \bibinfo
  {author} {\bibfnamefont {A.}~\bibnamefont {Parola}}, \bibinfo {author}
  {\bibfnamefont {D.}~\bibnamefont {Poilblanc}},\ and\ \bibinfo {author}
  {\bibfnamefont {F.}~\bibnamefont {Becca}},\ }\bibfield  {title} {\bibinfo
  {title} {{Gutzwiller projected states for the ${J}_{1}\ensuremath{-}{J}_{2}$
  Heisenberg model on the Kagome lattice: Achievements and pitfalls}},\ }\href
  {https://doi.org/10.1103/PhysRevB.104.144406} {\bibfield  {journal} {\bibinfo
   {journal} {Phys. Rev. B}\ }\textbf {\bibinfo {volume} {104}},\ \bibinfo
  {pages} {144406} (\bibinfo {year} {2021})}\BibitemShut {NoStop}%
\bibitem [{\citenamefont {Hu}\ \emph {et~al.}(2013)\citenamefont {Hu},
  \citenamefont {Becca}, \citenamefont {Parola},\ and\ \citenamefont
  {Sorella}}]{Hu-2013}%
  \BibitemOpen
  \bibfield  {author} {\bibinfo {author} {\bibfnamefont {W.-J.}\ \bibnamefont
  {Hu}}, \bibinfo {author} {\bibfnamefont {F.}~\bibnamefont {Becca}}, \bibinfo
  {author} {\bibfnamefont {A.}~\bibnamefont {Parola}},\ and\ \bibinfo {author}
  {\bibfnamefont {S.}~\bibnamefont {Sorella}},\ }\bibfield  {title} {\bibinfo
  {title} {{Direct evidence for a gapless ${Z}_{2}$ spin liquid by frustrating
  N\'eel antiferromagnetism}},\ }\href
  {https://doi.org/10.1103/PhysRevB.88.060402} {\bibfield  {journal} {\bibinfo
  {journal} {Phys. Rev. B}\ }\textbf {\bibinfo {volume} {88}},\ \bibinfo
  {pages} {060402} (\bibinfo {year} {2013})}\BibitemShut {NoStop}%
\bibitem [{\citenamefont {Ferrari}\ and\ \citenamefont
  {Becca}(2019)}]{Ferrari-2019}%
  \BibitemOpen
  \bibfield  {author} {\bibinfo {author} {\bibfnamefont {F.}~\bibnamefont
  {Ferrari}}\ and\ \bibinfo {author} {\bibfnamefont {F.}~\bibnamefont
  {Becca}},\ }\bibfield  {title} {\bibinfo {title} {{Dynamical Structure Factor
  of the ${J}_{1}\ensuremath{-}{J}_{2}$ Heisenberg Model on the Triangular
  Lattice: Magnons, Spinons, and Gauge Fields}},\ }\href
  {https://doi.org/10.1103/PhysRevX.9.031026} {\bibfield  {journal} {\bibinfo
  {journal} {Phys. Rev. X}\ }\textbf {\bibinfo {volume} {9}},\ \bibinfo {pages}
  {031026} (\bibinfo {year} {2019})}\BibitemShut {NoStop}%
\bibitem [{\citenamefont {Ferrari}\ \emph {et~al.}(2023)\citenamefont
  {Ferrari}, \citenamefont {Niu}, \citenamefont {Hasik}, \citenamefont {Iqbal},
  \citenamefont {Poilblanc},\ and\ \citenamefont {Becca}}]{Ferrari-2023}%
  \BibitemOpen
  \bibfield  {author} {\bibinfo {author} {\bibfnamefont {F.}~\bibnamefont
  {Ferrari}}, \bibinfo {author} {\bibfnamefont {S.}~\bibnamefont {Niu}},
  \bibinfo {author} {\bibfnamefont {J.}~\bibnamefont {Hasik}}, \bibinfo
  {author} {\bibfnamefont {Y.}~\bibnamefont {Iqbal}}, \bibinfo {author}
  {\bibfnamefont {D.}~\bibnamefont {Poilblanc}},\ and\ \bibinfo {author}
  {\bibfnamefont {F.}~\bibnamefont {Becca}},\ }\bibfield  {title} {\bibinfo
  {title} {{Static and dynamical signatures of Dzyaloshinskii-Moriya
  interactions in the Heisenberg model on the kagome lattice}},\ }\href
  {https://doi.org/10.21468/SciPostPhys.14.6.139} {\bibfield  {journal}
  {\bibinfo  {journal} {SciPost Phys.}\ }\textbf {\bibinfo {volume} {14}},\
  \bibinfo {pages} {139} (\bibinfo {year} {2023})}\BibitemShut {NoStop}%
\bibitem [{\citenamefont {Abrikosov}(1965)}]{Abrikosov-1965}%
  \BibitemOpen
  \bibfield  {author} {\bibinfo {author} {\bibfnamefont {A.~A.}\ \bibnamefont
  {Abrikosov}},\ }\bibfield  {title} {\bibinfo {title} {Electron scattering on
  magnetic impurities in metals and anomalous resistivity effects},\ }\href
  {https://doi.org/10.1103/PhysicsPhysiqueFizika.2.5} {\bibfield  {journal}
  {\bibinfo  {journal} {Physics}\ }\textbf {\bibinfo {volume} {2}},\ \bibinfo
  {pages} {5} (\bibinfo {year} {1965})}\BibitemShut {NoStop}%
\bibitem [{\citenamefont {Affleck}\ and\ \citenamefont
  {Marston}(1988)}]{Affleck-1988}%
  \BibitemOpen
  \bibfield  {author} {\bibinfo {author} {\bibfnamefont {I.}~\bibnamefont
  {Affleck}}\ and\ \bibinfo {author} {\bibfnamefont {J.~B.}\ \bibnamefont
  {Marston}},\ }\bibfield  {title} {\bibinfo {title} {{Large-$n$ limit of the
  Heisenberg-Hubbard model: Implications for high-${T}_{c}$ superconductors}},\
  }\href {https://doi.org/10.1103/PhysRevB.37.3774} {\bibfield  {journal}
  {\bibinfo  {journal} {Phys. Rev. B}\ }\textbf {\bibinfo {volume} {37}},\
  \bibinfo {pages} {3774} (\bibinfo {year} {1988})}\BibitemShut {NoStop}%
\bibitem [{\citenamefont {Marston}\ and\ \citenamefont
  {Affleck}(1989)}]{Affleck-1989}%
  \BibitemOpen
  \bibfield  {author} {\bibinfo {author} {\bibfnamefont {J.~B.}\ \bibnamefont
  {Marston}}\ and\ \bibinfo {author} {\bibfnamefont {I.}~\bibnamefont
  {Affleck}},\ }\bibfield  {title} {\bibinfo {title} {{Large-$n$ limit of the
  Hubbard-Heisenberg model}},\ }\href
  {https://doi.org/10.1103/PhysRevB.39.11538} {\bibfield  {journal} {\bibinfo
  {journal} {Phys. Rev. B}\ }\textbf {\bibinfo {volume} {39}},\ \bibinfo
  {pages} {11538} (\bibinfo {year} {1989})}\BibitemShut {NoStop}%
\bibitem [{\citenamefont {Baskaran}\ \emph {et~al.}(1987)\citenamefont
  {Baskaran}, \citenamefont {Zou},\ and\ \citenamefont
  {Anderson}}]{Baskaran-1987}%
  \BibitemOpen
  \bibfield  {author} {\bibinfo {author} {\bibfnamefont {G.}~\bibnamefont
  {Baskaran}}, \bibinfo {author} {\bibfnamefont {Z.}~\bibnamefont {Zou}},\ and\
  \bibinfo {author} {\bibfnamefont {P.~W.}\ \bibnamefont {Anderson}},\
  }\bibfield  {title} {\bibinfo {title} {{The resonating valence bond state and
  high-T$_c$ superconductivity — A mean field theory}},\ }\href
  {https://doi.org/https://doi.org/10.1016/0038-1098(87)90642-9} {\bibfield
  {journal} {\bibinfo  {journal} {Solid State Commun.}\ }\textbf {\bibinfo
  {volume} {63}},\ \bibinfo {pages} {973} (\bibinfo {year} {1987})}\BibitemShut
  {NoStop}%
\bibitem [{\citenamefont {Baskaran}\ and\ \citenamefont
  {Anderson}(1988)}]{Baskaran-1988}%
  \BibitemOpen
  \bibfield  {author} {\bibinfo {author} {\bibfnamefont {G.}~\bibnamefont
  {Baskaran}}\ and\ \bibinfo {author} {\bibfnamefont {P.~W.}\ \bibnamefont
  {Anderson}},\ }\bibfield  {title} {\bibinfo {title} {{Gauge theory of
  high-temperature superconductors and strongly correlated Fermi systems}},\
  }\href {https://doi.org/10.1103/PhysRevB.37.580} {\bibfield  {journal}
  {\bibinfo  {journal} {Phys. Rev. B}\ }\textbf {\bibinfo {volume} {37}},\
  \bibinfo {pages} {580} (\bibinfo {year} {1988})}\BibitemShut {NoStop}%
\bibitem [{\citenamefont {Affleck}\ \emph {et~al.}(1988)\citenamefont
  {Affleck}, \citenamefont {Zou}, \citenamefont {Hsu},\ and\ \citenamefont
  {Anderson}}]{Affleck-1988b}%
  \BibitemOpen
  \bibfield  {author} {\bibinfo {author} {\bibfnamefont {I.}~\bibnamefont
  {Affleck}}, \bibinfo {author} {\bibfnamefont {Z.}~\bibnamefont {Zou}},
  \bibinfo {author} {\bibfnamefont {T.}~\bibnamefont {Hsu}},\ and\ \bibinfo
  {author} {\bibfnamefont {P.~W.}\ \bibnamefont {Anderson}},\ }\bibfield
  {title} {\bibinfo {title} {{SU(2) gauge symmetry of the large-$U$ limit of
  the Hubbard model}},\ }\href {https://doi.org/10.1103/PhysRevB.38.745}
  {\bibfield  {journal} {\bibinfo  {journal} {Phys. Rev. B}\ }\textbf {\bibinfo
  {volume} {38}},\ \bibinfo {pages} {745} (\bibinfo {year} {1988})}\BibitemShut
  {NoStop}%
\bibitem [{\citenamefont {Maity}\ \emph
  {et~al.}(2026{\natexlab{b}})\citenamefont {Maity}, \citenamefont {Iqbal},\
  and\ \citenamefont {Samajdar}}]{mis}%
  \BibitemOpen
  \bibfield  {author} {\bibinfo {author} {\bibfnamefont {A.}~\bibnamefont
  {Maity}}, \bibinfo {author} {\bibfnamefont {Y.}~\bibnamefont {Iqbal}},\ and\
  \bibinfo {author} {\bibfnamefont {R.}~\bibnamefont {Samajdar}},\ }\bibfield
  {title} {\bibinfo {title} {{Fermionic parton theory of Rydberg ${{\mathbb{Z}
  }}_{2}$ quantum spin liquids}},\ }\href
  {https://doi.org/10.1038/s41467-026-75245-9} {\bibfield  {journal} {\bibinfo
  {journal} {Nat. Commun.}\ }\textbf {\bibinfo {volume} {17}},\ \bibinfo
  {pages} {8588} (\bibinfo {year} {2026}{\natexlab{b}})}\BibitemShut {NoStop}%
\bibitem [{\citenamefont {Polyakov}(1977)}]{Polyakov-1977}%
  \BibitemOpen
  \bibfield  {author} {\bibinfo {author} {\bibfnamefont {A.~M.}\ \bibnamefont
  {Polyakov}},\ }\bibfield  {title} {\bibinfo {title} {{Quark confinement and
  topology of gauge theories}},\ }\href
  {https://doi.org/https://doi.org/10.1016/0550-3213(77)90086-4} {\bibfield
  {journal} {\bibinfo  {journal} {Nucl. Phys. B}\ }\textbf {\bibinfo {volume}
  {120}},\ \bibinfo {pages} {429} (\bibinfo {year} {1977})}\BibitemShut
  {NoStop}%
\bibitem [{\citenamefont {Müller}\ \emph {et~al.}(2024)\citenamefont
  {Müller}, \citenamefont {Kiese}, \citenamefont {Niggemann}, \citenamefont
  {Sbierski}, \citenamefont {Reuther}, \citenamefont {Trebst}, \citenamefont
  {Thomale},\ and\ \citenamefont {Iqbal}}]{Mueller-2024}%
  \BibitemOpen
  \bibfield  {author} {\bibinfo {author} {\bibfnamefont {T.}~\bibnamefont
  {Müller}}, \bibinfo {author} {\bibfnamefont {D.}~\bibnamefont {Kiese}},
  \bibinfo {author} {\bibfnamefont {N.}~\bibnamefont {Niggemann}}, \bibinfo
  {author} {\bibfnamefont {B.}~\bibnamefont {Sbierski}}, \bibinfo {author}
  {\bibfnamefont {J.}~\bibnamefont {Reuther}}, \bibinfo {author} {\bibfnamefont
  {S.}~\bibnamefont {Trebst}}, \bibinfo {author} {\bibfnamefont
  {R.}~\bibnamefont {Thomale}},\ and\ \bibinfo {author} {\bibfnamefont
  {Y.}~\bibnamefont {Iqbal}},\ }\bibfield  {title} {\bibinfo {title}
  {{Pseudo-fermion functional renormalization group for spin models}},\ }\href
  {https://doi.org/10.1088/1361-6633/ad208c} {\bibfield  {journal} {\bibinfo
  {journal} {Rep. Prog. Phys.}\ }\textbf {\bibinfo {volume} {87}},\ \bibinfo
  {pages} {036501} (\bibinfo {year} {2024})}\BibitemShut {NoStop}%
\bibitem [{\citenamefont {Hering}\ \emph {et~al.}(2019)\citenamefont {Hering},
  \citenamefont {Sonnenschein}, \citenamefont {Iqbal},\ and\ \citenamefont
  {Reuther}}]{Hering-2019}%
  \BibitemOpen
  \bibfield  {author} {\bibinfo {author} {\bibfnamefont {M.}~\bibnamefont
  {Hering}}, \bibinfo {author} {\bibfnamefont {J.}~\bibnamefont
  {Sonnenschein}}, \bibinfo {author} {\bibfnamefont {Y.}~\bibnamefont
  {Iqbal}},\ and\ \bibinfo {author} {\bibfnamefont {J.}~\bibnamefont
  {Reuther}},\ }\bibfield  {title} {\bibinfo {title} {{Characterization of
  quantum spin liquids and their spinon band structures via functional
  renormalization}},\ }\href {https://doi.org/10.1103/PhysRevB.99.100405}
  {\bibfield  {journal} {\bibinfo  {journal} {Phys. Rev. B}\ }\textbf {\bibinfo
  {volume} {99}},\ \bibinfo {pages} {100405} (\bibinfo {year}
  {2019})}\BibitemShut {NoStop}%
\bibitem [{\citenamefont {Hering}\ \emph {et~al.}(2022)\citenamefont {Hering},
  \citenamefont {Noculak}, \citenamefont {Ferrari}, \citenamefont {Iqbal},\
  and\ \citenamefont {Reuther}}]{Hering-2022}%
  \BibitemOpen
  \bibfield  {author} {\bibinfo {author} {\bibfnamefont {M.}~\bibnamefont
  {Hering}}, \bibinfo {author} {\bibfnamefont {V.}~\bibnamefont {Noculak}},
  \bibinfo {author} {\bibfnamefont {F.}~\bibnamefont {Ferrari}}, \bibinfo
  {author} {\bibfnamefont {Y.}~\bibnamefont {Iqbal}},\ and\ \bibinfo {author}
  {\bibfnamefont {J.}~\bibnamefont {Reuther}},\ }\bibfield  {title} {\bibinfo
  {title} {{Dimerization tendencies of the pyrochlore Heisenberg
  antiferromagnet: A functional renormalization group perspective}},\ }\href
  {https://doi.org/10.1103/PhysRevB.105.054426} {\bibfield  {journal} {\bibinfo
   {journal} {Phys. Rev. B}\ }\textbf {\bibinfo {volume} {105}},\ \bibinfo
  {pages} {054426} (\bibinfo {year} {2022})}\BibitemShut {NoStop}%
\bibitem [{\citenamefont {Hu}\ \emph {et~al.}(2015)\citenamefont {Hu},
  \citenamefont {Zhu}, \citenamefont {Zhang}, \citenamefont {Gong},
  \citenamefont {Becca},\ and\ \citenamefont {Sheng}}]{Hu-2015}%
  \BibitemOpen
  \bibfield  {author} {\bibinfo {author} {\bibfnamefont {W.-J.}\ \bibnamefont
  {Hu}}, \bibinfo {author} {\bibfnamefont {W.}~\bibnamefont {Zhu}}, \bibinfo
  {author} {\bibfnamefont {Y.}~\bibnamefont {Zhang}}, \bibinfo {author}
  {\bibfnamefont {S.}~\bibnamefont {Gong}}, \bibinfo {author} {\bibfnamefont
  {F.}~\bibnamefont {Becca}},\ and\ \bibinfo {author} {\bibfnamefont {D.~N.}\
  \bibnamefont {Sheng}},\ }\bibfield  {title} {\bibinfo {title} {{Variational
  Monte Carlo study of a chiral spin liquid in the extended Heisenberg model on
  the kagome lattice}},\ }\href {https://doi.org/10.1103/PhysRevB.91.041124}
  {\bibfield  {journal} {\bibinfo  {journal} {Phys. Rev. B}\ }\textbf {\bibinfo
  {volume} {91}},\ \bibinfo {pages} {041124} (\bibinfo {year}
  {2015})}\BibitemShut {NoStop}%
\bibitem [{\citenamefont {Song}\ \emph {et~al.}(2019)\citenamefont {Song},
  \citenamefont {Wang}, \citenamefont {Vishwanath},\ and\ \citenamefont
  {He}}]{Song-2019}%
  \BibitemOpen
  \bibfield  {author} {\bibinfo {author} {\bibfnamefont {X.-Y.}\ \bibnamefont
  {Song}}, \bibinfo {author} {\bibfnamefont {C.}~\bibnamefont {Wang}}, \bibinfo
  {author} {\bibfnamefont {A.}~\bibnamefont {Vishwanath}},\ and\ \bibinfo
  {author} {\bibfnamefont {Y.-C.}\ \bibnamefont {He}},\ }\bibfield  {title}
  {\bibinfo {title} {Unifying description of competing orders in
  two-dimensional quantum magnets},\ }\href
  {https://doi.org/10.1038/s41467-019-11727-3} {\bibfield  {journal} {\bibinfo
  {journal} {Nat. Commun.}\ }\textbf {\bibinfo {volume} {10}},\ \bibinfo
  {pages} {4254} (\bibinfo {year} {2019})}\BibitemShut {NoStop}%
\bibitem [{\citenamefont {Song}\ \emph {et~al.}(2020)\citenamefont {Song},
  \citenamefont {He}, \citenamefont {Vishwanath},\ and\ \citenamefont
  {Wang}}]{Song-2020}%
  \BibitemOpen
  \bibfield  {author} {\bibinfo {author} {\bibfnamefont {X.-Y.}\ \bibnamefont
  {Song}}, \bibinfo {author} {\bibfnamefont {Y.-C.}\ \bibnamefont {He}},
  \bibinfo {author} {\bibfnamefont {A.}~\bibnamefont {Vishwanath}},\ and\
  \bibinfo {author} {\bibfnamefont {C.}~\bibnamefont {Wang}},\ }\bibfield
  {title} {\bibinfo {title} {{From Spinon Band Topology to the Symmetry Quantum
  Numbers of Monopoles in Dirac Spin Liquids}},\ }\href
  {https://doi.org/10.1103/PhysRevX.10.011033} {\bibfield  {journal} {\bibinfo
  {journal} {Phys. Rev. X}\ }\textbf {\bibinfo {volume} {10}},\ \bibinfo
  {pages} {011033} (\bibinfo {year} {2020})}\BibitemShut {NoStop}%
\bibitem [{\citenamefont {Budaraju}\ \emph {et~al.}(2023)\citenamefont
  {Budaraju}, \citenamefont {Iqbal}, \citenamefont {Becca},\ and\ \citenamefont
  {Poilblanc}}]{Budaraju-2023}%
  \BibitemOpen
  \bibfield  {author} {\bibinfo {author} {\bibfnamefont {S.}~\bibnamefont
  {Budaraju}}, \bibinfo {author} {\bibfnamefont {Y.}~\bibnamefont {Iqbal}},
  \bibinfo {author} {\bibfnamefont {F.}~\bibnamefont {Becca}},\ and\ \bibinfo
  {author} {\bibfnamefont {D.}~\bibnamefont {Poilblanc}},\ }\bibfield  {title}
  {\bibinfo {title} {{Piercing the Dirac spin liquid: From a single monopole to
  chiral states}},\ }\href {https://doi.org/10.1103/PhysRevB.108.L201116}
  {\bibfield  {journal} {\bibinfo  {journal} {Phys. Rev. B}\ }\textbf {\bibinfo
  {volume} {108}},\ \bibinfo {pages} {L201116} (\bibinfo {year}
  {2023})}\BibitemShut {NoStop}%
\bibitem [{\citenamefont {Hermele}\ \emph {et~al.}(2008)\citenamefont
  {Hermele}, \citenamefont {Ran}, \citenamefont {Lee},\ and\ \citenamefont
  {Wen}}]{Hermele-2008}%
  \BibitemOpen
  \bibfield  {author} {\bibinfo {author} {\bibfnamefont {M.}~\bibnamefont
  {Hermele}}, \bibinfo {author} {\bibfnamefont {Y.}~\bibnamefont {Ran}},
  \bibinfo {author} {\bibfnamefont {P.~A.}\ \bibnamefont {Lee}},\ and\ \bibinfo
  {author} {\bibfnamefont {X.-G.}\ \bibnamefont {Wen}},\ }\bibfield  {title}
  {\bibinfo {title} {{Properties of an algebraic spin liquid on the kagome
  lattice}},\ }\href {https://doi.org/10.1103/PhysRevB.77.224413} {\bibfield
  {journal} {\bibinfo  {journal} {Phys. Rev. B}\ }\textbf {\bibinfo {volume}
  {77}},\ \bibinfo {pages} {224413} (\bibinfo {year} {2008})}\BibitemShut
  {NoStop}%
\bibitem [{\citenamefont {Gemb\'e}\ \emph {et~al.}(2024)\citenamefont
  {Gemb\'e}, \citenamefont {Gresista}, \citenamefont {Schmidt}, \citenamefont
  {Hickey}, \citenamefont {Iqbal},\ and\ \citenamefont {Trebst}}]{Gembe-2024}%
  \BibitemOpen
  \bibfield  {author} {\bibinfo {author} {\bibfnamefont {M.}~\bibnamefont
  {Gemb\'e}}, \bibinfo {author} {\bibfnamefont {L.}~\bibnamefont {Gresista}},
  \bibinfo {author} {\bibfnamefont {H.-J.}\ \bibnamefont {Schmidt}}, \bibinfo
  {author} {\bibfnamefont {C.}~\bibnamefont {Hickey}}, \bibinfo {author}
  {\bibfnamefont {Y.}~\bibnamefont {Iqbal}},\ and\ \bibinfo {author}
  {\bibfnamefont {S.}~\bibnamefont {Trebst}},\ }\bibfield  {title} {\bibinfo
  {title} {{Noncoplanar orders and quantum disordered states in maple-leaf
  antiferromagnets}},\ }\href {https://doi.org/10.1103/PhysRevB.110.085151}
  {\bibfield  {journal} {\bibinfo  {journal} {Phys. Rev. B}\ }\textbf {\bibinfo
  {volume} {110}},\ \bibinfo {pages} {085151} (\bibinfo {year}
  {2024})}\BibitemShut {NoStop}%
\end{thebibliography}%
\end{document}